\documentclass{aa}

\usepackage{graphicx}
\usepackage{natbib}
\usepackage{lscape}
\usepackage{color}

\usepackage{txfonts}
\begin{document}

   \title{Searching for signatures of planet formation in stars with circumstellar debris discs
    \thanks{
     Based on observations collected  at the Centro Astron\'omico Hispano
     Alem\'an (CAHA) at Calar Alto, operated jointly by the Max-Planck Institut
     f\"{u}r Astronomie and the Instituto de Astrof\'isica de Andaluc\'ia (CSIC);
     observations made with the Italian Telescopio Nazionale Galileo (TNG) operated
     on the island of La Palma by the Fundaci\'on Galileo Galilei of the INAF
     (Istituto Nazionale di Astrofisica);
     observations made with the Nordic Optical Telescope, operated
     on the island of La Palma jointly by Denmark, Finland, Iceland,
     Norway, and Sweden, in the Spanish Observatorio del Roque de los
     Muchachos of the Instituto de Astrofisica de Canarias; observations
     made at the Mercator Telescope, operated on the island of La Palma
     by the Flemish Community; and
     data obtained from the ESO Science Archive Facility.}\fnmsep
   \thanks{
     Tables~\ref{parameters_table},~\ref{lista_unificada_lines},
     and ~\ref{abundance_table_full} are only available in the electronic version of the paper or
     at the CDS via anonymous ftp to cdsarc.u-strasbg.fr (130.79.128.5)
     or via http://cdsweb.u-strasbg.fr/cgi-bin/qcat?J/A+A/}
     }    


   \author{J. Maldonado 
          \inst{1}
          \and  C. Eiroa
          \inst{2}
          \and  E. Villaver
          \inst{2}  
          \and  B. Montesinos
          \inst{3}  
          \and  A. Mora
          \inst{4}
          }

             \institute{INAF - Osservatorio Astronomico di Palermo,
              Piazza Parlamento 1, 90134 Palermo, Italy
             \and
              Universidad Aut\'onoma de Madrid, Dpto. F\'isica Te\'orica, M\'odulo 15,
              Facultad de Ciencias, Campus de Cantoblanco, 28049 Madrid, Spain
             \and
              Department of Astrophysics,  Centro de Astrobiolog\'ia (CAB, CSIC-INTA), ESAC Campus,  
              PO Box 78, 28691 Villanueva de la Ca\~{n}ada, Madrid, Spain
             \and
              ESA-ESAC Gaia SOC, PO Box 78, 28691 Villanueva de la Ca\~{n}ada, Madrid, Spain 
             }

   \offprints{J. Maldonado \\ \email{jmaldonado@astropa.inaf.it}}
   \date{Received ...; Accepted ....}

 
  \abstract
   {Tentative correlations between the presence of dusty circumstellar debris
    discs and low-mass planets have been recently presented. In parallel, detailed
    chemical abundance studies have reported different trends between samples of planet and
    non-planet hosts. Whether these chemical differences are indeed related to the presence
    of planets is still strongly debated.
   }
   {
    We aim to test whether solar-type stars with debris discs 
    show any chemical peculiarity that could be related
    to the planet formation process.
   }
   {
    We determine in a homogeneous way the metallicity, [Fe/H], and abundances of individual elements
    of a sample of 251 stars including stars with known debris discs, stars harbouring simultaneously
    debris discs and planets, stars hosting exclusively planets, and a comparison sample
    of stars without known discs nor planets. High resolution \'echelle spectra 
    (R$\sim$ 57000) from 2-3 m class telescopes are used. Our methodology includes
    the calculation of the fundamental stellar parameters (T$_{\rm eff}$, $\log g$,
    microturbulent velocity, and metallicity) by applying the iron ionisation and
    equilibrium conditions to several isolated Fe~{\sc i} and Fe~{\sc ii} lines,
    as well as, individual abundances of 
    C, O, Na,
    Mg, Al, Si, S, Ca, Sc, Sc,
    Ti, V, Cr, Mn, Co, Ni,
    Cu, and Zn.
    } 
   {
    No significant differences have been found in metallicity, individual abundances
    or abundance-condensation temperature trends between stars with debris discs and stars with neither debris
    nor planets. Stars with debris discs and planets have the same metallicity behaviour as stars
    hosting planets, and they also show a similar  <[X/Fe]>-T$_{\rm C}$ trend.
    Different behaviour in the  <[X/Fe]>-T$_{\rm C}$ trends is found between the 
    samples of stars without planets and the samples of planet hosts. 
    In particular, when considering only refractory elements, negative slopes
    are shown in cool giant planet hosts, whilst positive ones are shown 
    in stars hosting low-mass planets.
    The statistical significance of the derived slopes is however low,
    a fact that can be due to the wide range of stellar parameters of our samples. 
    Stars hosting exclusively close-in giant planets behave in a different
    way, showing higher metallicities and positive <[X/Fe]>-T$_{\rm C}$ slope.
    A search for correlations between the <[X/Fe]>-T$_{\rm C}$ slopes and the stellar properties
    reveals a moderate but significant correlation with the stellar radius and 
    as well as a weak correlation with the stellar age, which remain even if
    Galactic chemical evolution effects are considered.
    No correlation between the  <[X/Fe]>-T$_{\rm C}$ slopes and the disc/planet properties
    are found.
   }
    {
      The fact that stars with debris discs
     and stars with low-mass planets do not show neither metal enhancement
     nor a different  <[X/Fe]>-T$_{\rm C}$ trend
     might indicate a correlation between the presence of debris discs and
     the presence of low-mass planets. 
     We extend results from previous works based mainly in solar analogues which
     reported differences in the <[X/Fe]>-T$_{\rm C}$ trends between planet hosts and 
     non hosts to a wider range of parameters. However, these differences
     tend to be present only when the star hosts a cool distant planet
     and not in stars hosting exclusively low-mass planets. 
     The interpretation of these differences as 
     a signature of planetary formation should
     be considered with caution since
     moderate correlations between the T$_{\rm C}$ slopes with the stellar radius and
     the stellar age are found, suggesting that an evolutionary effect might be at work.
     }

  \keywords{techniques: spectroscopic - stars: abundances -stars: late-type -stars: planetary systems}

  \maketitle

\section{Introduction}
\label{introduccion}

  Main sequence stars are often surrounded not only by one or several planets but also by
  faint dusty circumstellar discs usually known as debris discs
  \citep[e.g.][]{1993prpl.conf.1253B}.
  The evidence of  debris discs comes from the presence of 
  flux excesses over the stellar photospheric emission at IR wavelengths,
  thought to arise from dust particles continuously produced by the collision,
  disruption and/or sublimation of planetesimals 
  \citep[for reviews, see e.g.][]{2013pss3.book..431M,2014arXiv1401.0743M}.  
  Our own Solar System is an example of a planetary system which also harbours
  a debris disc produced by collisions of minor
  bodies like asteroids, comets, and Kuiper belt objects
  \citep{2009and..book...53J}.

  Initially discovered around early-type 
  stars \citep[e.g. Vega,][]{1984ApJ...278L..23A}, subsequent studies
  have shown that debris discs are quite common. In fact, it has been
  established that more than  
  33\% of the A-type stars show IR excess at 70 $\mu$m  \citep{2006ApJ...653..675S},
  whilst recent {\it Herschel} data show that the frequency of debris discs 
  around mature solar-type stars is $\sim$ 20\%  \citep{2013A&A...555A..11E}. 
  Although rare, several M dwarfs are also known to harbour
  a debris discs \citep[e.g.][]{2012A&A...548A..86L}. 
  Further, some evolved stars are also known to be associated with debris discs 
  \citep[e.g.][]{2014MNRAS.437.3288B}.
  In addition, 
  observations of polluted white dwarfs with heavy elements in their atmospheres
  are  also thought to be related with the presence of  planetesimals belts 
  \citep[e.g.][]{2012MNRAS.424..333G}. 
  This observational evidence  reveals 
  that the presence of planetesimals is ubiquitous. 

  Planetesimals constitute the raw material from which planets are formed
  and therefore a correlation between discs and planets should be expected.
  Indeed, debris discs and planets are known to coexist in around 32 stars.
  However, the so-long sought for relationship between debris discs and planets 
  still remains elusive. 
  First, the incidence of debris discs does not seem to be higher
  around planet hosts \citep{2009ApJ...700L..73K}.
  In addition, no clear correlation between the presence of discs
  and the stellar properties has been found
  \citep{2005ApJ...622.1160B,2006A&A...452..921C,2006MNRAS.366..283G,2007ApJ...658.1312M,  
  2009ApJ...705.1226B,2009ApJ...700L..73K}
  although \citet[][hereafter MA12]{2012A&A...541A..40M} suggest the presence
  of a ``deficit'' of stars with discs at low metallicities
  ([Fe/H] $\le$ -0.10) when compared to stars
  without detected discs.
  The lack of a relation between the presence of debris discs and planets might
  suggest the existence of a mechanism that excludes the presence of both at the same time.
  \cite{2007ApJ...658.1312M,2015arXiv150103813M}  argued that dynamically active gas giant planets may clear out
  part of an initially massive debris disc by grinding or ejecting away planetesimals,
  a result also predicted by simulations \citep{2011A&A...530A..62R,2012A&A...541A..11R}. 
  Along these lines, a hint of lower fractional luminosity of the dust values,
  L$_{\rm dust}$/L$_{\star}$,
  in systems
  with high eccentricity planets was found in MA12.

  Most of our current knowledge of the disc-planet connection is still
  based on detections of  gas-giant planets.
  This situation is rapidly evolving, as a new population
  of low-mass planets (M$_{\rm p}$$\sin i$ $\lesssim$ 30M$_{\oplus}$) is being discovered. 
  Recent results from microlensing surveys \citep{2012Natur.481..167C}, as well as
  long term monitoring programmes from the ground \citep[e.g.][]{2011arXiv1109.2497M},
  seems to suggest that like planetesimals, low-mass planets may be abundant.
  As stars with debris discs, stars hosting low-mass planets   
  do not show the metal-rich signature seen in gas-giant main sequence planet
  hosts \citep{2010ApJ...720.1290G,2011arXiv1109.2497M,2011A&A...533A.141S,2012Natur.486..375B}.
  From the theoretical point of view,  
  a strong correlation between the presence of cold
  dusty discs and low-mass planets is predicted \citep{2011A&A...530A..62R,2012A&A...541A..11R}. 

  Significant improvements have also been made in the detection of debris discs, specially
  around late-type stars thanks to the unprecedented sensitivity provided by the
  {\it Herschel Space Observatory}. 
  In particular, \cite{2012MNRAS.424.1206W} suggested a possible correlation
  between the presence of debris discs and low-mass planets, based on a sample of the nearest
  60 G-type stars.
  Further analysis of the {\it Herschel} data by \cite{2014A&A...565A..15M} 
  in a sample of 37 solar-type exoplanet host reveals a correlation between 
  the presence of dust, low-mass planets, and low stellar metallicities.
  However, the detailed statistical analysis of 204 FGK stars by  
  \cite{2015arXiv150103813M} does not find evidence of debris
  discs being more common around low-mass planet hosts, although the authors
  caution about possible contamination of the control sample by 
  (possible) undetected low-mass planets and relatively small sample sizes. 

  In parallel, significant efforts have been made to identify 
  which stellar properties have a larger influence (and how) in planet formation.  
  Detailed chemical abundances of planet hosts, specially in solar analogues, have
  suggested different trends on abundance-condensation temperature  
  \citep[e.g.][]{2009ApJ...704L..66M,2009A&A...508L..17R,2010A&A...521A..33R,
   2014A&A...561A...7R,2010MNRAS.407..314G,2011MNRAS.416L..80G},
  although their interpretation as a chemical fingerprint of the planet
  formation process has been questioned, and other works rather point towards chemical
  evolution effects \citep{2010ApJ...720.1592G,2013A&A...552A...6G,2011ApJ...732...55S}
  or an inner Galactic origin of the planet hosts \citep[e.g.][]{2014A&A...564L..15A}
  as their possible causes.
  
  In this paper a detailed analysis of the chemical abundances
  of a large sample of stars known to harbour debris discs, and a sample
  of stars hosting simultaneously debris discs and planets is presented.
  We aim to test whether these stars show any chemical peculiarity, and
  to unravel their origin (disc, planet, or other).
  This works follows our previous chemical
  analysis of stars with debris discs in MA12
  where we focused exclusively on metallicities but now we extend it to the
  individual abundances of other 17 elements besides iron,
  including an analysis of possible trends between the abundances and
  the elemental condensation temperature. 
  The paper is organised as follows:
  Section~\ref{secction_observations} describes the stellar samples analysed in this work,
  the spectroscopic observations and how stellar parameters and abundances are
  obtained.
  The distribution of abundances are presented in
  Section~\ref{section_analysis}. 
  The results are
  discussed at length in Section~\ref{seccion_discussion}.
  Our conclusions follow in Section~\ref{conclusions} .

\section{Observations}
\label{secction_observations}
\subsection{The stellar sample}
\label{stellar_sample}

  A sample of solar-type stars with known debris discs, SWDs, was built using as
  a reference the stars listed in MA12. It contains 107 solar-type stars
  with debris discs discovered by the {\it IRAS}, {\it ISO}, and {\it Spitzer} telescopes,
  most of them detected at MIPS 70 $\mu$m, with fractional dust luminosities,
  L$_{\rm dust}$/L$_{\star}$, of the order of 10$^{-5}$ and higher \citep{Trilling08}.
  From the MA12 list we retain for study those stars for which we have been able to
  obtain high-resolution spectra (See Section~\ref{spectroscopic_observations}).
  To the list we have added six new stars,
  namely HIP 17420, HIP 29271, HIP 51459, HIP 71181, HIP 73100, and
  HIP 92043,
  recently identified as new excess
  sources by the DUNES\footnote{http://www.mpia-hd.mpg.de/DUNES/}
  {\it Herschel Space Observatory} OTKP  \citep{2010A&A...518L.131E,2013A&A...555A..11E}.
  The total number of stars in this sample amounts to 68:
  19 F-type stars, 29 G-type stars, and 20 K-type stars.  

  The {\it comparison} sample, SWODs (stars without discs), is also taken from MA12. It contains
  145 stars (we have spectra for 86 of them) in which IR-excesses were not found at 24 and 70 $\mu$m by 
  {\it Spitzer}. 
  Since {\it Spitzer} is limited up to fractional luminosities
  of L$_{\rm dust}$/L$_{\star}$ $\ge$ 10$^{-5}$ , we cannot rule out the possibility
  that some of these stars have fainter discs. Indeed, three out of the new
  SWD stars were listed in MA12 as SWODs.
  In addition, we have complemented the SWOD sample with 32 stars from the DUNES survey
  showing no-IR excess at any of the {\it Herschel}-PACS wavelengths. In this case,
  the higher sensitivity of {\it Herschel} with respect to {\it Spitzer} allows us
  to rule out the presence of discs brighter than $\sim$ 10$^{-6}$
  \citep{2013A&A...555A..11E}.
  The total number of stars included in the SWOD sample amounts to 119:
  22 F-type stars, 68 G-type stars, and 29 K-type stars.

  To elucidate the possible effects that planet formation might have, 
  planet-hosting stars have not been included
  neither in the SWD nor the SWOD sample.
  To identify these stars, the 
  Extrasolar Planets Encyclopedia\footnote{http://exoplanet.eu/} and the
  Exoplanet Orbit Database\footnote{http://exoplanets.org/}
  have been carefully checked. Nevertheless, we can not rule out the presence
  of non-detected planets around some of the stars, 
  specially low-mass planets, which are expected to be
  common around solar-type stars 
  \citep{2011arXiv1109.2497M}.  

\subsection{Stars with known debris discs and planets}
\label{debris_and_planets}

 In a similar way the sample of stars known to host simultaneously
 a dusty debris discs and at least one planet, SWDPs, has been updated
 with respect to MA12.
 Three new stars with discs and planets have been added: 
 HIP 27887  and HIP 109378  which are new excess sources
 identified by {\it Herschel} \citep{2013A&A...555A..11E,2014A&A...565A..15M};
 HIP 80902 has been added to the list since the suggested planet around this 
 star  has been recently confirmed
 \citep{2012A&A...545A..55B}.
 The substellar companion around HIP 107350, has an
 estimated minimum 
 mass of 16 M$_{\rm Jup}$ \citep{2007ApJ...654..570L} and therefore
 has not been included in the SWDP sample. 

 Two evolved stars with planets are known to show IR-excess,
 HIP 58576 ({\sc Hipparcos}' spectral-type K0-IV),
 and  HIP 75458 (K2 III). While the position of HIP 58576
 in a colour-magnitude diagram suggests it is rather
 a main-sequence star,
 HIP 75458 is clearly a giant.
 \cite{IWSSL}
 found that when applying an
 homogeneous procedure nearby main sequence and giant stars
 show a common metallicity scale. However,
 tidal interactions  in the star-planet system as the star evolves off the MS, can lead to
 variations in the planetary orbits and to the engulfment of close-in planets
 \citep{2009ApJ...705L..81V,2014ApJ...794....3V}, a process which can alter the
 photospheric abundances of the host star in a short time scale when the star is not fully
 convective yet. 
 We therefore exclude HIP 75458 star from the chemical
 analysis that follows.
 The final number of
 SWDPs analyzed is 31:
 4 F-type stars, 18 G-type stars, and 9 K-type stars.

 For completeness, we also include in this work those stars known to host
 at least one planet but not debris disc\footnote{
 As listed at September 18, 2014,
 in the Extrasolar Planets Encyclopedia},
 hereafter SWPs. Since the properties of these
 stars (in particular the metallicity) are the subject of a significant
 large number of studies, we do not try to get additional spectra, showing
 only the data we used in MA12.
  The number of stars included in the SWP sample amounts to 32:
  17 stars hosting exclusively cool Jupiters, 5 stars harbouring hot Jupiters,
  7 stars hosting low-mass planets, and 3 stars with both low-mass and gas 
  giant planets.
 
\subsection{Spectroscopic observations}
\label{spectroscopic_observations}

 The high-resolution spectra used in this work comes from several
 spectrographs and telescopes and have already been used 
 in some of our previous works
 \citep{2010A&A...521A..12M,2012A&A...541A..40M,2013A&A...554A..84M,2010A&A...520A..79M} 
 which can be consulted for details concerning the observing
 runs and the reduction procedure. 
 Summarising, the data were taken with the following instruments: 
 i) FOCES \citep{foces}  at the  2.2 meter telescope of the Calar Alto
 observatory (CAHA, Almer\'ia, Spain);
 ii) SARG \citep{sarg} at the
 Telescopio Nazionale Galileo (TNG, 3.58 m),
 La Palma (Canary Islands,  Spain);
 iii) FIES \citep{1999anot.conf...71F} at the
 Nordic Optical Telescope  (NOT, 2.56 m), La Palma;
 and iv) HERMES \citep{2011A&A...526A..69R} at the MERCATOR telescope
 (1.2 m), also in La Palma. 
 We also used additional spectra  from  the  public
 library  ``S$^{4}$N'' \citep{s4n}, which  contains spectra  taken with
 the 2dcoud\'{e}  spectrograph at McDonald Observatory and the FEROS
 instrument at  the ESO  1.52 m  telescope in La  Silla;  from the
 ESO/ST-ECF  Science Archive Facility  \footnote{http://archive.eso.org/cms/},
 as well as the pipeline processed FEROS and HARPS data archive
 \footnote{http://archive.eso.org/wdb/wdb/eso/repro/form}. 
 The spectral range
 and resolving power of each of the spectrographs
 are listed in Table~\ref{tabla_espectrografos}.
 Further details concerning the use of ESO Archive are given in
 Appendix~\ref{apendiceA}.  

 Ideally  all  our
 targets  should have  been observed  with the  same  spectrograph
 using the same configuration. 
 Furthermore, the fact that the sample
 considered here spans a large range of stellar parameters
 (e.g. $\sim$ 2000 K in T$_{\rm eff}$) prevent us from performing a
 differential analysis.
 Nevertheless, all the spectra used in this work
 have a similar resolution
 (with the exception of HARPS which provides a better one),
 high signal-to-noise ratio
 (median value $\sim$ 140 at 6050\AA) 
 and cover a wide spectral-range with 
 enough lines to provide
 a high-quality abundance determination, enough for the purposes
 of this work. 


\begin{table}
\centering
\caption{
Properties of the different spectrographs used in this work.
}
\label{tabla_espectrografos}
\begin{scriptsize}
\begin{tabular}{lccc}
\hline\noalign{\smallskip}
Spectrograph &  Spectral range (\AA)  & Resolving power  & $N$ stars \\
\hline 
FOCES        &  3470-10700            &  57000           & 58 \\
SARG         &  5500-10100            &  57000           & 10 \\
FIES         &  3640-7360             &  67000           & 20 \\
HERMES       &  3800-9000             &  85000           & 37 \\
FEROS        &  3500-9200             &  42000           & 56 \\
McDonald     &  3400-10900            &  60000           & 48 \\
HARPS        &  3780-6910             & 115000           & 22 \\
\hline
\end{tabular}
\end{scriptsize}
\end{table}

\subsection{Stellar parameters}
\label{stellar_parameters}

  Basic stellar parameters T$_{\rm eff}$, $\log g$, microturbulent
  velocity $\xi_{\rm t}$, and [Fe/H] are determined using the code
  {\sc TGVIT}\footnote{http://optik2.mtk.nao.ac.jp/\textasciitilde{}takeda/tgv/}
  \citep{2005PASJ...57...27T}, which implements the iron ionisation
  and excitation equilibrium  conditions, a methodology
  which has been proved successful when applied to solar-like stars,
  spectral types F5/K2.

  Iron abundances are computed for a well-defined
  set of 302 {\rm Fe~{I}} and 28 {\rm Fe~{II}} lines.
  {\sc TGVIT} iteratively modifies the
  basic stellar parameters  searching the global minimum of the function
  \citep{2002PASJ...54..451T}:

  \begin{equation}
  \centering 
  D^2 = (\sigma^{2}_{1} + c_{1}\sigma^{2}_{2}) + c_{2}(\left\langle A_{1} \right\rangle -
         \left\langle A_{2} \right\rangle + c_{3})^2
  \end{equation}
  
  \noindent where $A_{1}$ and $A_{2}$ are the mean iron abundances computed using {\rm Fe~{I}} and {\rm Fe~{II}}
  lines respectively, 
  $\sigma_{1}$, $\sigma_{2}$ the corresponding standard deviations, and
  $c_{i}$ are weighting coefficients than the user can modify. 
  Forcing a minimum in $\sigma_{1}$ is equivalent to   
  searching for no correlation between the {\rm Fe~{I}} abundances with either the excitation potential
  or the reduced EW. 
  The surface gravity is obtained  by forcing  $A_{1}$ and $A_{2}$ 
  to be the same.
  Since {\rm Fe~{II}} lines are significantly less abundant than the {\rm Fe~{I}} lines
  in the spectra
  of late-type stars, the weighting coefficients
  $c_{1}$ and $c_{3}$ were set to zero.

  The line list as well as the adopted parameters (excitation potential,
  $\log (gf)$ values, solar EWs) can be found on Y. Takeda's web page. This
  code makes use of ATLAS9, plane-parallel, LTE atmosphere models
  \citep{1993KurCD..13.....K}. The assumed solar Fe abundance
  is A$_{\odot}$ = 7.50, as in \cite{2005PASJ...57...27T}.
  Uncertainties in the stellar parameters
  are computed by progressively changing each stellar
  parameter from the converged solution
  to a value in which any of the aforementioned conditions 
  (excitation equilibrium, match of the curve of grow, ionisation
  equilibrium)  are no longer
  fulfilled \citep[see for details][Section~5.2]{2002PASJ...54..451T}.
  Uncertainties in the iron abundances are computed by
  propagating the errors in T$_{\rm eff}$,  $\log g$, and $\xi_{t}$.
  As discussed in \cite{2002PASJ...54..451T,2002PASJ...54.1041T}
  this procedure only evaluates ``statistical''
  errors, since 
  other systematic sources of uncertainties, such as the choice of model atmosphere,
  the adopted atomic parameters, or the list lines used, are
  not taken into account. 

  In order to avoid errors due to uncertainties in the damping parameters,
  only lines with EWs $<$ 120 m\AA \space were considered \citep[e.g.][]{2008PASJ...60..781T}.
  Stellar EWs are measured using the automatic code {\sc ARES} \citep{2007A&A...469..783S},
  adjusting the {\it reject} parameter according to the S/N ratio of the spectra 
  as described in \cite{2008A&A...487..373S}.
  The estimated stellar parameters and iron abundances are  given in
  Table~\ref{parameters_table}.

\begin{table*}
\centering
\caption{
 Spectroscopic parameters with uncertainties
 for the stars measured in this work. Columns 7 and 9
 give the mean iron abundance derived from Fe~{\sc I} and Fe~{\sc  II} lines,
 respectively, while columns 8 and 10 give the corresponding number of lines.
 The rest of the columns are self-explanatory. Only the first five lines are
 shown here; the full version of the table is available on line.
 }
\label{parameters_table}
\begin{tabular}{llcccccccccl}
\hline\noalign{\smallskip}
HIP &  HD &  T$_{\rm eff}$ &  $\log g$      & $\xi_{t}$     & [Fe/H] &  $\left\langle A({\rm Fe}~{\rm I}) \right\rangle$ & n$_{\rm I}$ &
 $\left\langle A({\rm Fe}~{\rm II}) \right\rangle$  & n$_{\rm II}$ &  Spec.$^{\dag}$ \\
    &     &   (K)   & (${\rm cm s^{-2}}$)  & (${\rm km s^{-1}}$) & dex  &      &      &      &      &       \\
 (1)& (2) &  (3)    & (4)                  & (5)                 & (6)  & (7)  & (8)  & (9)  & (10) & (11)  \\
\hline\noalign{\smallskip}
\multicolumn{2}{c}{Sun} & 5784 $\pm$ 15 & 4.51 $\pm$ 0.03 & 1.01 $\pm$ 0.09 &  0.02 $\pm$ 0.01  & 7.52 $\pm$ 0.02 & 253 & 7.52 $\pm$ 0.02 & 26 & 5 \\
\hline\noalign{\smallskip}
\multicolumn{11}{c}{Stars with known debris discs.}\\
\hline\noalign{\smallskip}
 171	 &	224930	&	5354	$\pm$	15	&	4.32	$\pm$	0.04	&	0.74	$\pm$	0.12	&	-0.83	$\pm$	0.01	&	6.67	$\pm$	0.02	&	208	&	6.67	$\pm$	0.02	&	18	&	5	\\
490	 &	105	&	5967	$\pm$	35	&	4.52	$\pm$	0.08	&	1.76	$\pm$	0.26	&	-0.16	$\pm$	0.03	&	7.35	$\pm$	0.04	&	151	&	7.34	$\pm$	0.05	&	19	&	7	\\
544	 &	166	&	5584	$\pm$	20	&	4.73	$\pm$	0.05	&	1.22	$\pm$	0.16	&	0.15	$\pm$	0.02	&	7.65	$\pm$	0.03	&	256	&	7.65	$\pm$	0.03	&	21	&	5	\\
1598	 &	1562	&	5768	$\pm$	20	&	4.56	$\pm$	0.04	&	1.12	$\pm$	0.13	&	-0.27	$\pm$	0.02	&	7.23	$\pm$	0.02	&	205	&	7.23	$\pm$	0.03	&	17	&	1	\\
1599	 &	1581	&	5877	$\pm$	20	&	4.25	$\pm$	0.04	&	1.13	$\pm$	0.14	&	-0.24	$\pm$	0.02	&	7.26	$\pm$	0.02	&	221	&	7.26	$\pm$	0.03	&	23	&	6	\\
\noalign{\smallskip}\hline\noalign{\smallskip}
\end{tabular}
\tablefoot{$^{\dag}$Spectrograph: {\bf(1)} CAHA/FOCES; {\bf(2)} TNG/SARG; {\bf(3)} NOT/FIES; {\bf(4)} MERCATOR/HERMES; {\bf(5)} S$^{4}$N-McD;
{\bf(6)} S$^{4}$N-FEROS; {\bf(7)} ESO/FEROS; {\bf(8)} ESO/HARPS}
\end{table*}

\subsection{Photometric parameters and comparison with previous works}
\label{previous_works}

\begin{figure}
\centering
\includegraphics[angle=270,scale=0.45]{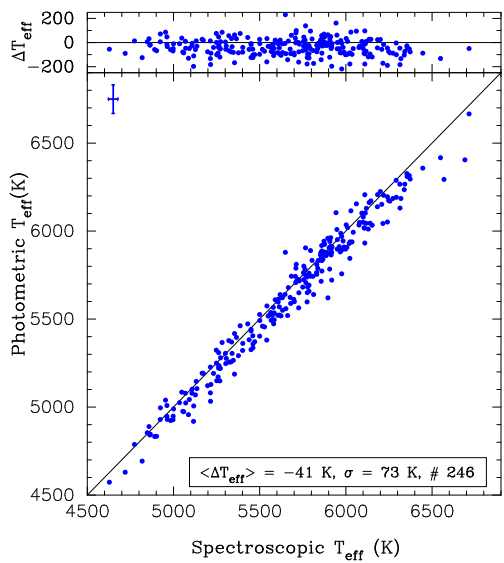}
\caption{
Comparison between our spectroscopically derived T$_{\rm eff}$
and those obtained from $(B-V)$ colours. The upper panel shows the
differences between the photometric and the spectroscopic values.
Mean uncertainties in the derived temperatures are also shown.
}
\label{temperaturas_fotometricas}
\end{figure}

 Photometric effective temperatures are derived from the 
 {\sc Hipparcos} $(B-V)$ colours \citep{1997ESASP1200.....P} by using
 the calibration provided by \citet[][Table~4]{2010A&A...512A..54C}.
 Since all our targets are nearby (all but two within 80 pc) colours have
 not been de-reddened. 
 The comparison between the photometric derived temperatures and the
 spectroscopic ones is illustrated in Figure~\ref{temperaturas_fotometricas}.
 We note that 
 the spectroscopic estimates tend to be 
 slightly larger than the photometric temperatures. Nevertheless, the mean
 value of $\Delta$T$_{\rm eff}$ $=$ T$_{\rm eff}^{\rm phot}$
 - T$_{\rm eff}^{\rm spec}$
 is small, only -41 K, with an RMS standard
 deviation of 73 K. A similar trend was found when applying this relationship
 to a sample of evolved (subgiant and red giant) stars
 \citep{2013A&A...554A..84M}.


  Evolutionary values of gravities are computed from 
  {\sc Hipparcos} $V$ magnitudes and the revised parallaxes provided
  by \cite{Leeuwen}.  L. Girardi's code 
  {\sc PARAM}\footnote{http://stev.oapd.inaf.it/cgi-bin/param}
  \citep{2006A&A...458..609D} has been used together with the new {\it PARSEC}
  isochrones from \cite{2012MNRAS.427..127B}.
  The code also estimates the stellar evolutionary parameters of age, mass, and radius 
  of the star.
  Our derived spectroscopic T$_{\rm eff}$ and metallicities are used as inputs
  for {\sc PARAM}.


\begin{figure}
\centering
\includegraphics[angle=270,scale=0.55]{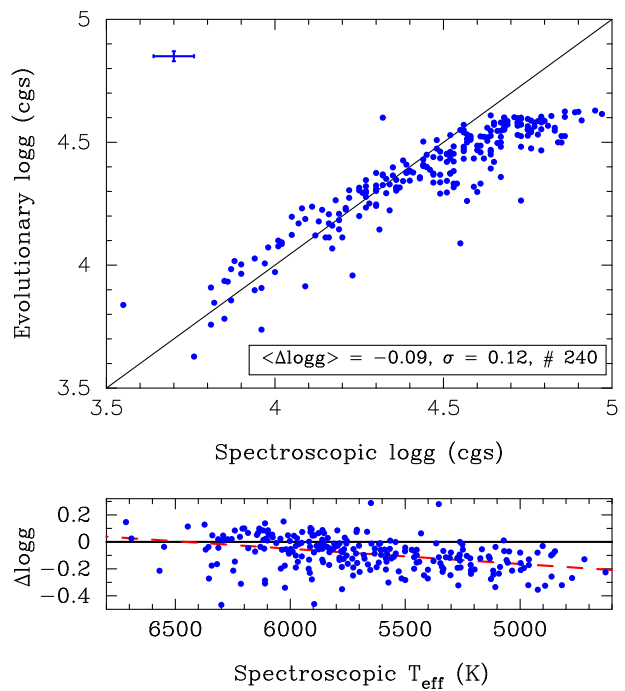}
\caption{
Top panel: comparison between our spectroscopically derived $\log g$ values and
$\log g$ estimates based on {\sc Hipparcos} parallaxes. 
Mean uncertainties in $\log g$ values are shown.
Bottom panel: differences in $\log g$ (defined as
evolutionary - spectroscopic) as a function of the spectroscopic
T$_{\rm eff}$. A linear fit is shown (dashed line).
}
\label{hipparcos_parallaxes}
\end{figure}

  The comparison between the spectroscopic and evolutionary $\log g$ values is shown
  in Figure~\ref{hipparcos_parallaxes}.
  Although the differences are low, mean value only -0.09 (cgs) with a RMS deviation
  of only 0.12 (cgs), it is clear from the figure that evolutionary
  and spectroscopic values do not always compare well, in particular for
  high spectroscopic $\log g$ values.
  This discrepancy 
  has already been discussed by several authors
  \citep[e.g.][and references therein]{2007ApJ...664.1190S,2012ApJ...757..161T,2013A&A...555A.150T}.
  The dependence of $\Delta\log g$ $=$ $\log g_{\rm spec}$ - $\log g_{\rm evol}$ is
  explored in the bottom panel of  Figure~\ref{hipparcos_parallaxes}. Albeit a significant scatter, a
  trend with the effective temperature can be easily recognised.
  To our knowledge, the origin of this discrepancy remains unknown.
  Several explanations have been put forward like departures from
  LTE or granulation and activity effects, but it might as well
  have an origin related to the relatively small number of 
  Fe~{\sc ii} lines present in the spectra of cool dwarfs
  \citep[e.g.][and references therein]{2013A&A...555A.150T}.

 The discrepancy between $\log g_{\rm spec}$ and $\log g_{\rm evol}$
 should not affect the other stellar parameters.
 Temperatures and metallicities derived using the ionisation and
 excitation equilibrium of iron have been shown to be mostly
 independent of the adopted surface gravity
 \citep{2012ApJ...757..161T}.
 Abundances derived from neutral lines are mostly independent of the surface gravity
 whilst abundances from single ionised atoms are known to scale with gravity 
 \citep{2008oasp.book.....G}.
 We therefore do not expect surface gravity to introduce
 significant effects on the computation of individual chemical abundances
 from non-ionised species. 
 In this line, \cite{2013A&A...558A.106M}  computed chemical
 abundances of 90 stars with transiting exoplanets using spectroscopic $\log g$
 values and $\log g$ estimates derived using the stellar density
 determined from the light curve. They found that only the abundances from ionised
 species are significantly affected.

\begin{figure}
\centering
\includegraphics[angle=270,scale=0.45]{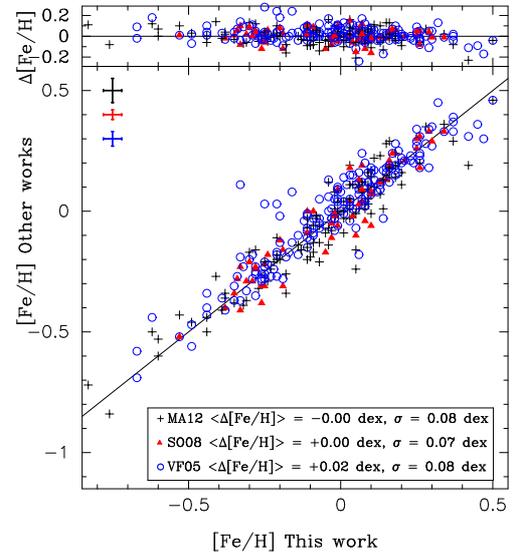}
\caption{
[Fe/H] values, this work, versus literature estimates.
Mean uncertainties in the metallicities are shown in the upper left corner
of the figure. The upper panel shows the differences
between the metallicities derived in this work and the values
given in the literature.
}
\label{comparacion_metalicidades}
\end{figure}

 We finally compare our metallicities with those already reported in the literature.
 Values for the comparison are taken from purely spectroscopic works: MA12 as a 
 consistency double check; the studies of
 \citet[][hereafter SO08]{2008A&A...487..373S,2011A&A...526A..99S,2011A&A...533A.141S}
 which use a similar approach (iron ionisation and excitation conditions) to this paper;
 and from \citet[][hereafter VF05]{2005ApJ...622.1102F} whose parameters are determined
 by fitting the observed spectra to synthetic models. 
 The comparison is shown in Figure~\ref{comparacion_metalicidades}. 

 Our sample contains 116 stars in common with MA12 and we note that
 the mean difference between our metallicities and those reported in
 MA12 is -0.00 dex ($\sigma$ = 0.08 dex). 
 56 stars are in common with SO08, being the mean difference +0.00 dex with a RMS
 standard deviation of 0.08 dex.
 Finally, for the comparison with VF05, we obtain a mean $\Delta$[Fe/H] = +0.02 dex, with
 $\sigma$ = 0.08 dex (173 stars in common).  
 In addition, there are no significant differences between our metallicity scale
 (zero point, slope) and those defined in MA12, SO08, and VF05 works.
 We, therefore, conclude that there are no systematic differences between our derived metallicities
 and other spectroscopic estimates in the literature.  

\subsection{Abundances}
\label{abundancias}

  Chemical abundance of individual elements,
  C, O, Na,
  Mg, Al, Si, S, Ca, Sc, 
  Ti, V, Cr,  Mn, Co, Ni,
  Cu, and Zn 
  are obtained using the 2014 version of the 
  code {\sc MOOG}\footnote{http://www.as.utexas.edu/\textasciitilde{}chris/moog.html}
  \citep{1973PhDT.......180S} together with ATLAS9 atmosphere models
  \citep{1993KurCD..13.....K}.
  Abundances of Sc, Ti, and Cr, were obtained by using lines of the neutral atom
  as well as from lines of the single ionized atoms.
  As it is common in the literature, through this paper we will use X~{\sc i} to refer to the  abundances of X computed
  from lines of the neutral atom, while  X~{\sc ii} means abundance of
  X derived from lines of the single ionised species.   
  The measured equivalent widths of a list of narrow, non-blended lines for each
  of the aforementioned species are used as inputs. The selected
  lines are taken from the lists provided by
  \citet[][Table~2]{2009A&A...497..563N},
  \citet[][Table~4]{2014A&A...561A...7R} for C, O, S, and Cu, and
  \citet[][Table~1]{2005PASJ...57...65T} in the case of Zn. 
  For completeness the line list used here is reproduced in 
  Table~\ref{lista_unificada_lines}. 

\begin{table}
\centering
\caption{Wavelength, excitation potential (EP), and oscillator strength $\log (gf)$
for the lines selected in the present work. The full version
of the table is only available in the on line version.}
\label{lista_unificada_lines}
\begin{tabular}{lccccc}
\hline\noalign{\smallskip}
 Ion      &   Wavelength (\AA) & EP (eV)  &  $\log (gf)$ & Ref. \\
\hline\noalign{\smallskip}
C~{\sc i}       & 6587.61        & 8.54  & -1.021        & RA14 \\
C~{\sc i}       & 7111.47        & 8.64  & -1.074        & RA14 \\
C~{\sc i}       & 7113.18        & 8.65  & -0.762        & RA14 \\
\multicolumn{5}{l}{...................} \\
Zn~{\sc i}      & 4810.54        & 4.08  & -0.29         & TA05 \\
Zn~{\sc i}      & 6362.35        & 5.79  &  0.09         & TA05 \\
\noalign{\smallskip}\hline\noalign{\smallskip}
\end{tabular}
\tablefoot{RA14: \cite{2014A&A...561A...7R}, NE09: \cite{2009A&A...497..563N}
 VALD: \cite{1995A&AS..112..525P,1999A&AS..138..119K},
 TA05: \cite{2005PASJ...57...65T} }
\end{table}

  The O~{\sc i} triplet lines at 777 nm are known to be severely 
  affected by departures from LTE \citep[e.g.][]{1993A&A...275..269K,2001NewAR..45..559K}.
  To account for non-LTE effects the prescriptions given by 
  \cite{2003A&A...402..343T} were followed. These corrections
  are essentially determined by the line EW and the stellar parameters
  T$_{\rm eff}$, $\log g$, and $\xi_{t}$. Although they do not contain
  an explicit dependence on the stellar metallicity, we note that
  stars with less metals are expected to have weaker EWs. 
 
  Hyperfine structure (HFS) was taken into account for 
  V~{\sc i}, 
  Co~{\sc i}, and Cu~{\sc i}, 
  using the {\sc MOOG} driver {\it blends} with
  the wavelengths and relative $\log gf$ values listed in
  \citet[][Table~4]{2014A&A...561A...7R}. Wavelengths
  and  $\log gf$ values for the ``unresolved line''
  are from the VALD\footnote{http://vald.astro.univie.ac.at/\textasciitilde{}vald/php/vald.php}
  database \citep{1995A&AS..112..525P,1999A&AS..138..119K}. 
  Another element whose abundance is known to be affected by HFS effects is Mn~{\sc i}.
  \cite{2014A&A...561A...7R} provides HFS data for two Mn~{\sc i} lines. 
  We note, however, a significant offset between the HFS abundance of Mn~{\sc i}
  derived from the 4502.20 \space \AA \space line and the abundance obtained from
  the 6021.80 \space \AA \space line; while the latter gives abundances which are in
  agreement with the non-HFS derived ones, HFS abundances derived from the 4502.20 \space \AA \space
  line are systematically lower by $\sim$ 0.4 (e.g.,
  $\log\epsilon_{\rm Mn I 4502.20 \odot}$  = 4.89,
  $\log\epsilon_{\rm Mn I 6021.80 \odot}$  = 5.46).
  Because of this difference we prefer not to take into account the HFS corrections for Mn~{\sc i}.

  In general, there is a good agreement between the abundances of a given element
  computed from lines of the neutral atom, and those computed using lines
  of the single ionised species, 
  although we note a tendency of abundances from neutral ions to be slightly
  shifted towards higher values for Sc and Ti. This behaviour is not reproduced in the
  abundances of Chromium where Cr~{\sc i} and Cr~{\sc ii} are found to be essentially
  the same at low values ($\log\epsilon_{\rm Cr}$  $\lesssim$ 5.7) while at higher abundances
  there seems to be a trend of slightly larger Cr~{\sc ii} abundances. 

  The solar spectrum provided in the ``S$^{4}$N'' \citep{s4n} library
  has been used to derive our own solar reference abundances which are
  given in Table~\ref{solar_abundances}.
  Our derived solar abundances are in reasonable agreement with recent
  determinations \citep[e.g.][]{2009ARA&A..47..481A,2014arXiv1405.0279S,
  2014arXiv1405.0287S,2014arXiv1405.0288G}, with the only exception of Zn~{\sc I} for which we obtain
  a significantly higher abundance. 

\begin{table}
\centering
\caption{Derived solar abundances ($\log\epsilon_{\rm X\odot}$),
with their corresponding line-to-line scatter error ($\sigma$/$\sqrt{N}$), and number
of lines ($N$).} 
\label{solar_abundances}
\begin{tabular}{lccc}
\hline\noalign{\smallskip}
 Ion                       & $\log\epsilon_{\rm X\odot}$ & Error &  $N$ \\
\hline\noalign{\smallskip}
C~{\sc I}        &  8.53 & 0.04   &   5  \\
O~{\sc I} (nLTE) &  8.79 & 0.04   &   3  \\
Na~{\sc I}       &  6.38 & 0.01   &   3  \\
Mg~{\sc I}       &  7.62 & 0.02   &   2  \\
Al~{\sc I}       &  6.48 & 0.01   &   2  \\
Si~{\sc I}       &  7.60 & 0.01   &   17 \\
S~{\sc I}        &  7.20 & 0.09   &   3  \\
Ca~{\sc I}       &  6.42 & 0.02   &   12 \\
Sc~{\sc I}       &  3.16 & 0.01   &   2  \\
Sc~{\sc II}      &  3.20 & 0.03   &   6  \\
Ti~{\sc I}       &  5.02 & 0.01   &   28 \\
Ti~{\sc II}      &  5.04 & 0.01   &   8  \\
V~{\sc I} (HFS)  &  3.91 & 0.03   &   14 \\
Cr~{\sc I}       &  5.68 & 0.01   &   20 \\
Cr~{\sc II}      &  5.67 & 0.01   &   2  \\
Mn~{\sc I}       &  5.43 & 0.01   &   5  \\
Co~{\sc I} (HFS) &  4.95 & 0.03   &   4  \\
Ni~{\sc I}       &  6.29 & 0.01   &   43 \\
Cu~{\sc I} (HFS) &  4.29 & 0.07   &   2  \\
Zn~{\sc I}       &  4.75 & 0.05   &   2  \\
\noalign{\smallskip}\hline\noalign{\smallskip}
\end{tabular}
\end{table}

  We have selected four ``representative'' stars covering the whole
  T$_{\rm eff}$ range in order to
  provide an estimate on how the uncertainties in the atmospheric
  parameters propagate into the abundance calculation, namely
  HIP 23311 (4848 K), HIP 77408 (5340 K), HIP 113044 (5976 K),
  and HIP 28767 (6241 K). These stars have been selected since
  their T$_{\rm eff}$ are similar to the
  10\%, 25\%, 75\%, and 99\% percentiles of the
  temperature distribution. 
  Abundances for each of these four stars were recomputed
  using atmosphere models with 
  T$_{\rm eff}$ + $\Delta$T$_{\rm eff}$, T$_{\rm eff}$ - $\Delta$T$_{\rm eff}$,
  and similarly for $\log g$ and $\xi_{t}$. Results are given in
  Table~\ref{abundance_sensitivity}.
   As final uncertainties for the derived abundances, we give the quadratic
  sum of the uncertainties due to the propagation of the errors in the stellar
  parameters, plus
  the line-to-line scatter errors
  (assuming a Gaussian statistics, they are 
  computed as $\sigma/\sqrt{N}$, where $\sigma$ is the standard deviation of the derived individual
  abundances from the $N$ lines). 
  We would like to point out that even these uncertainties should be considered as
  lower limits, given that the errors in the stellar parameters are only
  statistical (as explained in Section~\ref{stellar_parameters}), and the abundance estimates are affected by
  systematics which are not
  taken into account in line-to-line errors
  (e.g. atomic data or uncertainties in the atmosphere models).
  Our obtained final abundances are given in Table~\ref{abundance_table_full},
  which is only available on line.

\begin{table}
\centering
\caption{
Abundance sensitivities. It shows
how the derived abundances change when each stellar parameter
is perturbed by its corresponding uncertainty. 
}
\label{abundance_sensitivity}
\begin{scriptsize}
\begin{tabular}{lrrrrrr}
\hline\noalign{\smallskip}
        &  \multicolumn{3}{c}{HIP 23311}                              & \multicolumn{3}{c}{{HIP 77408} } \\
Ion     &  \multicolumn{3}{c}{\hrulefill}                             & \multicolumn{3}{c}{\hrulefill}   \\
   & $\Delta$T$_{\rm eff}$ &  $\Delta$$\log g$  &  $\Delta$$\xi_{t}$  & $\Delta$T$_{\rm eff}$  & $\Delta$$\log g$   & $\Delta$$\xi_{t}$   \\
   & $\pm$43               &  $\pm$0.11         &  $\pm$0.41          & $\pm$25                & $\pm$0.06          & $\pm$0.17           \\
   &  (K)                  &  (cms$^{\rm -2}$)  &  (kms$^{\rm -1}$)   & (K)                    &  (cms$^{\rm -2}$)  &  (kms$^{\rm -1}$)   \\
\hline\noalign{\smallskip}
C~{\sc I}                 & 0.06    &       0.03    &       $<$0.01 &       0.02    &       0.02    &       $<$0.01 \\
O~{\sc I}                 & 0.07    &       0.03    &       0.01    &       0.03    &       0.02    &       0.01    \\
Na~{\sc I}                & 0.04    &       0.04    &       0.04    &       0.02    &       0.01    &       0.01    \\
Mg~{\sc I}                & 0.01    &       0.02    &       0.03    &       0.01    &       0.01    &       0.02    \\
Al~{\sc I}                & 0.02    &       0.02    &       0.04    &       0.01    &       $<$0.01 &       0.01    \\
Si~{\sc I}                & 0.02    &       0.02    &       0.01    &       0.01    &       0.01    &       0.01    \\
S~{\sc I}                 & 0.05    &       0.03    &       $<$0.01 &       0.02    &       0.02    &       $<$0.01 \\
Ca~{\sc I}                & 0.04    &       0.03    &       0.05    &       0.02    &       0.02    &       0.03    \\
Sc~{\sc I}       	  & 0.05    &       0.02    &       0.09    &       0.03    &       $<$0.01 &       0.01    \\    
Sc~{\sc II}     	  & 0.01    &       0.04    &       0.06    &       $<$0.01 &       0.02    &       0.03    \\
Ti~{\sc I}       	  & 0.05    &       0.03    &       0.10    &       0.03    &       $<$0.01 &       0.04    \\
Ti~{\sc II}       	  & 0.01    &       0.04    &       0.06    &       $<$0.01 &       0.02    &       0.03    \\
V~{\sc I}                 & 0.06    &       0.02    &       0.09    &       0.03    &       $<$0.01 &       0.01    \\
Cr~{\sc I}         	  & 0.03    &       0.02    &       0.07    &       0.02    &       0.01    &       0.03    \\
Cr~{\sc II}       	  & 0.03    &       0.04    &       0.05    &       0.01    &       0.02    &       0.03    \\
Mn~{\sc I}                & 0.02    &       0.03    &       0.07    &       0.02    &       $<$0.01 &       0.04    \\
Co~{\sc I}                & $<$0.01 &       0.04    &       0.01    &       0.01    &       0.01    &       $<$0.01 \\
Ni~{\sc I}                & $<$0.01 &       0.01    &       0.05    &       0.01    &       0.01    &       0.02    \\
Cu~{\sc I}                & 0.02    &       0.03    &       0.04    &       $<$0.01 &       0.02    &       $<$0.01 \\
Zn~{\sc I}                & 0.02    &       0.01    &       0.06    &       0.01    &       0.01    &       0.03    \\
\hline\noalign{\smallskip}          &               &               &               &               &               \\
               &  \multicolumn{3}{c}{HIP 113044}                   & \multicolumn{3}{c}{HIP 28767} \\
Ion            &  \multicolumn{3}{c}{\hrulefill}                   & \multicolumn{3}{c}{\hrulefill} \\
 & $\Delta$T$_{\rm eff}$ & $\Delta$$\log g$   & $\Delta$$\xi_{t}$  & $\Delta$T$_{\rm eff}$  & $\Delta$$\log g$  & $\Delta$$\xi_{t}$  \\
 & $\pm$20               & $\pm$0.05          &  $\pm$0.12         & $\pm$25                & $\pm$0.05         &  $\pm$0.16         \\
 &  (K)                  &  (cms$^{\rm -2}$)  &  (kms$^{\rm -1}$)  & (K)                    & (cms$^{\rm -2}$)  &  (kms$^{\rm -1}$)  \\
\hline\noalign{\smallskip}
C~{\sc I}              & 0.01    &       0.01    &       $<$0.01 &       0.01    &       0.01    &       $<$0.01 \\
O~{\sc I}              & 0.02    &       0.02    &       0.02    &       0.01    &       0.01    &       0.01    \\
Na~{\sc I}             & 0.01    &       0.01    &       0.01    &       0.01    &       0.01    &       0.01    \\
Mg~{\sc I}             & 0.01    &       0.01    &       0.02    &       0.01    &       0.01    &       0.02    \\
Al~{\sc I}             & 0.01    &       $<$0.01 &       0.01    &       0.01    &       $<$0.01 &       0.01    \\
Si~{\sc I}             & $<$0.01 &       $<$0.01 &       0.01    &       0.01    &       $<$0.01 &       0.01    \\
S~{\sc I}              & 0.01    &       0.01    &       0.01    &       0.01    &       0.01    &       0.01    \\
Ca~{\sc I}             & 0.01    &       0.01    &       0.03    &       0.02    &       0.01    &       0.03    \\
Sc~{\sc I}             & 0.02    &       $<$0.01 &       $<$0.01 &       0.02    &       $<$0.01 &       $<$0.01 \\
Sc~{\sc II}            & $<$0.01 &       0.02    &       0.03    &       $<$0.01 &       0.02    &       0.03    \\
Ti~{\sc I}             & 0.02    &       $<$0.01 &       0.01    &       0.02    &       $<$0.01 &       0.01    \\
Ti~{\sc II}            & $<$0.01 &       0.02    &       0.03    &       $<$0.01 &       0.02    &       0.04    \\
V~{\sc I}              & 0.02    &       $<$0.01 &       $<$0.01 &       0.02    &       $<$0.01 &       $<$0.01 \\
Cr~{\sc I}             & 0.01    &       $<$0.01 &       0.02    &       0.02    &       $<$0.01 &       0.03    \\
Cr~{\sc II}    	       & 0.01    &       0.02    &       0.04    &       $<$0.01 &       0.02    &       0.05    \\
Mn~{\sc I}             & 0.01    &       $<$0.01 &       0.03    &       0.02    &       $<$0.01 &       0.03    \\
Co~{\sc I}             & 0.02    &       $<$0.01 &       $<$0.01 &       0.02    &       $<$0.01 &       $<$0.01 \\
Ni~{\sc I}             & 0.01    &       $<$0.01 &       0.02    &       0.02    &       $<$0.01 &       0.01    \\
Cu~{\sc I}             & 0.01    &       $<$0.01 &       $<$0.01 &       0.01    &       $<$0.01 &       $<$0.01 \\
Zn~{\sc I}             & $<$0.01 &       $<$0.01 &       0.04    &       0.01    &       0.01    &       0.04    \\
\hline\noalign{\smallskip}
\end{tabular}
\end{scriptsize}
\end{table}

\section{Analysis}
\label{section_analysis}

\subsection{Metallicity distributions} 
\label{analysis_metallicity_distributions}

\begin{figure}
\centering
\includegraphics[angle=270,scale=0.45]{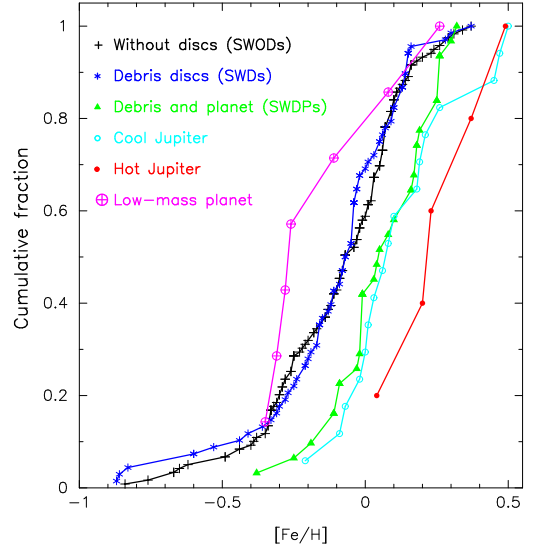}
\caption{
Histogram of [Fe/H] cumulative frequencies for the different samples
studied in this work. The SWP sample is divided into stars hosting exclusively
cool distant Jupiters, stars hosting hot close-in planets, and stars with orbiting
exclusively low-mass planets. Stars with both low-mass and gas giant planets are not
shown.}
\label{distribuciones_acumuladas}
\end{figure}

   The cumulative distribution function of the metallicity for
   the different samples analysed in this work is presented in 
   Figure~\ref{distribuciones_acumuladas}. For guidance some
   statistical diagnostics are also given in Table~\ref{metal_statistics}. 
   We note that statistics corresponding at the different samples of planet hosts 
   should be considered with caution given their small sizes.  
 
   The SWP sample has been divided into stars hosting 
   exclusively cool distant Jupiters (semimajor axes, $a$ $>$ 0.1 au,  17 stars), and stars
   hosting hot close-in planets ($a$ $<$ 0.1 au,  5 stars)
   given the higher frequency of planets with
   $a$ $\lesssim$ 0.07 au shown in the semimajor axis distribution of close-in
   gas giant planets, see
   \citet[][Figure~9]{2009ApJ...693.1084W} and
   \citet[][Figure~1]{2009ApJ...694L.171C}.
    Stars 
   harbouring only  low-mass planets
  (with M$_{\rm p}\sin i$ values below $\sim$ 30 M$_{\oplus}$,  7 stars)
   have also been considered as a different subsample,
   since their host stars seem to show different properties with 
   respect to stars hosting gas giant planets 
   (see Section~\ref{introduccion}).
   Low-mass and gas giant planets might coexist. Indeed, two of our stars
   harbouring a low-mass planet do also host at least one cool Jupiter.
   A remarkable case is HIP 43587 (55 Cnc), 
   a high-metallicity  star (+0.42 dex)  harbouring
   a five planetary system
   including three hot Jupiters, one low-mass planet, and one
   cool Jupiter. 

   As in MA12, we find the metallicity distribution of SWDs and SWODs to
   be similar. Indeed, a two sample Kolmogorov-Smirnov test (hereafter K-S test)
   \footnote{Performed with  the IDL Astronomy User's Library
   routine {\sc kstwo}, see  http://idlastro.gsfc.nasa.gov/}  
   shows that both distributions are quite similar
   ($p$-value 51\%). 
   Results are given Table~\ref{ksteststable}
   which provides the value of the K-S statistic ($D$), its significance
   level ($p$) and the effective size (n$_{\rm eff}$). Further
   details regarding the K-S test can be found in MA12 (Appendix~A). 

   In MA12 
   when comparing the metallicity distribution of SWDs and SWODs,
   a ``deficit'' of stars with debris discs at metallicities below approximately
   -0.1 dex was found. We certainly do not reproduce this result in this work but we 
   caution that the sample sizes analysed here are smaller than in MA12.
    Further observations would be required to clarify this point.  
   
   We find that the metallicity distribution of SWDPs is clearly different
   from the one of SWDs and similar to the one of the stars harbouring
   cool giant planets. A  K-S test confirms that the metallicity distribution of SWDPs 
   differ within a confidence level greater than 98\% from those of SWDs and SWODs,
   while the K-S test reveals the distributions of SWDPs
   is very similar to the one of cool giant hosts
   ($p$-value 81\%), see Table~\ref{ksteststable}.
   Eight SWDPs host at least one
   low-mass planets. We note that 
   the metallicities of these stars are $\lesssim$ +0.05 dex
   with only one exception, HIP 1499.

   We therefore conclude that the SWDP sample reproduces the known behaviour
   of the planet hosts, showing the
   metal-rich signature only when the planet is a gas giant one
   \citep[e.g.][]{1997MNRAS.285..403G,2004A&A...415.1153S,2005ApJ...622.1102F}
   and not in the case of exclusively low-mass planets
   \citep{2010ApJ...720.1290G,2011arXiv1109.2497M,2011A&A...533A.141S,2012Natur.486..375B}.
   
   We note that there seems to be a scarcity of hot Jupiters in SWDPs
   with only two stars harbouring simultaneously a debris disc and a hot
   Jupiter. 
   Regarding the metallicity distribution of hot Jupiters hosts,
   we find these stars to be
   more metal-rich than stars hosting exclusively cool distant planets,
   as already noted in MA12.
   However, we caution that the K-S test produces inconclusive results;
   the probability of both samples showing similar distributions being 20\%. 
   This trend has been  previously discussed in
   \citet{1998A&A...334..221G,2000A&A...354...99Q,2004MNRAS.354.1194S}
   and more recently in \cite{2013A&A...560A..51A}.
   A more extensive discussion is provided
   in Appendix~\ref{apendiceB}.

\begin{table}
\centering
\caption{[Fe/H] statistics of the stellar samples.}
\label{metal_statistics}
\begin{scriptsize}
\begin{tabular}{lcccccc}
\hline\noalign{\smallskip}
$Sample$  &  $Mean$ & $ Median$ & $Deviation$  &  $Min$&  $Max$ & $N$ \\
\hline\noalign{\smallskip}
 SWDs             &  -0.11  & -0.05  & 0.25 & -0.87 & +0.37 &  68 \\
 SWODs            &  -0.09  & -0.07  & 0.24 & -0.84 & +0.37 & 119 \\
 SWDPs            &  +0.06  & +0.05  & 0.17 & -0.38 & +0.32 &  31 \\
\noalign{\smallskip}\hline\noalign{\smallskip}
 Cool Jupiters    &  +0.13  & +0.08  & 0.20 & -0.21 & +0.50 &  17 \\
 Hot Jupiters     &  +0.27  & +0.23  & 0.15 & +0.04 & +0.49 &   5 \\
 Low-mass planets &  -0.14  & -0.26  & 0.21 & -0.35 & +0.26 &   7 \\
 Low-mass + Cool  &  +0.23  & +0.25  & 0.02 & +0.21 & +0.25 &   2 \\
 Low + Cool + Hot &  +0.42  &        &      &       &       &   1 \\ 
\noalign{\smallskip}\hline\noalign{\smallskip}
 Disc + Cool Jupiters        &  +0.11  & +0.16  & 0.14 & -0.11 & +0.32 &  21 \\
 Disc + Hot Jupiters         &  +0.15  & +0.26  & 0.11 & +0.04 & +0.26 &   2 \\
 Disc + Low-mass planets     &  -0.08  & -0.01  & 0.17 & -0.38 & +0.16 &   8 \\
\noalign{\smallskip}\hline\noalign{\smallskip}
\end{tabular}
\end{scriptsize}
\end{table}


\begin{table}
\centering
\caption{Results of the K-S tests performed in this work. We consider a 
confidence level of 98\% in order to reject the null hypothesis H$_{0}$
(both samples coming from the same underlying continuous
distribution).}
\label{ksteststable}
\begin{scriptsize}
\begin{tabular}{llcccccc}
\hline\noalign{\smallskip}
Sample 1 & Sample 2 & n$_{1}$ & n$_{2}$ & n$_{\rm eff}$ & $D$  & $p$   & H$_{0}^{\ddag}$ \\
\hline 
SWDS   &  SWODs   &   68  & 119  &  43  & 0.12  &  0.51             & 0 \\
\hline
SWDPs  &  SWODs   &   31  & 119  &  25  & 0.31  & 0.01              & 1 \\
SWDPs  &  SWDs    &   31  &  68  &  21  & 0.39  & $\sim$ 10$^{-3}$  & 1 \\
SWDPs  &  Cool    &   31  &  17  &  11  & 0.18  & 0.81              & 0 \\
\hline
Cool   &  SWODs   &   17  & 119  &  15  & 0.37  & 0.02              & 0 \\
Cool   &  SWDs    &   17  &  68  &  14  & 0.47  & $\sim$ 10$^{-3}$  & 1 \\
\hline
Hot       &  Cool    &    5  &  17  &   4  & 0.51  & 0.19             & 0 \\
Low-mass  &  Cool    &    7  &  17  &   5  & 0.66  & 0.01             & 1 \\
\hline
\end{tabular}
\tablefoot{  
 $D$ is the maximum deviation between the empirical distribution
 function of the samples 1 and 2. $p$ corresponds to the estimated
 likelihood of the null hypothesis, a  value which is known to be reasonable
 accurate for sample sizes for which n$_{\rm eff}$ $\ge$ 4.  
 $^{\ddag}$ (0): Accept null hypothesis; (1): Reject null hypothesis.}
\end{scriptsize}
\end{table}

\subsection{Other chemical signatures}

\begin{figure}
\centering
\includegraphics[angle=270,scale=0.45]{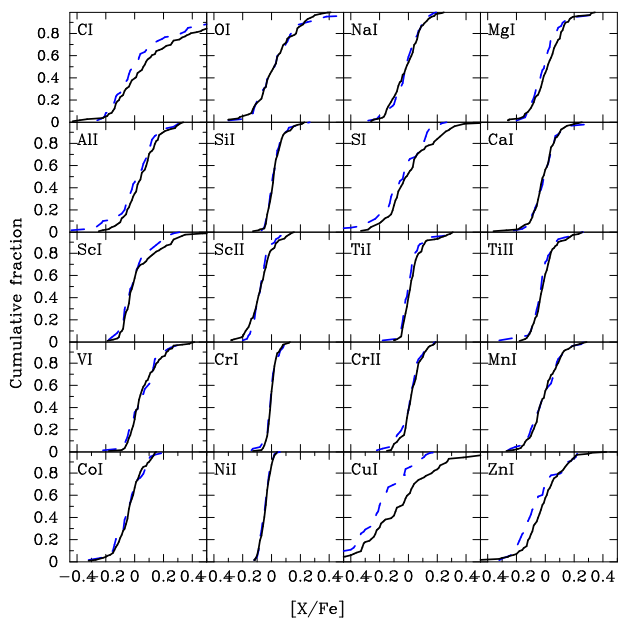}
\caption{
$[X/Fe]$ cumulative fraction of
SWDs (blue dashed line)  and SWODs (black continuous line).}
\label{distribuciones_acumuladas_other}
\end{figure}

  In order to find differences in the abundances of other chemical
  elements besides iron, the cumulative distribution [X/Fe] comparing
  the abundances between SWDs and SWODs is shown in Figure~\ref{distribuciones_acumuladas_other}.
  Some statistical diagnostics are also presented in Table~\ref{abund_table},
  where the results of a K-S test for each ion are also listed.
  For each star, abundances with large errors (uncertainties greater than 0.30 dex)
  were excluded from this exercise. 

  Similar behaviour between stars with debris discs and stars without
  known discs is found. Indeed from the 20 chemical species analysed, the
  K-S accepts the null hypothesis (i.e. SWDs SWODs distributions
  being drawn from the same parent population) in 19.
  The only exception is the  Cu~{\sc i} abundance for which the
  K-S test returns a probability of the null hypothesis 
  lower than 0.01. 
  We also note that the K-S probability corresponding to the
  Zn~{\sc i} in significantly low, of the order of 0.04
  (although the null hypothesis is not rejected).
  Nevertheless, we caution that 
  abundances of Cu~{\sc i}, and Zn~{\sc i} are based on only
  two lines. Furthermore Cu~{\sc i} abundance is severely
  affected by HFS effects. 


\begin{table}[!htb]
\centering
\caption{Comparison between the elemental abundances of SWODs and SWDs.}
\label{abund_table}
\begin{scriptsize}
\begin{tabular}{lccccccc}
\hline\noalign{\smallskip}
               &  \multicolumn{2}{c}{\textbf{SWODs}}  & \multicolumn{2}{c}{\textbf{SWDs}}   & \multicolumn{3}{c}{\bf {K-S test}}       \\
 $[X/Fe]$      &  \multicolumn{2}{c}{\hrulefill}      & \multicolumn{2}{c}{\hrulefill}      & \multicolumn{3}{c}{\hrulefill}           \\
               &  Median       &  Deviation           & Median      & Deviation             & $D$       & $p$-value &  n$_{\rm eff}$   \\
\hline\noalign{\smallskip}
 C~{\sc i}      &      0.07   &      0.34    &          0.01   &      0.33  &       0.15  &       0.30  &      38 \\
 O~{\sc i}      &      0.02   &      0.15    &          0.03   &      0.19  &       0.08  &       0.96  &      36 \\
 Na~{\sc i}     &     -0.01   &      0.11    &         -0.01   &      0.10  &       0.09  &       0.82  &      43 \\
 Mg~{\sc i}     &      0.02   &      0.12    &          0.00   &      0.12  &       0.18  &       0.10  &      43 \\
 Al~{\sc i}     &      0.05   &      0.12    &          0.03   &      0.15  &       0.12  &       0.55  &      39 \\
 Si~{\sc i}     &      0.02   &      0.06    &          0.01   &      0.07  &       0.11  &       0.63  &      43 \\
 S~{\sc i}      &     -0.01   &      0.18    &         -0.03   &      0.17  &       0.18  &       0.25  &      32 \\
 Ca~{\sc i}     &     -0.01   &      0.10    &          0.00   &      0.10  &       0.06  &    $>$0.99  &      43 \\
 Sc~{\sc i}     &     -0.01   &      0.20    &         -0.01   &      0.11  &       0.11  &       0.80  &      32 \\
 Sc~{\sc ii}    &     -0.07   &      0.09    &         -0.07   &      0.07  &       0.12  &       0.54  &      43 \\
 Ti~{\sc i}     &      0.02   &      0.08    &          0.00   &      0.08  &       0.19  &       0.08  &      43 \\
 Ti~{\sc ii}    &     -0.01   &      0.09    &         -0.02   &      0.09  &       0.14  &       0.36  &      42 \\
 V~{\sc i}      &      0.04   &      0.11    &          0.05   &      0.11  &       0.09  &       0.88  &      43 \\    
 Cr~{\sc i}     &      0.00   &      0.04    &          0.00   &      0.04  &       0.12  &       0.56  &      43 \\
 Cr~{\sc ii}    &      0.02   &      0.07    &          0.02   &      0.08  &       0.11  &       0.64  &      42 \\
 Mn~{\sc i}     &      0.00   &      0.11    &         -0.01   &      0.11  &       0.08  &       0.91  &      42 \\
 Co~{\sc i}     &     -0.03   &      0.10    &         -0.03   &      0.10  &       0.10  &       0.83  &      38 \\   
 Ni~{\sc i}     &     -0.04   &      0.04    &         -0.04   &      0.04  &       0.09  &       0.86  &      43 \\
 Cu~{\sc i}     &     -0.04   &      0.27    &         -0.20   &      0.23  &       0.30  &    $<$0.01  &      33 \\
 Zn~{\sc i}     &     -0.02   &      0.17    &         -0.07   &      0.15  &       0.21  &       0.04  &      42 \\       
\noalign{\smallskip}\hline\noalign{\smallskip}
\end{tabular}
\end{scriptsize}
\end{table}

\subsection{[X/Fe]-T$_{\rm C}$ trends in SWDPs}

  Another way of searching for chemical differences is to study
  possible trends between the abundances and the elemental condensation temperature,
  T$_{\rm C}$. For each of the stars analyzed in this work, the [X/Fe] trend
  as a function of the T$_{\rm C}$  was obtained. T$_{\rm C}$ values correspond
  to a 50\% equilibrium condensation temperature for a solar system composition gas  
  \citep{2003ApJ...591.1220L}. Each trend
  is characterised by a linear fit, weighting each abundance by its
  corresponding uncertainty
  \footnote{Note that our abundances are given with respect to the Sun,
   while other works compute the abundance difference Star - Sun
   \citep[e.g.][]{2009ApJ...704L..66M,2013A&A...552A...6G}.}.
  Given the relatively low number of volatile elements considered in this work
   and the fact that their 
  abundances are in general more difficult to obtain accurately 
  (few lines that can be blended, non-LTE effects), we compute the slope
  of the [X/Fe] vs. T$_{\rm C}$ fit considering all (refractories and volatiles)
  elements (T$^{\rm all}_{\rm C}$-slope) and considering only refractories
  (T$^{\rm refrac}_{\rm C}$-slope).
  Following the discussion in \citet[][Section~5.3]{2010A&A...521A..33R} we consider as volatile those elements
  with T$_{\rm C}$ lower than 900 K, namely C, O, S, and Zn.

\begin{figure*}
\centering
\begin{minipage}{0.47\linewidth}
\includegraphics[angle=270,scale=0.45]{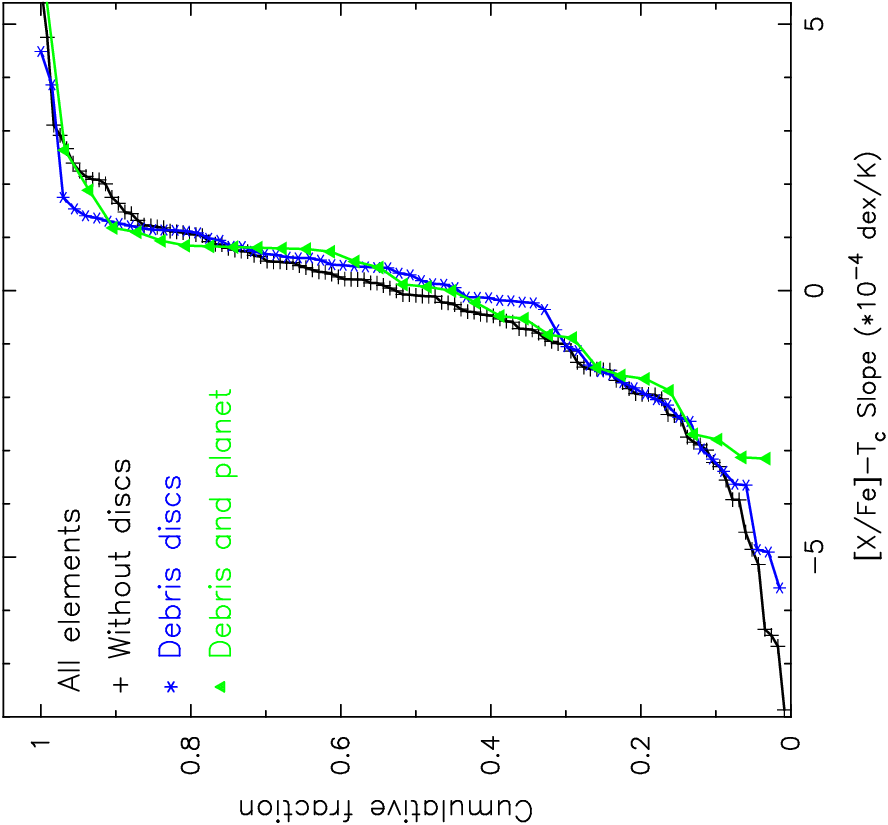}
\end{minipage}
\begin{minipage}{0.47\linewidth}
\includegraphics[angle=270,scale=0.45]{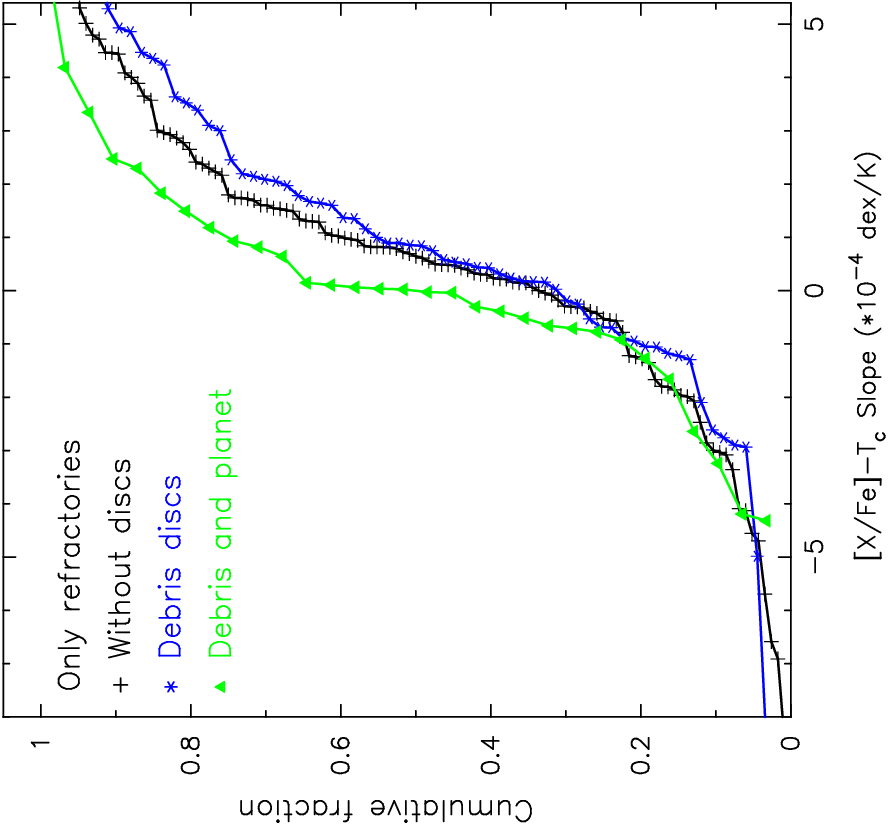}
\end{minipage}
\caption{
 Histogram of [X/Fe]-T$_{\rm C}$ slopes  derived when all elements (volatiles plus refractories)
 are taken into account (left) and when only refractories are considered (right). 
}
\label{tc_distribuciones}
\end{figure*}

  The cumulative distribution functions of 
  T$^{\rm all}_{\rm C}$-slope, and T$^{\rm refrac}_{\rm C}$-slope
  are shown in Figure~\ref{tc_distribuciones} for the SWOD sample
  (black crosses), the SWD sample (blue asterisks), and the
  the SWDP sample (green triangles). The T$^{\rm all}_{\rm C}$-slope distribution
  is on the left panel whilst the distribution of T$^{\rm refrac}_{\rm C}$-slope
  is shown on the right one.
  Several interesting trends emerge from this figure.
  First, when considering the cumulative distribution of T$^{\rm all}_{\rm C}$-slope
  (left), all the samples considered here, SWDs, SWODs, and SWDPs, show similar
  distributions.
  However, if we consider only refractory elements (right), the SWDP sample 
  seems to behave in a different way with respect to the SWOD and SWD samples.
  It can be seen that SWDPs seem to have slightly more negative values of T$^{\rm refrac}_{\rm C}$-slope
  than the other two samples, in particular at slopes above
  -1$\times$10$^{-\rm 4}$ dex/K.
  Statistically, this different behaviour of SWDPs in T$^{\rm refrac}_{\rm C}$-slope seems
  to be significant  
  with K-S $p$-probabilities
  of the order of $\sim$ 0.03 when comparing SWDPs and SWODs, and of the order of
  $\sim$ 0.01 when SWDPs are compared to SWDs.
 
\begin{figure*}
\centering
\begin{minipage}{0.49\linewidth}
\includegraphics[angle=270,scale=0.375]{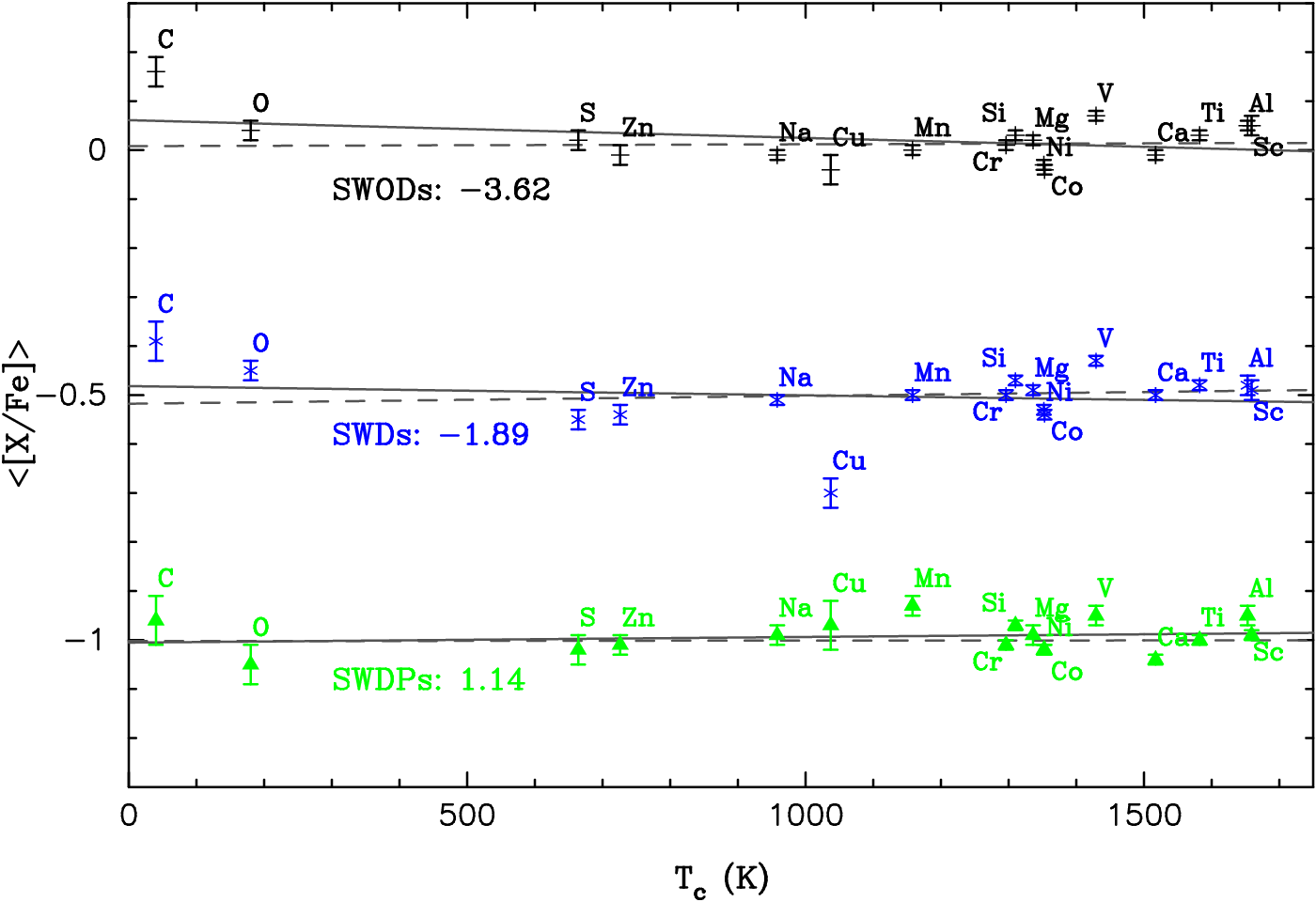}
\end{minipage}
\begin{minipage}{0.49\linewidth}
\includegraphics[angle=270,scale=0.375]{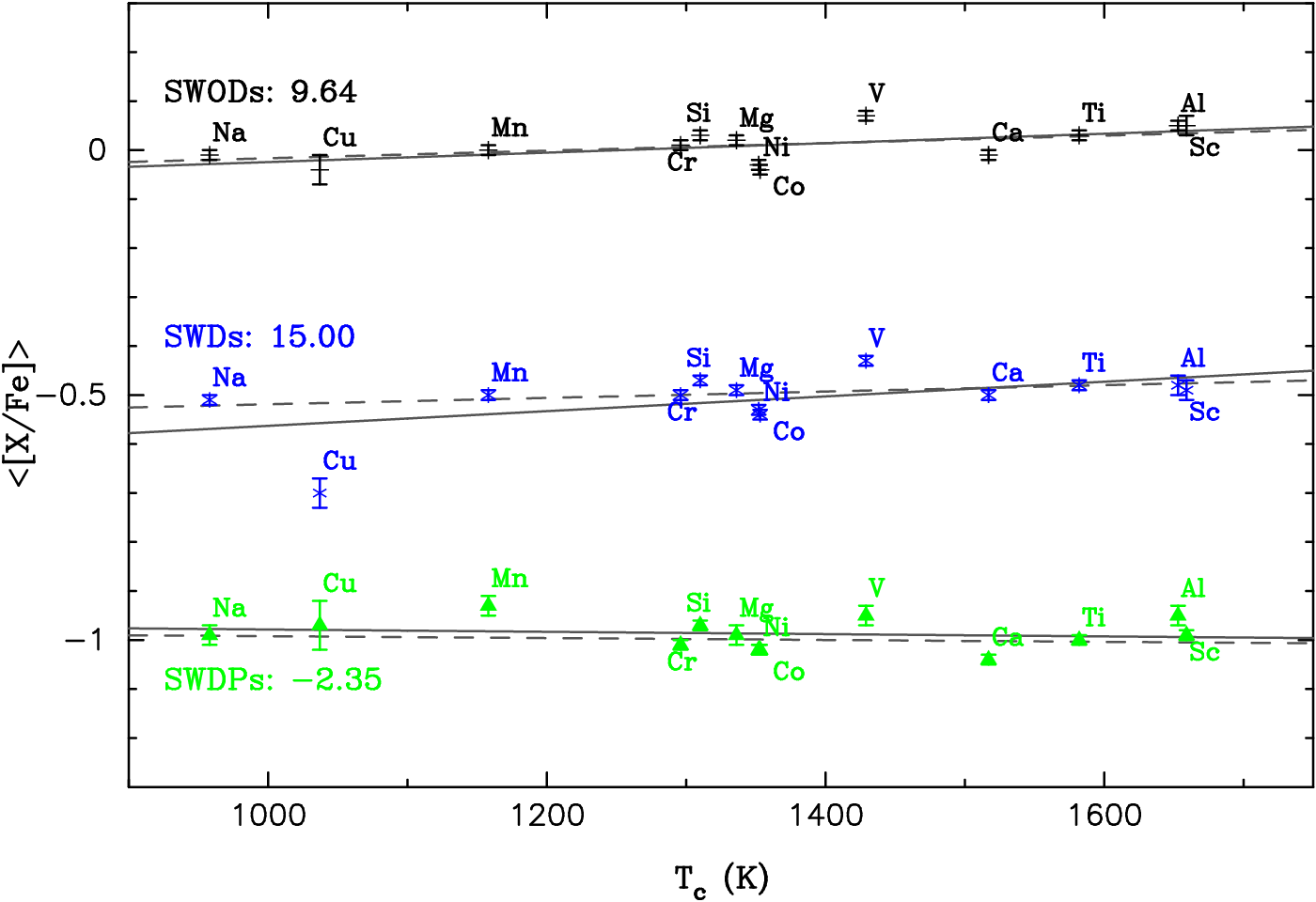}
\end{minipage}
\caption{
  <[X/Fe]>-T$_{\rm C}$ trends for the SWOD, SWD, and SWDP samples when all elements
 (volatiles and refractories) are taken into account (left) and when only refractories are considered (right).
  For the sake of clarity, an offset of -0.50 dex was applied between the samples.
  Unweighted fits are shown by continuous lines, while weighted
  fits are plotted in dashed lines. For guidance, the derived slopes from the unweighted fits are shown
  in the plots (units of 10$^{\rm -5}$ dex/K). 
}
\label{tc_plots}
\end{figure*}

  Mean abundances for each of the samples (SWDS, SWODs, SWDPs)
  were also computed, and T$^{\rm all}_{\rm C}$-slope, T$^{\rm refrac}_{\rm C}$-slope
  derived. 
  As errors we considered the star-to-star scatter.
  As explained above, each trend is characterised by a linear fit. Two different fits were performed, 
  one weighting each element by its corresponding error and another one without weighting.
  The corresponding plots are shown in Figure~\ref{tc_plots}, and
  a summary of the fits is shown in Table~\ref{linearfits}.
  To give a significance for the derived slopes a Monte Carlo simulation was carried out.
  We created 10000 series of simulated
  random abundances and errors,  keeping the media and the standard deviation of the original data.
  For each series of simulated data the corresponding  T$_{\rm C}$-slope was derived. Assuming
  that the distribution of the simulated slopes follows a Gaussian function we then compute
  the probability that the simulated slope takes the value found when fitting the original data.

  In the left panel, when all elements (volatiles and refractories) are considered,
  the unweighted fits (continuous line) reveal a different behaviour of SWDPs with respect to the
  samples of stars without known planetary companions. SWDP shows a slightly positive slope 
   while the slope of the SWODs and SWDs seems to be negative.
  The analysis of the significance of the slopes does however introduce a word of caution
  since these trends seems to be at least tentative (see Table~\ref{linearfits}).
  In fact, when the linear fit is done by weighting each element by its corresponding error (dashed lines)
  the suggested trends tend to disappear and SWODs, SWDs, and SWDPs seem to show
  similar positive slopes. The significance of the weighted fits are rather moderate
  (probability of the slope ``being by chance'' $\le$ 23\%).

  When only refractory elements are considered (right panel), in the unweighted fits (continuous
  line) SWDs and SWODs show a positive trend,
  whilst SWDPs follow a slightly negative tendency. 
  In this case, the different sign of SWDPs with respect to SWDs and SWODs is also
  present in the weighted fits (dashed lines). 
  The significance of the  T$^{\rm refrac}_{\rm C}$-slope fits are in all cases 
  (weighted and weighted fits) moderate (probability of the slope ``being by chance'' between 8 and 20\%)
  with the only exception of the unweighted fit of the SWDP sample (62\%).  
   

\begin{table}
\centering
\caption{Results of the  <[X/Fe]>-T$_{\rm C}$ linear fits. For each fit its 
 probability of slope ``being by chance'' (prob.) is also given.}
\label{linearfits}
\begin{scriptsize}
\begin{tabular}{lcccc}
\hline\noalign{\smallskip}
       & \multicolumn{4}{c}{\bf All elements}  \\
Sample & \multicolumn{2}{c}{Weighted fit}  & \multicolumn{2}{c}{Unweighted fit}  \\
       & slope ($\times$10$^{\rm -5}$dex/K)       & prob.   & slope ($\times$10$^{\rm -5}$dex/K)  &  prob.   \\
\hline
 SWODs    &        0.38 &        0.13 &   -3.62 &        0.31  \\ 
 SWDs     &        1.60 &        0.10 &   -1.89 &        0.62  \\ 
 SWDPs    &        0.09 &        0.23 &    1.14 &        0.59  \\ 
\hline
 Cool     &        3.83 &        0.18 &    3.18 &        0.32  \\ 
 Low-mass &       -0.59 &        0.16 &   -4.73 &        0.34  \\ 
 Hot      &       20.62 &        0.04 &    6.92 &        0.50  \\
 Low-mass + Cool & -11.65 &      0.10 &   -1.60 &        0.67  \\  
\hline
 Disc + Cool &    -0.14 &        0.22 &    2.74 &        0.33  \\
 Disc + Low  &    -0.65 &        0.31 &   -3.43 &        0.32  \\
 Disc + Hot  &    -3.68 &        0.04 &   -0.47 &        0.69  \\
\hline
       & \multicolumn{4}{c}{\bf Only refractory}  \\
Sample & \multicolumn{2}{c}{Weighted fit}  & \multicolumn{2}{c}{Unweighted fit}  \\
       & slope ($\times$10$^{\rm -5}$dex/K)       & prob.   & slope ($\times$10$^{\rm -5}$dex/K)  &  prob.   \\
\hline
SWODs     &        7.69 &        0.10 &     9.64 &        0.12  \\  
SWDs      &        6.50 &        0.08 &    15.00 &        0.19  \\  
SWDPs     &       -1.92 &        0.17 &    -2.35 &        0.62  \\  
\hline
Cool      &        2.33 &        0.19 &    -0.75 &        0.70  \\  
Low-mass  &       15.62 &        0.13 &     6.87 &        0.53  \\  
Hot       &       23.03 &        0.03 &    14.36 &        0.55  \\
Low-mass + Cool & -24.75 &       0.08 &   -19.32 &        0.26  \\   
\hline
 Disc + Cool &    -7.09 &        0.15 &    -3.87 &        0.50  \\ 
 Disc + Low  &     6.96 &        0.22 &     4.17 &        0.60  \\
 Disc + Hot  &    -5.40 &        0.12 &   -24.79 &        0.12  \\
\hline
\end{tabular}
\end{scriptsize}
\end{table}

\subsection{Comparison with planet hosts}

\begin{figure*}
\centering
\begin{minipage}{0.49\linewidth}
\includegraphics[angle=270,scale=0.375]{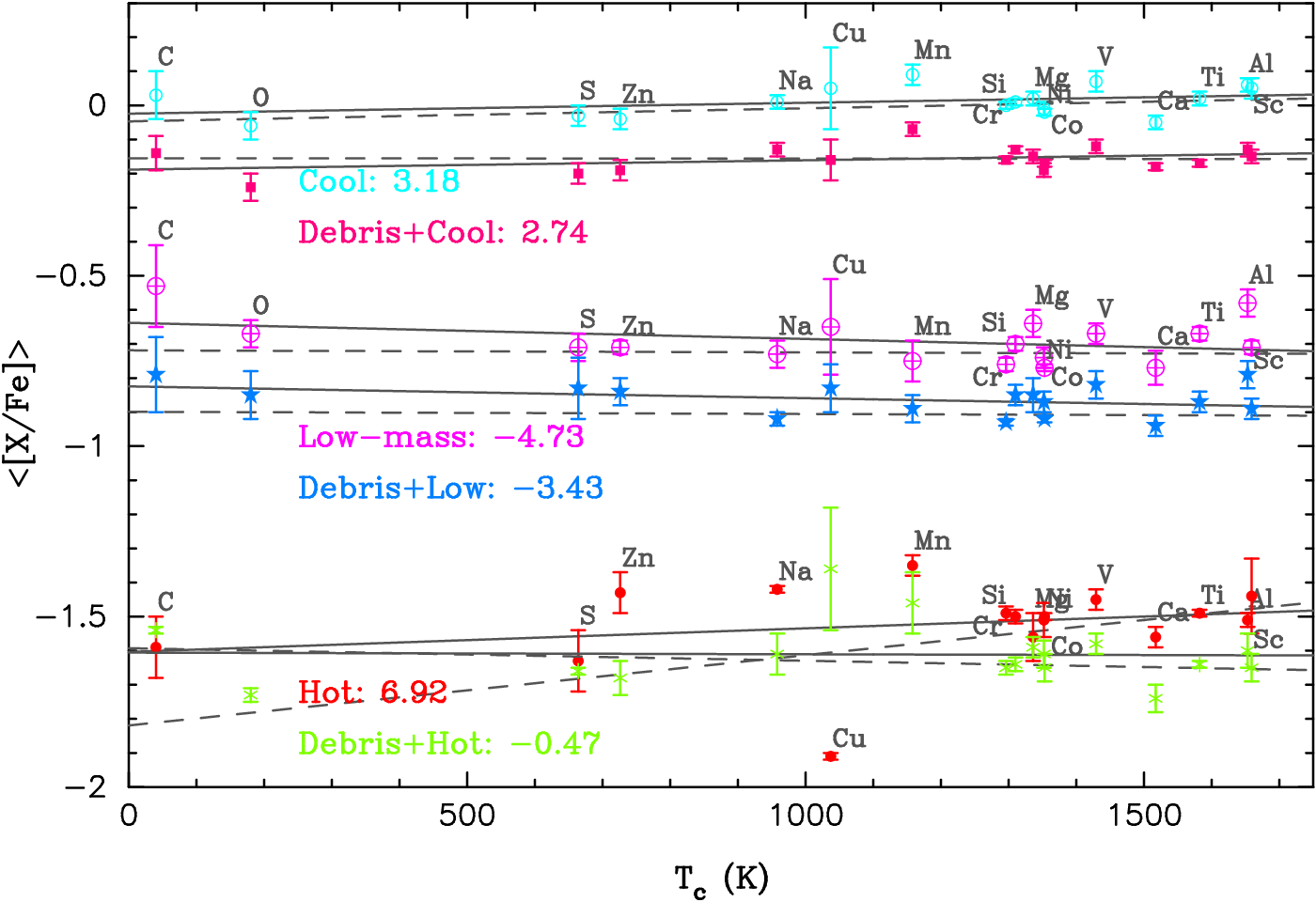}
\end{minipage}
\begin{minipage}{0.49\linewidth}
\includegraphics[angle=270,scale=0.375]{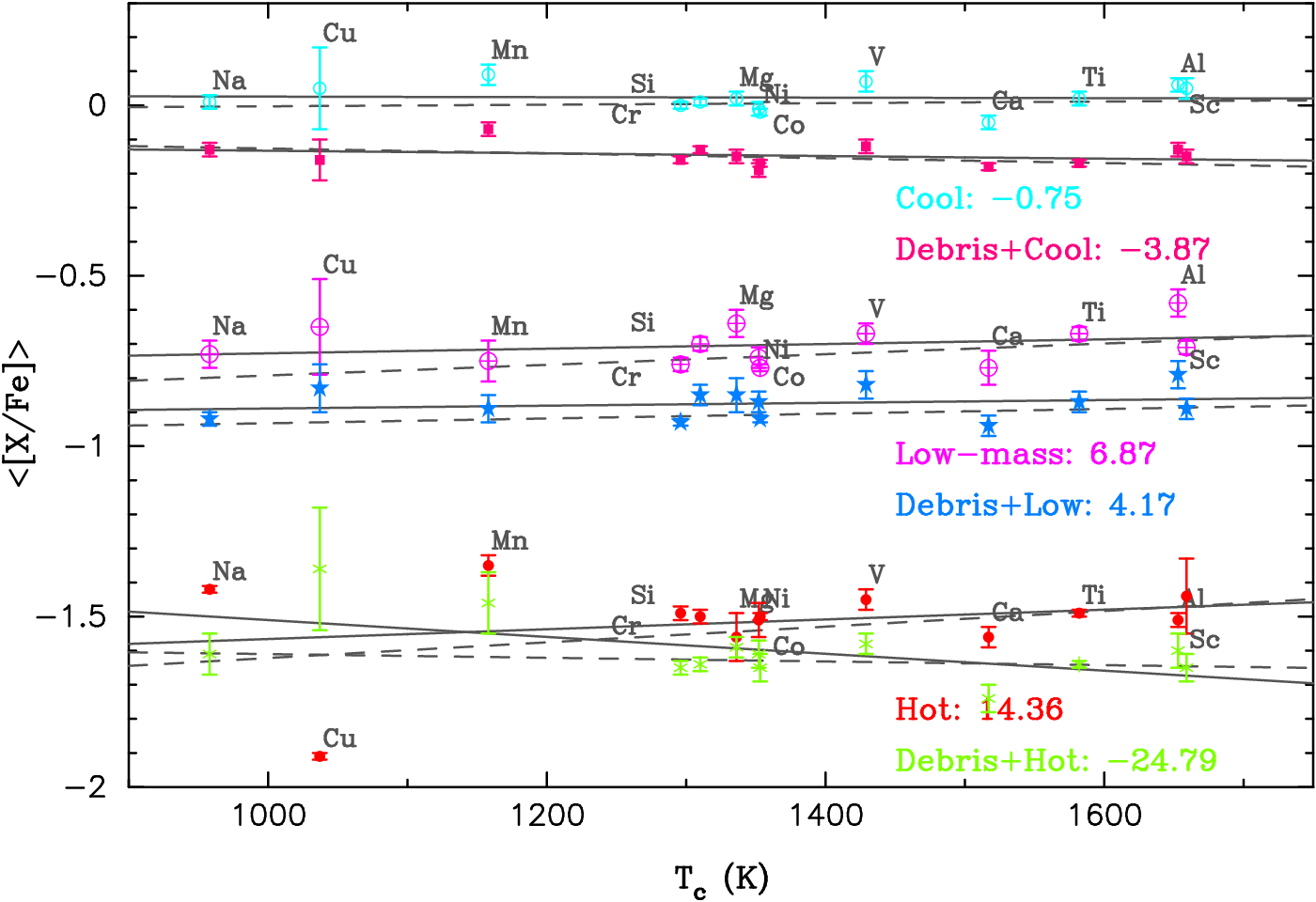}
\end{minipage}
\caption{
  <[X/Fe]>-T$_{\rm C}$ trends for planet host stars. The stars are divided into six categories, three
  of them corresponding to stars with known planets but no debris discs, namely, stars hosting cool Jupiters
  (light-blue open circles), low-mass planet hosts (pink earth symbols), and stars hosting
  hot Jupiters (red filled circles).  
  The SWDP sample is divided into the same categories: stars with debris discs and cool Jupiters
  (pink filled squares), debris discs and low-mass planets (cyan filled stars), and stars harbouring
  debris discs and hot Jupiters (light green asterisks). 
  Each planet host subsample is shown against its corresponding SWDP subsample
  (e.g, stars with cool Jupiters versus stars with discs and Cool Jupiters)  
  with an offset of -0.15 between the samples for the sake of clarity.
  The offset between the samples of cool, low-mass, and hot Jupiters hosts is -0.75.
  Unweighted fits are shown by continuous lines, while weighted
  fits are plotted in dashed lines. For guidance, the derived slopes from the unweighted fits are shown
  in the plots (units of 10$^{\rm -5}$ dex/K).
  The left panel shows the   <[X/Fe]>-T$_{\rm C}$ trends 
  when all elements
  (volatiles and refractories) are taken into account whilst the
  right one shows the <[X/Fe]>-T$_{\rm C}$ trends when only refractories are considered.
}
\label{tc_plots_planets}
\end{figure*}

  The <[X/Fe]>-T$_{\rm C}$ of the  SWDP sample can also be compared with the  
  results from our sample of stars hosting cool giant planets, hot close-in Jupiters,
  and low-mass planets.
  For this purpose the SWDP sample has been divided into stars with discs and
  cool giant planets, stars with discs and low-mass planets, and stars harbouring
  discs and hot Jupiters. 
  The corresponding trends are shown in Figure~\ref{tc_plots_planets} where
  each planet host subsample (cool, low-mass, or hot planets)
  is compared with its corresponding SWDP subsample
  (disc and cool, disc and low-mass, disc and hot planets).
  The fits results are given in Table~\ref{linearfits}. 
  Several conclusions can be drawn from this analysis.
   
  We first note that SWDP seem to behave in a similar way as stars with known planets.
  It can be easily seen in the left panel of Figure~\ref{tc_plots_planets}
  (i.e. when all the elements are considered) that this 
  statement holds for stars hosting cool Jupiters and low-mass planets.
  A similar behaviour is found when only refractory elements are considered
  (right panel).

   Second, there seems to be a hint for low-mass planet hosts to show
   a different behaviour
   in the unweighted fits in comparison with
   stars with gas giant planets. This is true for both
   T$^{\rm all}_{\rm C}$ (low-mass planet hosts show negative slopes whilst
   cool Jupiters hosts show positive slopes)
   and  T$^{\rm ref}_{\rm C}$ analysis (positive slopes for stars with low-mass
   planets, negatives for stars with cool Jupiters).
   This trend seems to be also present 
   in the weighted fits for all elements.  

\begin{figure*}
\centering
\begin{minipage}{0.49\linewidth}
\includegraphics[angle=270,scale=0.375]{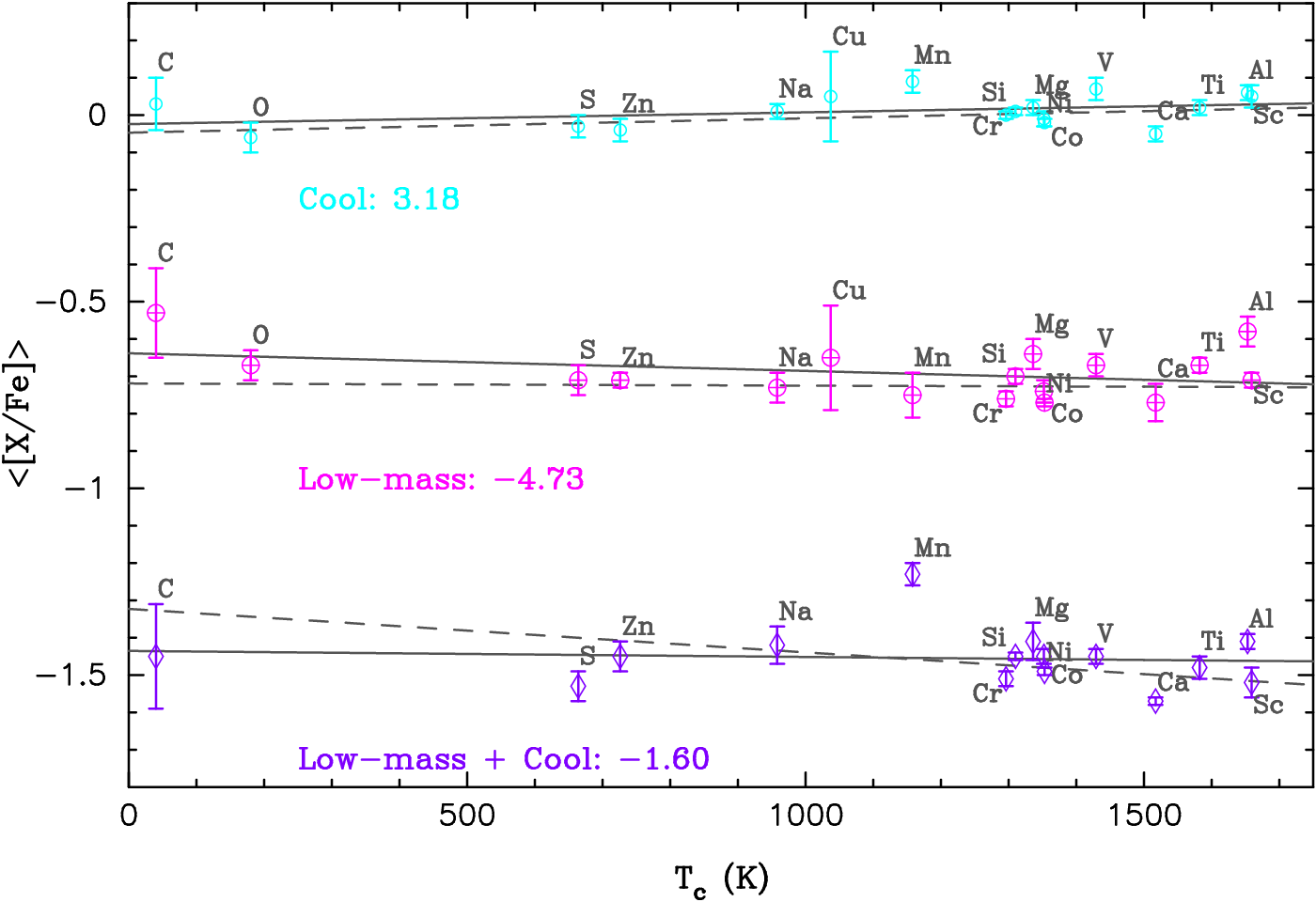}
\end{minipage}
\begin{minipage}{0.49\linewidth}
\includegraphics[angle=270,scale=0.375]{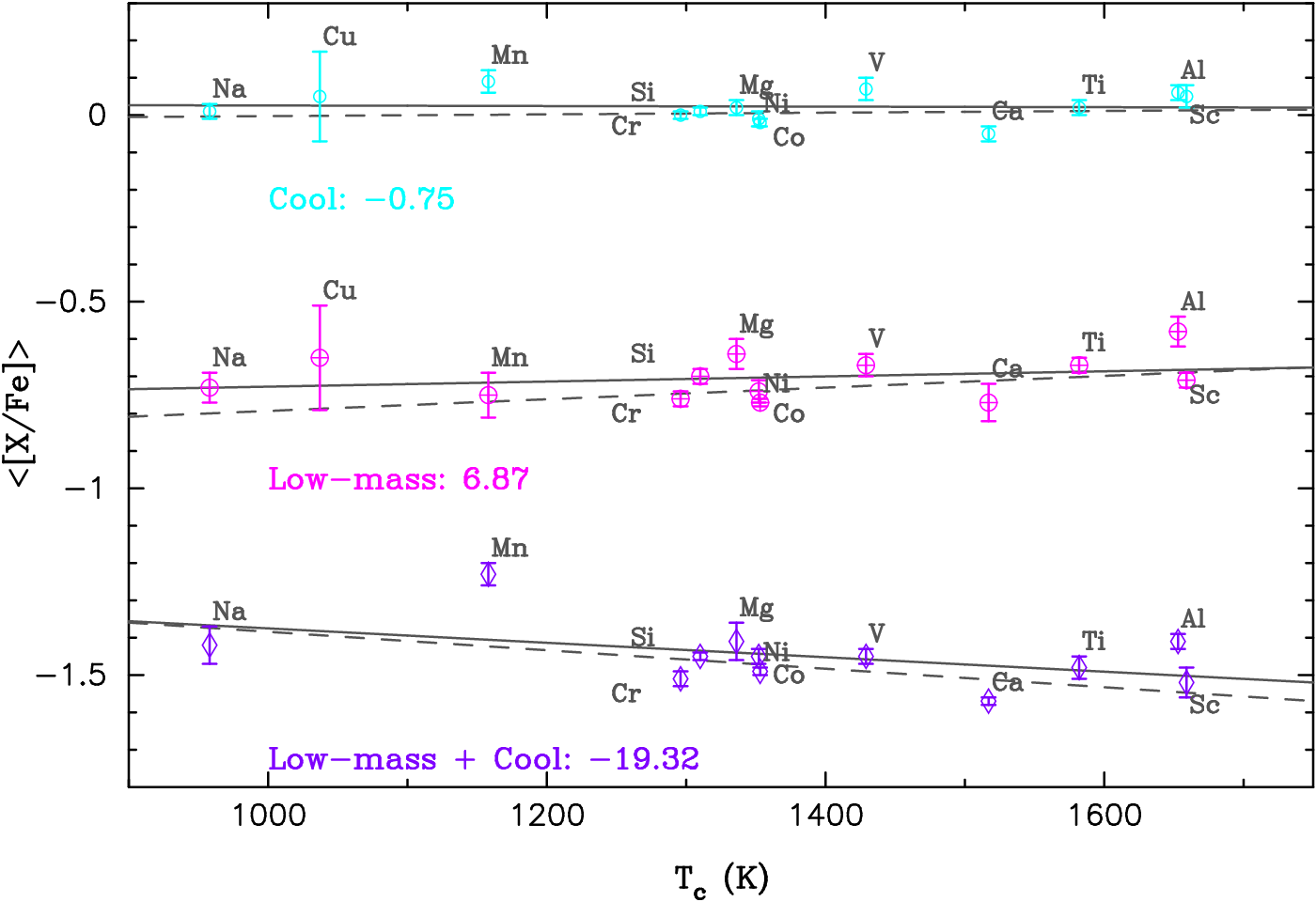}
\end{minipage}
\caption{
  <[X/Fe]>-T$_{\rm C}$ trends for stars with cool Jupiters (light-blue open circles), stars with exclusively
  low-mass planets (pink earth symbols), and stars with low-mass planets and 
  cool Jupiters (purple open diamonds). For the sake of clarity, an offset of -0.75 dex was applied between the samples.
  For guidance, the derived slopes from the unweighted fits are shown
  in the plots (units of 10$^{\rm -5}$ dex/K).
  The left panel shows the   <[X/Fe]>-T$_{\rm C}$ trends
  when all elements
  (volatiles and refractories) are taken into account whilst the
  right one shows the <[X/Fe]>-T$_{\rm C}$ trends when only refractories are considered.
}
\label{tc_plots_low_mass}
\end{figure*}

   We note at this point that we have classified as stars with low-mass
   planets those stars hosting exclusively these kind of planets
   (i.e., without giant Jupiters). 
   We have two stars harbouring simultaneously low-mass and cool Jupiters
   \footnote{HIP 43587 is not considered since it also hosts several
   hot Jupiters.}. Their <[X/Fe]>-T$^{\rm all}_{\rm C}$ trends are 
   compared with cool Jupiters hosts and low-mass planets hosts
   in Figure~\ref{tc_plots_low_mass}. It can be seen that stars
   with low-mass and cool planets do not seem to behave as stars
   with exclusively low-mass planets. In particular, in the
   T$^{\rm ref}_{\rm C}$ analysis, where stars with low-mass and cool
   planets show a clear negative slope whilst stars with only low-mass
   planets show a positive one.

   Setting together the results from the  analysis of Figures~\ref{tc_plots},
   \ref{tc_plots_planets}, and \ref{tc_plots_low_mass} it seems
   that the slopes in the unweighted T$^{\rm all}_{\rm C}$ analysis tend
   to be negative, unless the star host a Jupiter (stars with cool Jupiters,
   hot Jupiters, and debris plus cool Jupiters). On the other hand,
   in the unweighted T$^{\rm ref}_{\rm C}$ fits the slopes tend to be positive
   in all samples, but negative in those samples hosting a cool gas giant planet
   (stars with cool Jupiters, with debris plus cool Jupiters, with low-mass
   planets and cool Jupiters, and also stars with debris and hot Jupiters). In other words,
   the samples of stars with cool giant planets seem to behave in a different way
   with respect to the non planet hosts samples.

   Finally, and  despite our low statistics (only five stars), it is worth to 
   mention that  the sample of
   stars hosting hot Jupiters (Figure~\ref{tc_plots_planets}) show in all cases (all elements, only
   refractories, weighted, and unweighted fits) a clear positive slope
   (only SWDs and SWODs in the unweighted T$^{\rm ref}_{\rm C}$ analysis
   show positive slopes of similar values).  
   There is however a clear disagreement between stars with hot Jupiters and
   the behaviour
   of the stars hosting simultaneously a debris discs and hot Jupiters.
   The reasons of this discrepancy can be found in the low number of
   stars in both samples, but perhaps, as well as in their clearly different
   mean abundance of copper. We also note, that none of the stars
   in the hot Jupiters host sample have a reliable oxygen abundance.

\section{Discussion}
\label{seccion_discussion}
 
  In a recent work \citet[][hereafter ME09]{2009ApJ...704L..66M}, reports a ``deficit''
  of refractory elements in the Sun with respect to other solar
  twins. After discussing several possible origins, the authors
  conclude that the most likely explanation is related to the formation
  of planetary systems like our own, in particular to the formation 
  of rocky low-mass planets. 
  A similar conclusion
  was reached by \cite{2009A&A...508L..17R}, and \cite{2010MNRAS.407..314G}.
  Although very appealing,  
  ME09 hypothesis has however been challenged. Instead other works point rather towards
  Galactic Chemical Evolution effects 
  as the cause of the detected
  small chemical depletion's \citep{2010ApJ...720.1592G,2013A&A...552A...6G}
  or/to an age/Galactic birth place explanation \citep{2014A&A...564L..15A}.

  In the previous section several interesting (although some certainly tentative)
  trends in planet hosts have been found:
  {\it i)} There seems to be no chemical differences between SWDs and SWODs;
  {\it ii)} SWDPs behave as stars with planets;
  {\it iii)} Stars with low-mass planets
  do not seem to behave in different way with respect to the SWD
  and SWOD samples;
  {\it iv)}
  The samples of stars with cool Jupiters seem to be
  the ones that might follow a different trend with respect to the SWD and
  SWOD samples; and 
  {\it v)} Stars hosting hot Jupiters seem to show positive slopes. 
  At this point, in order to understand the origin and significance of these
  findings, two main questions should be discussed:
  {\it a)}
  Might the  <[X/Fe]>-T$_{\rm C}$ trends
  be influenced by effects of metallicity, age, or galactocentric
  distance?, and
  {\it b)}  Do these <[X/Fe]>-T$_{\rm C}$ trends fit in the
  framework of ME09 hypothesis?.

\subsection{Age, metallicity
            and Galactocentric distance effects}

   Abundance patterns may be affected by Galactic Chemical Evolution
   effects (GCE effects). \cite{2013A&A...552A...6G} accounts for these effects by fitting
   straight lines to the [X/Fe] versus [Fe/H] plots. The obtained trends are then
   removed from the original [X/Fe] data.    
   \cite{2014A&A...561A...7R}, however, argues 
   that correcting from GCE effects in this manner may 
   prevent us from finding elemental depletions due to planet formation.

\begin{figure*}
\centering
\begin{minipage}{0.47\linewidth}
\includegraphics[angle=270,scale=0.45]{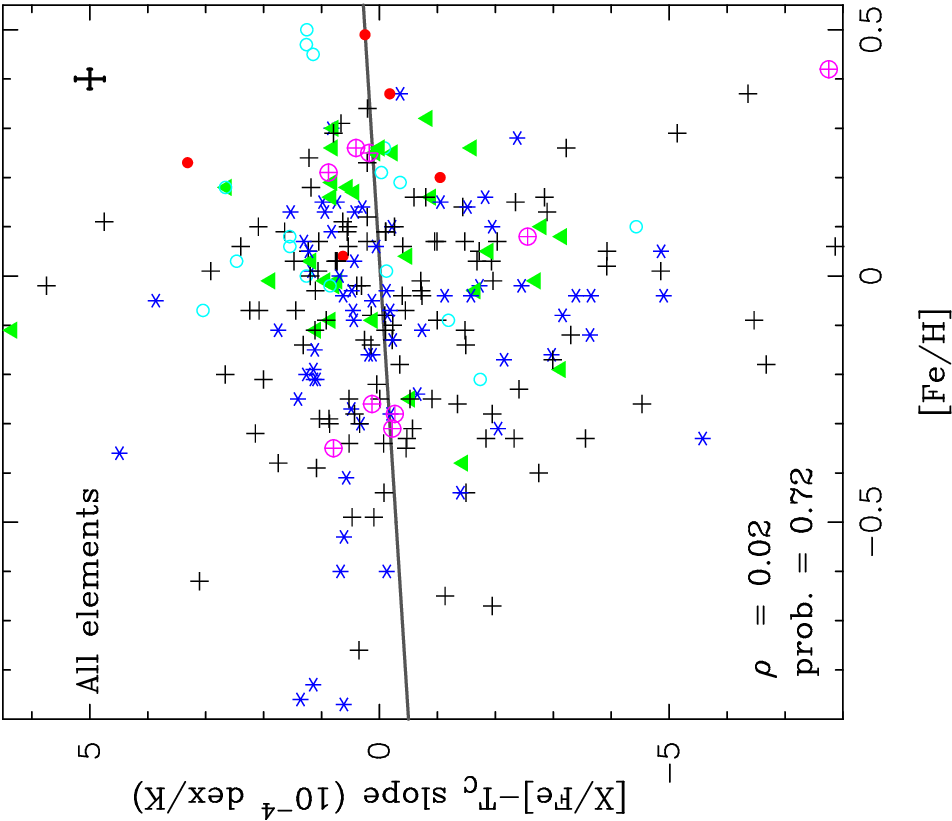}
\end{minipage}
\begin{minipage}{0.47\linewidth}
\includegraphics[angle=270,scale=0.45]{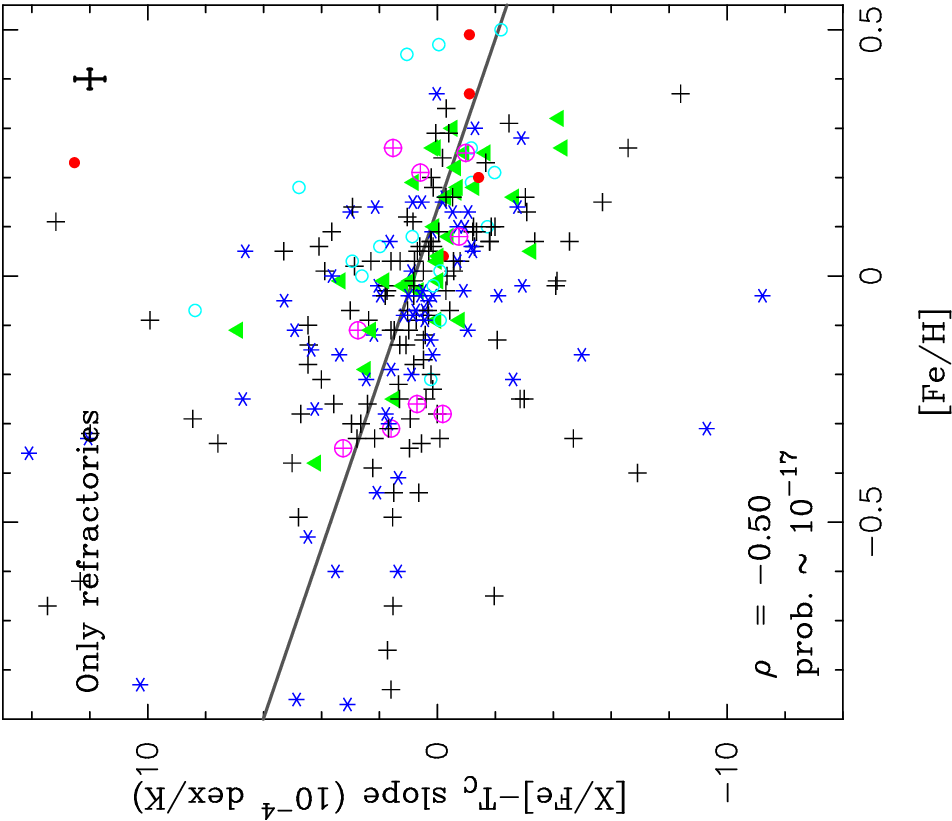}
\end{minipage}
\caption{
  T$^{\rm all}_{\rm C}$-slope 
  (left) and T$^{\rm refrac}_{\rm C}$-slope (right) as a function
  of the stellar metallicity. A linear fit to the data is shown
  by the grey continuous line. Results from the Spearman's
  correlation test are shown in the lower left corner of the plot,
  while typical error bars are shown in the upper right corner.
  Colours and symbols are as in Figure~\ref{distribuciones_acumuladas}.
}
\label{tc_metallicity}
\end{figure*}

   A way to disentangle the effects due to chemical evolution from 
   those related to fractionated accretion is to analyze the dependence
   of the T$_{\rm C}$ slope as a function of the stellar metallicity.
   When considering all elements, abundances of C~{\sc i} and O~{\sc i}
   tend to rise towards lower metallicities, producing negative slopes
   in the abundance vs. T$_{\rm C}$ plot for stars with low metallicities
   \citep[e.g.][and references therein]{2006A&A...449..809E}.
   The T$^{\rm all}_{\rm C}$-slope  vs [Fe/H] plot for our stars is explored in
   Figure~\ref{tc_metallicity} (left) where it can be seen that no clear
   trend is found. We recall at this point, that our abundance ratios as a function
   of the stellar metallicity are consistent with previous works
   (see Appendix~\ref{apendiceC}). 

   However, when considering the T$^{\rm ref}_{\rm C}$-slope
   (Figure~\ref{tc_metallicity}, right) a trend of
   of decreasing slopes towards high metallicities seems to be present
   (negative in this case since the abundances of C~{\sc i} and O~{\sc i} are not
   considered). A  Spearman’s correlation test shows that the correlation,
   although moderate ($\rho$= -0.50), seems to be highly significant
   (prob $\sim$ 10$^{\rm -17}$). 
   The possibility that GCE effects affect our abundance analysis can therefore
   not be ruled out.

  The observed correlation between the presence of gas-giant planets
  and enhanced metallicity has been widely debated in the context
  of two different scenarios of planet formation, core-accretion
  and disc instabilities. Little attention has been paid to other
  lines of argument. In particular, \cite{2009ApJ...698L...1H} 
  suggested a possible inner-disc origin of the planet hosts as
  a possible explanation. 
  Recently, \cite{2014A&A...564L..15A} found correlations between
  the T$_{\rm C}$-slope and the stellar age, the surface gravity, and
  the mean Galactocentric distance of the star, R$_{\rm mean}$, suggesting that the
  age and the Galactic birth place (and not the presence of planets) are
  likely the parameters that determine the chemical properties of the stars.

  A similar search for correlations was performed in our sample of planet 
  hosts. 
  Stellar ages from MA12 or derived in the same
  manner were used. These ages are based on the $\log R'_{HK}$
  activity index following the prescriptions of \cite{2008ApJ...687.1264M}.
  As a consistency check, the T$_{\rm C}$-age relationship was also
  studied by using the isochrone ages provided by the
  {\it PARAM} code \citep{2006A&A...458..609D}.
  Regarding the Galactic parameters, values for the mean Galactocentric radii
  were taken from \cite{2011A&A...530A.138C}.

\begin{figure*}
\centering
\begin{minipage}{0.47\linewidth}
\includegraphics[angle=270,scale=0.45]{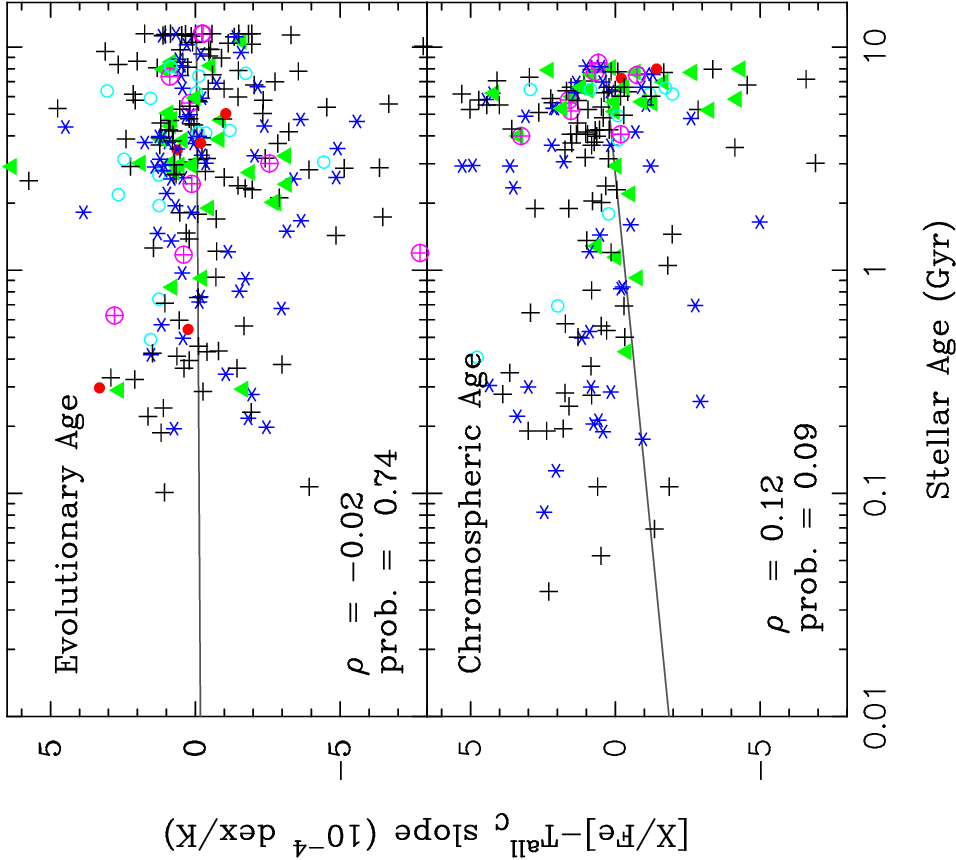}
\end{minipage}
\begin{minipage}{0.47\linewidth}
\includegraphics[angle=270,scale=0.45]{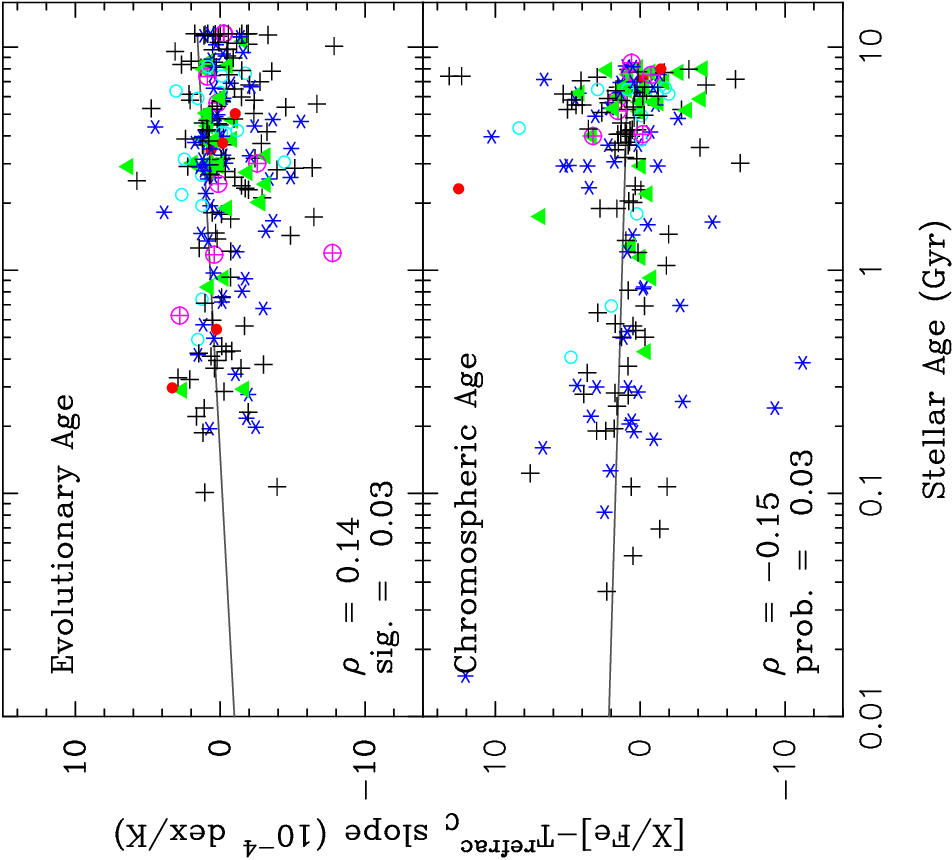}
\end{minipage}
\caption{
  T$^{\rm all}_{\rm C}$-slope 
  (left) and T$^{\rm refrac}_{\rm C}$-slope (right) as a function
  of the $\log R'_{HK}$-derived ages (top panels) and
  isochrone ages (bottom panels).
  A linear fit to the data is shown
  by the grey continuous line. Results from the Spearman's
  correlation test are also shown in the plot. 
  Colours and symbols are as in Figure~\ref{distribuciones_acumuladas}.
}
\label{tc_age}
\end{figure*}

  The results (see, Figure~\ref{tc_age}) show a weak but significant
  correlation with the stellar age.    
  Results from a Spearman’s correlation test provide:
  $\rho$= -0.02, prob $\sim$ 0.74 for T$^{\rm all}_{\rm C}$-isochrone age;
  $\rho$=  0.14, prob $\sim$ 0.03 for T$^{\rm refrac}_{\rm C}$-isochrone age;
  $\rho$=  0.12, prob $\sim$ 0.09 for T$^{\rm all}_{\rm C}$-chromospheric age; 
  $\rho$= -0.15, prob $\sim$ 0.03 for T$^{\rm refrac}_{\rm C}$-chromospheric age.
  Our results confirm the findings by \cite{2014A&A...564L..15A} in the
  sense that a correlation between T$_{\rm C}$-slope and age is likely to be present,
  although we find it to be relatively weak.

\begin{figure*}
\centering
\begin{minipage}{0.47\linewidth}
\includegraphics[angle=270,scale=0.45]{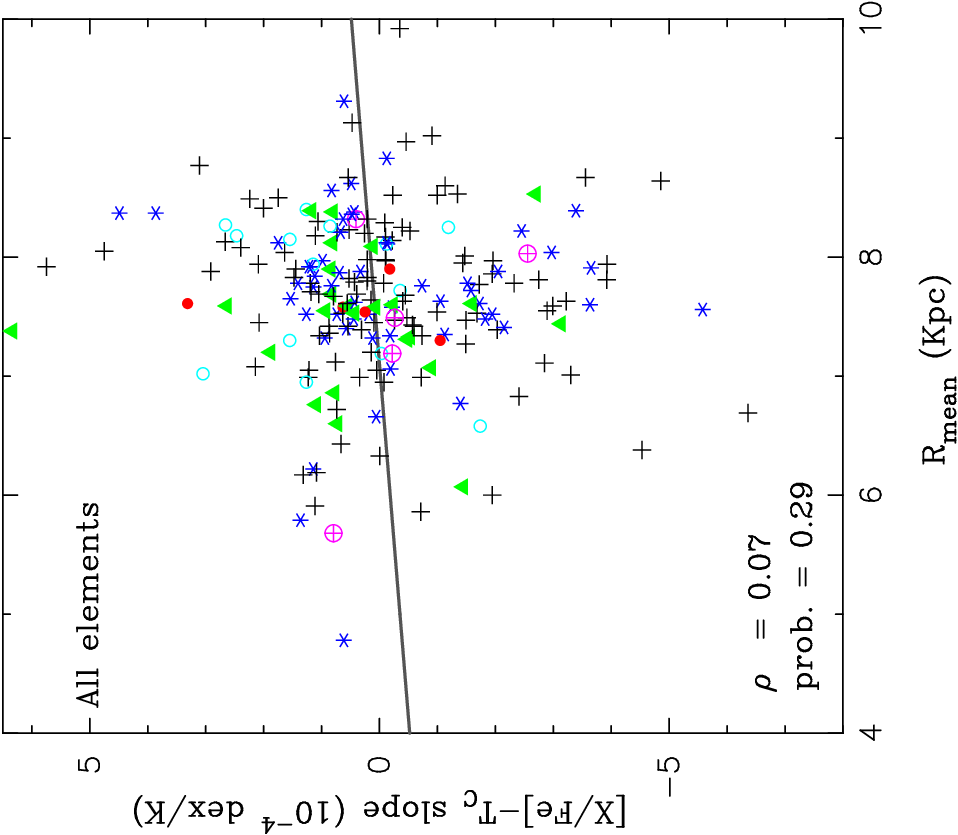}
\end{minipage}
\begin{minipage}{0.47\linewidth}
\includegraphics[angle=270,scale=0.45]{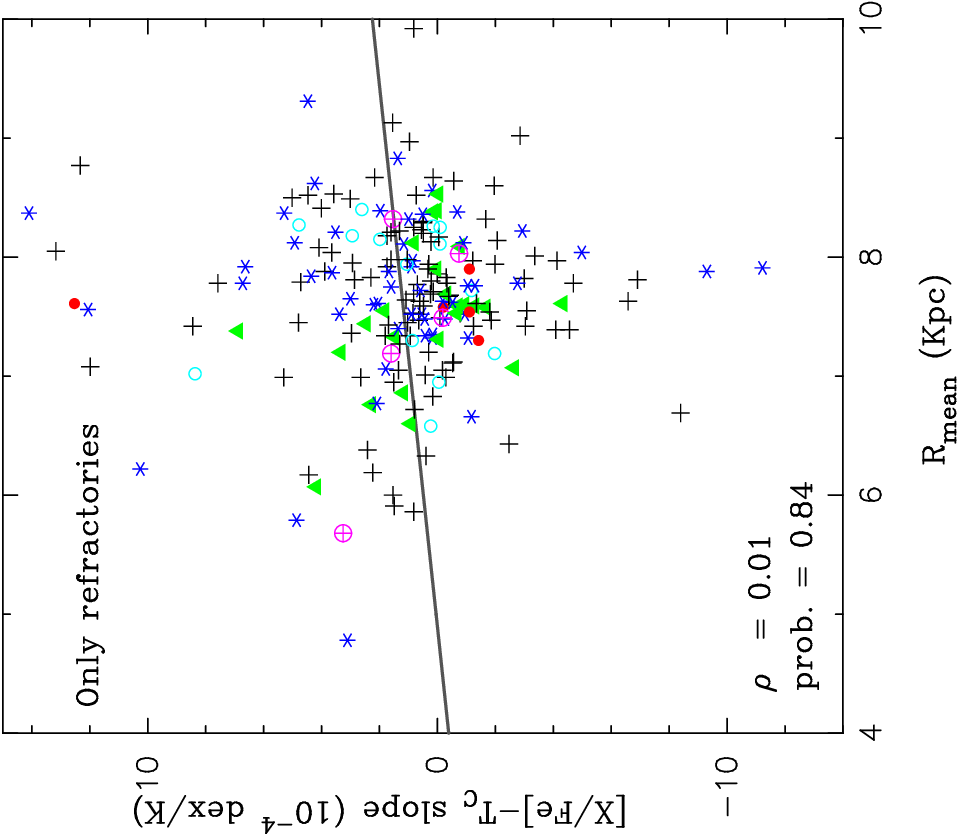}
\end{minipage}
\caption{
  T$^{\rm all}_{\rm C}$-slope 
  (left) and T$^{\rm refrac}_{\rm C}$-slope (right) as a function
  of the mean Galactocentric distance of the stars.
  A linear fit to the data is shown
  by the grey continuous line. Results from the Spearman's
  correlation test are also shown in the plot.
  Colours and symbols are as in previous figures. 
}
\label{tc_rmean}
\end{figure*}

  We find, however, no clear correlation between T$_{\rm C}$-slope
  and R$_{\rm mean}$ in disagreement with \cite{2014A&A...564L..15A}
  although these authors only suggest a tentative evidence (and not
  a strong correlation). Furthermore, our abundance ratios are not
  corrected from possible galactic chemical evolution effects.
  The plot of T$_{\rm C}$-slope vs. R$_{\rm mean}$ is shown in 
  Figure~\ref{tc_rmean}.
 
  In order to test whether our results are affected or not by 
  chemical galactic effects (GCE), our abundances were corrected
  following the procedure of \cite{2013A&A...552A...6G} although we  fitted
  the [X/H] versus [Fe/H] plane (see Figure~\ref{swds_swods_xh_vs_feh}),
  instead of the [X/Fe]-[Fe/H] plane (our Figure~\ref{swds_swods_xfe_vs_feh}).
  The reason for doing so is the larger scatter found in the
  [X/Fe] versus metallicity plane 
  probably due to the large number
  of stars as well as the broader range of stellar parameters covered
  in this work. 
  Besides the expected trend of higher [X/H] values as we move
  towards higher metallicities and a larger scatter in the elements
  whose abundances are based on a smaller number of lines, 
  no other significant trend are revealed by the  
  [X/H] versus [Fe/H] plots.  
   
  As before, values of T$^{\rm all}_{\rm C}$-slope and T$^{\rm refrac}_{\rm C}$-slope 
  were computed for each individual star and a search for correlations with the
  stellar radii, metallicity, age, and mean Galactocentric distance was performed.
  A summary of the results from  the Spearman's correlation tests are given in 
  Table~\ref{gce_corrected_trends}.
  It can be seen that our results do not change significantly.
  The GCE-corrected T$^{\rm refrac}_{\rm C}$-slope does not show any clear
  correlation with any of the mentioned stellar properties. As for
  the GCE-corrected T$^{\rm all}_{\rm C}$-slope, it shows a weak but
  significant correlation with the stellar age (but only when considering
  chromospheric ages) and a moderate ($\rho$ = 0.54) significant correlation 
  with the stellar radius. These two correlations were also found
  without correcting for GCE effects. 


\begin{table}
\centering
\caption{Results from the Spearman's correlation tests between
the galactic chemical evolution effects corrected T$_{\rm C}$-slope
and different stellar properties.}
\label{gce_corrected_trends}
\begin{tabular}{lcccc}
\hline\noalign{\smallskip}
 Property         & \multicolumn{2}{c}{All elements}     & \multicolumn{2}{c}{Only refractories} \\
                  &  $\rho$     &          prob.         &  $\rho$     &          prob.          \\
\hline
Radius            &   0.54      &  $\sim$10$^{\rm -19}$  &   0.06   &  0.39 \\
$[Fe/H]$          &  -0.00      &   0.98                 &  -0.06   &  0.32 \\
Chromospheric Age &   0.20      &  $\sim$10$^{\rm -3}$   &  -0.07   &  0.35 \\
Evolutionary Age  &   0.03      &   0.63                 &  -0.00   &  0.92 \\
R$_{\rm mean}$    &   0.03      &   0.67                 &  -0.02   &  0.82 \\
\hline
\end{tabular}
\end{table}

\subsection{Abundance patterns and the presence of discs and planets}

   The two first observational results of this work, the lack
   of a chemical difference in SWDs with respect to SWODs,
   and the fact that SWDPs behave in a similar
   way as stars with planets
    (showing 
   different <[X/Fe]>-T$_{\rm C}$ trends
   when the planet is a cool giant one, but not when the star hosts
   exclusively low-mass planets),
   indicate that it is the presence of
   planets, and not the presence of
   discs, the factor which reveals the chemical behaviour of the corresponding
   star. 

   This was first established in MA12, although only on the basis
   of metallicity distributions.  
   This result fits well in the framework of the core-accretion
   model of planet formation 
   \citep[e.g.][]{1996Icar..124...62P,2004ApJ...616..567I,2005Icar..179..415H,
   2009A&A...501.1139M,2012A&A...541A..97M}, where the conditions for the formation
   of debris are more easily met than the conditions for the formation of gas-giant
   planets.
  
   MA12 noticed that among the SWDP sample there were a significant fraction
   of stars hosting low-mass planets, mainly in multiplanet systems. 
   \cite{2012MNRAS.424.1206W} suggested that stars
   with low-mass planets might be more likely to have detectable debris discs,
   arguing that the same processes that lead to the formation of low-mass
   planets may result in high levels of outer debris.
   The lack of a metallicity enhancement in SWDs with respect to SWODs, and the
   lower metallicities of
   stars with low-mass planets with respect to stars hosting gas-giant planets
   do certainly support this hypothesis.
   In fact, additional evidence of the correlation between the presence
   of dust, low-mass planets, and lower stellar metallicities has been
   recently found by \cite{2014A&A...565A..15M}. However,
   we caution that there are several biases that might prevent
   us from finding a clear statistically significant correlation
   \citep{2015arXiv150103813M}.

\begin{figure*}[!htb]
\centering
\begin{minipage}{0.32\linewidth}
\includegraphics[angle=270,scale=0.40]{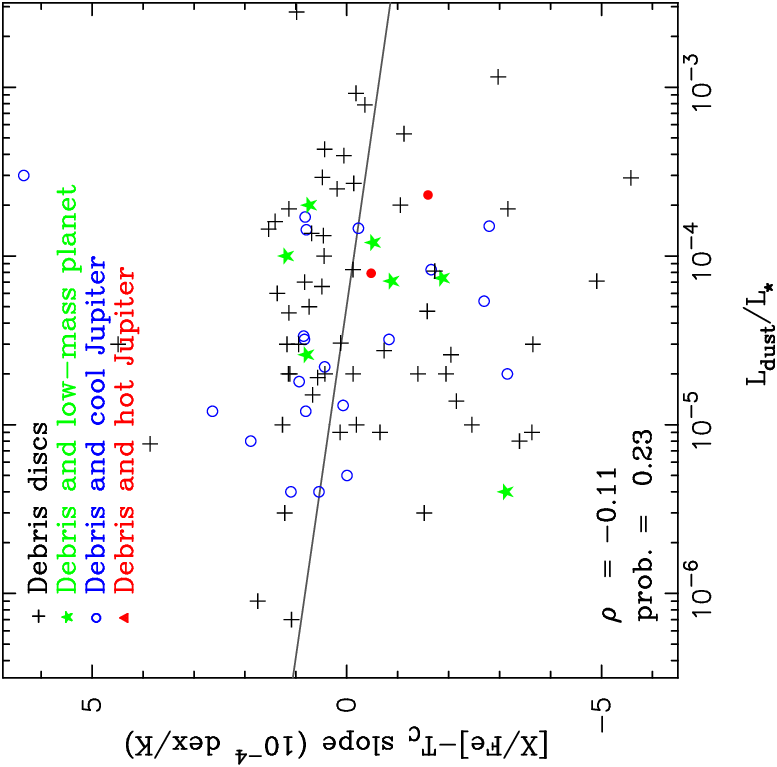}
\end{minipage}
\begin{minipage}{0.32\linewidth}
\includegraphics[angle=270,scale=0.40]{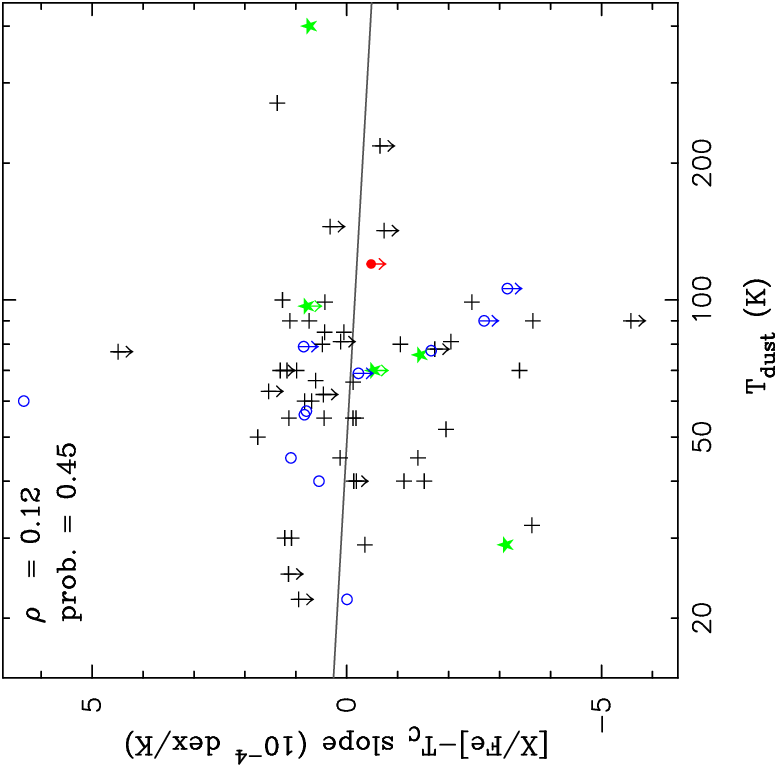}
\end{minipage}
\begin{minipage}{0.32\linewidth}
\includegraphics[angle=270,scale=0.40]{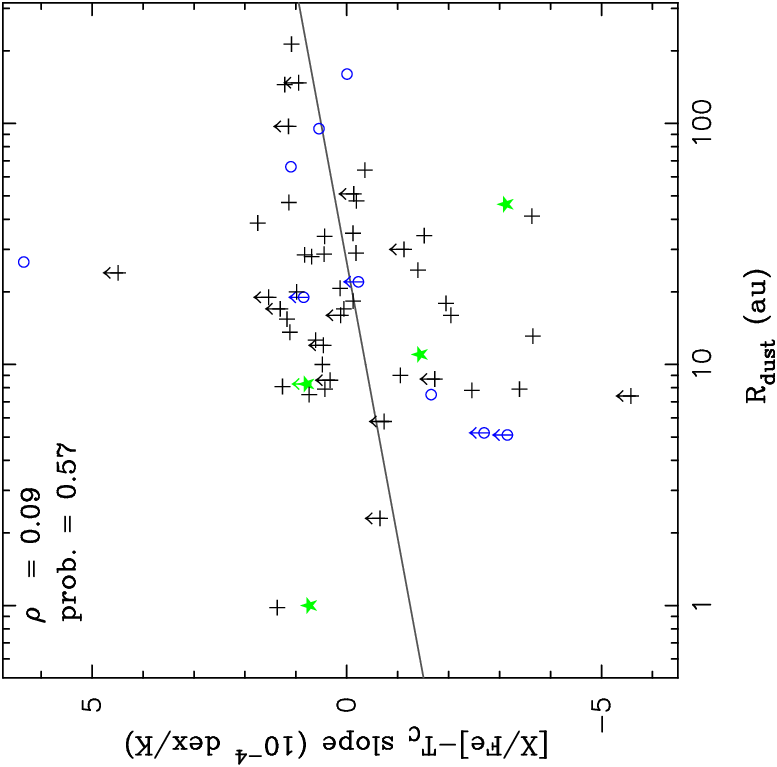}
\end{minipage}
\caption{
 T$^{\rm all}_{\rm C}$-slope as a function of the dust parameters,
 fractional dust luminosity L$_{\rm dust}$/L$_{\star}$ (left panel),
 dust temperature T$_{\rm dust}$ (middle panel), and
 disc radius R$_{\rm disc}$.
 A linear fit to the data is shown
 by the grey continuous line. Results from the Spearman's
 correlation test are also shown in the plot
 (lower/upper limits are not considered in the statistics).
 }
\label{tc_fdust_properties}
\end{figure*}

   In the framework of ME09 hypothesis, stars hosting low-mass planets
   should show negative <[X/Fe]>-T$_{\rm C}$ slopes. 
   Since low-mass planets might be common in SWDS, 
   a search for correlations between the properties of the dust
   and T$^{\rm all}_{\rm C}$-slope of the stars with discs (SWDs
   and SWDPs) was performed. 
   Basic physical parameters of the discs, fractional dust luminosity
   L$_{\rm dust}$/L$_{\star}$, dust temperature T$_{\rm dust}$, and disc
   radius R$_{\rm disc}$ are taken from \cite{2013A&A...555A..11E} when possible
   since it constitutes an homogeneous
   database of debris disc parameters.
   Otherwise, these values are taken from the literature 
   (see MA12 and references therein).
   The results from the statistical tests, see Figure~\ref{tc_fdust_properties},
   show however no clear correlation
   between the properties of the discs and the chemical composition
   of the star.
   In addition, no significant difference in any of the considered
   properties between the stars with debris discs and cool Jupiters
   and the SWDs has been found\footnote{There are very few
   discs and low-mass/hot Jupiters planets hosts with data
   for performing a K-S test. Stars with lower/upper limits
   have not been considered in the statistics.}.

   We should note at this point that the fact that we do not find a clear
   chemical fingerprint of low-mass
   planet formation in the SWDs as  a whole does not contradict 
   the idea that low-mass planets and discs might be a correlated
   phenomenon. 
   First, ME09 interpretation should be further confirmed. 
   Second, there are several biases in our analysis since the
   values of the dust properties are taken from different sources.
   Furthermore values of the disc radius are usually computed by
   assuming black body emission which is known to underestimate
   the radial distance of the dust from the star by a factor of up
   to four around G stars 
   \citep{2011A&A...529A.117M,2012MNRAS.424.1206W}.    
   Finally, the chemical depletions we are looking for are small ($\sim$ 0.08 dex)
   and in most cases have been found by a differential analysis
   between stars with very similar parameters (and mostly solar
   twins). 
   A more detailed differential analysis of individual SWD
   stars with respect to their corresponding SWODs twins is deferred to
   a forthcoming
   paper. 

   The second observational result of this work,
   i.e, the slopes of stars hosting low-mass planets do not seem
   to support ME09 hypothesis.
   We do find low-mass planets hosts to show negative slopes, but only
   when all elements are considered and, more important, SWDs and SWODs
   do also show negative slopes in this case. 
   Moreover, our data suggests that the stars hosting cool Jupiters are 
   the ones that show a different <[X/Fe]>-T$_{\rm C}$ trend with respect
   to the non planet host samples. This is true when all elements are considered
   and also when the analysis is restricted to refractory elements. In this case,
   the slope of the samples hosting cool Jupiters are negative. 

   Finally, the positive slope in stars hosting hot Jupiters should be interpreted
   with caution given the low statistics of our sample, but also because
   SWODs and SWDs also show clear positive slopes in the  T$^{\rm ref}_{\rm C}$ analysis. 
   We simply note that a significant positive slope
   in the framework of ME09 hypothesis can be interpreted as an indication of the non
   presence of low-mass planets. 

\subsection{Signatures of pollution}

 A correlation between elemental abundances and condensation temperature
 is a natural prediction of the self-enrichment hypothesis for the
 gas-giant planet metallicity correlation \citep{1997MNRAS.285..403G}.
 This is because the accretion of material by a star is expected
 to occur close to the star, a rather high temperature environment.
 Therefore, refractory elements might be added preferentially
 when compared to volatile elements. 
 The abundance pattern of hotter dwarfs constitutes
 an important test for this scenario. These stars 
 are known to have narrower convective envelopes (i.e., to experience less
 mixing) and, therefore, the chemical signature in the <[X/Fe]>-T$_{\rm C}$
 trend suggested in planet hosts should be more significant in these stars than
 in late-type stars.

 The stellar
 radii is used here as a proxy of the convective envelope' size. 
 In main-sequence stars it is larger for early type stars whilst
 it diminishes for late-type stars. 
 The stellar radii have been computed as explained in Section~\ref{previous_works}.
 Our sample of planet hosts, i.e, the sum of the SWP and the SWDP 
 samples (irrespective of the planet's type) has been
 divided into three categories according to their radii:
 {\it i) F-stars}, with R$_{\star}$ $>$ 1.12 R$_{\odot}$;  
 {\it ii) G-stars}, with  0.91 R$_{\odot}$ $<$ R$_{\star}$ $<$ 1.12 R$_{\odot}$;
 and {\it iii) K-stars}, with 0.91 R$_{\odot}$ $<$ R$_{\star}$
 \citep[see][Table~B.1]{2008oasp.book.....G}.

\begin{figure*}
\centering
\begin{minipage}{0.47\linewidth}
\includegraphics[angle=270,scale=0.45]{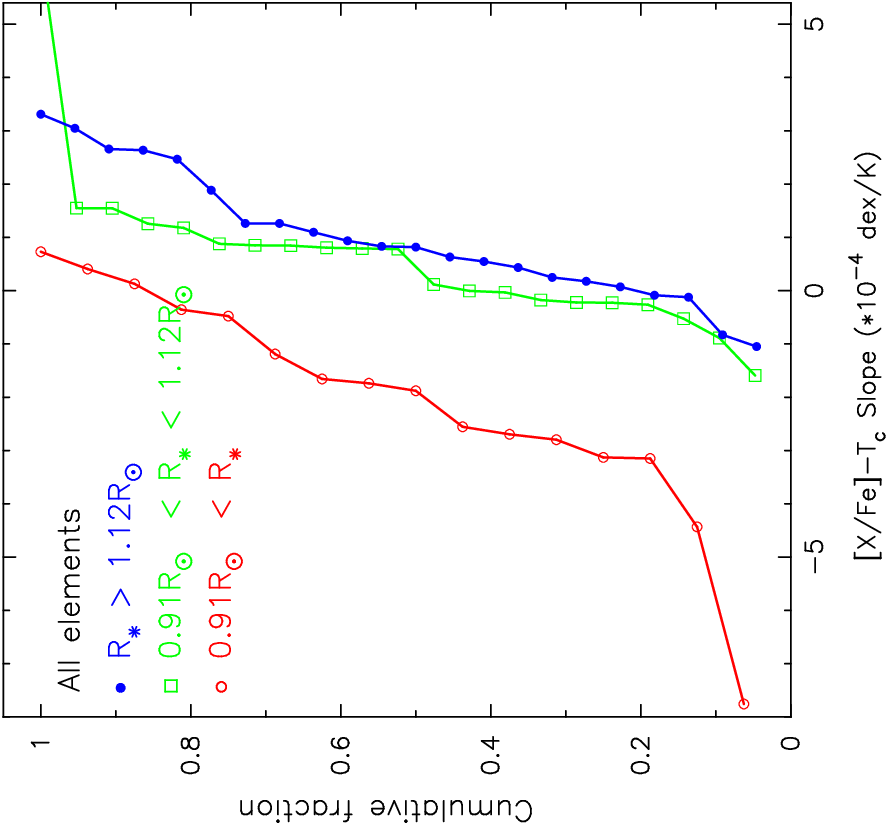}
\end{minipage}
\begin{minipage}{0.47\linewidth}
\includegraphics[angle=270,scale=0.45]{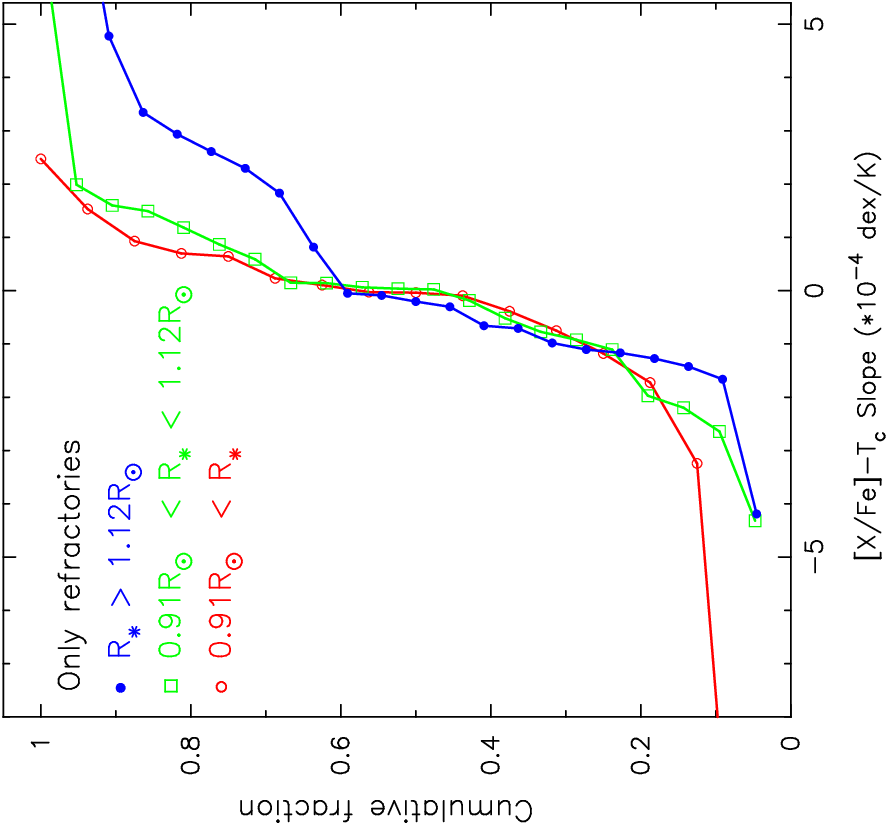}
\end{minipage}
\caption{
 Histogram of [X/Fe]-T$_{\rm C}$ slopes for all planet hosts as a function
 of the stellar radius
 when all elements (volatiles plus refractories)
 are taken into account (left) and when only refractories are considered (right).
}
\label{tc_distribuciones_radious}
\end{figure*}

 The cumulative distribution functions of
 T$^{\rm all}_{\rm C}$-slope, and T$^{\rm refrac}_{\rm C}$-slope of
 these three subsamples are compared in Figure~\ref{tc_distribuciones_radious}.
 It is clear from this figure that the only significant difference is in the distribution
 of T$^{\rm all}_{\rm C}$-slope,
 where K-stars show more negative slopes than the other two samples
 (with K-S $p$-values of $\sim$ 10$^{\rm -5}$ when comparing K  with
 F stars, and  $\sim$ 10$^{\rm -4}$ when comparing K with G stars).
 This point argues against the pollution hypothesis. When considering all
 elements, negative slopes are signatures of a possible  deficit of refractory element with respect to
 volatiles. If this deficit is not primordial but rather due to the later accretion
 of refractory-depleted material, stars with narrower convective zones (early F)
 should show higher levels of depletion, or in other words, more negative slopes.
 Furthermore, no difference in the T$^{\rm refrac}_{\rm C}$-slope distributions
 is found between the F, G, and K stars samples.
 And thus the pollution hypothesis is not sustained with the data at hand.
    
\subsection{Trends with planetary properties}
\label{seccion_trends_dust_planets}

\begin{figure}
\centering
\includegraphics[angle=270,scale=0.55]{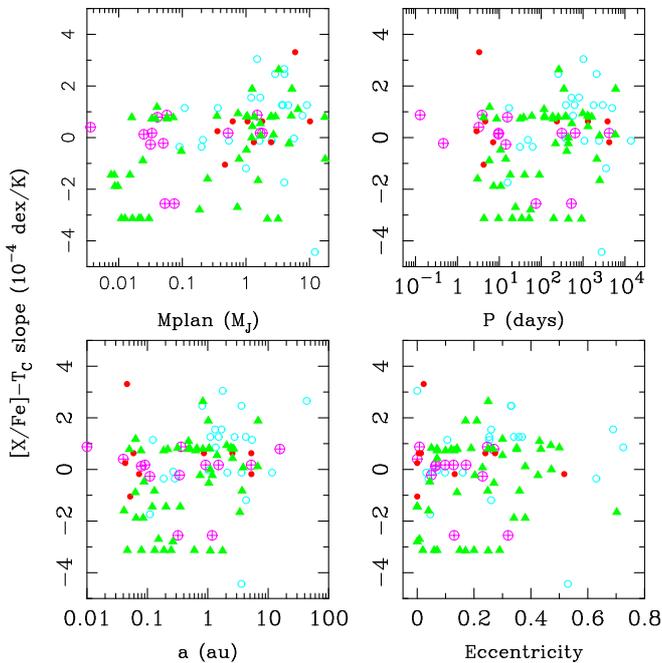}
\caption{
  T$^{\rm all}_{\rm C}$-slope as a function of
  the planetary properties. For multiplanet
  systems, all planets are plotted.
  Colours and symbols are as in Figure~\ref{distribuciones_acumuladas}.
}
\label{tc_vs_planetary_properties}
\end{figure}

  Finally, the stellar [X/Fe] vs. T$^{\rm all}_{\rm C}$-slopes are 
  plotted in Figure~\ref{tc_vs_planetary_properties}
  as a function of the planetary properties,
  minimum mass, period, semimajor axis, and eccentricity. 
  No clear trend has been identified, in agreement with previous
  works \citep[e.g.][]{2006A&A...449..809E}.
  
\section{Summary}
\label{conclusions}

  In this work a detailed chemical analysis of stars with dusty debris discs
  has been presented. Their chemical abundances have been compared 
  to those of stars with planets, stars harbouring debris discs and planets,
  and stars with neither debris nor planets. 

  No clear difference have been found in
  metallicity, individual abundances or <[X/Fe]>-T$_{\rm C}$ trends
  between SWDs and SWODs. The behaviour of SWDPs seems to be driven by the type of planet
  (cool Jupiter or low-mass planet).
  This is in agreement with the core-accretion model for planet formation 
  in which the conditions required to form debris discs are more easily
  met than the conditions to form gas-giant planets.
  The fact that SWDs as stars with low-mass planets do not show metal
  enhancement might indicate a correlation between both phenomena.
  Giant planets in eccentric orbits might produce dynamical instabilities
  which can clear out the inner and outer parts of a debris discs a fact that 
  might explain the lack of a clear correlation between debris discs
  and more massive planets.

  We find tentative different behaviours in  <[X/Fe]>-T$_{\rm C}$ trends 
  between the samples of stars with planets and the samples of stars without planets.
  Stars with cool giant planets seem to behave in a different way
  with respect to the samples of stars without planets. 
  This result holds independently of whether all elements or only refractories
  are considered. 
  In particular, when only refractory elements are considered stars with cool
  planets show negative slopes.
  Regarding stars with exclusively low-mass planets, we find them to behave
  as the non planet hosts samples. 
  Despite our low statistics,
  stars hosting exclusively close-in giant planets seem to show higher metallicities
  than stars harbouring more distant planets. Furthermore, they show positive
  T$_{\rm C}$ slopes although this trend should be further investigated. 

  Finally, we should put some words of caution about the interpretation
  of the negative slopes as a signature of 
  planet formation
  since the derived trends show relatively low statistical significance levels
  and the T$_{\rm C}$ slopes show a moderate but significant
  correlation with the stellar metallicity. 
  Even after correction for these possible effect, a relatively
  weak correlation between T$_{\rm C}$-slope with the stellar age and
  a moderate one with the stellar radius remains.

\begin{acknowledgements}

  J. M. acknowledges support from the Italian Ministry of Education,
  University, and Research  through the  
  \emph{Premiale HARPS-N} research project under grant
  \emph{Ricerca di pianeti intorno a stelle di piccola massa}.
  Additional founding from the Spanish Ministerio
  de Econom\'ia y Competitividad under grant
  \emph{AYA2011-26202} is also acknowledged. 
  E. V. acknowledges support from the Spanish Ministerio
  de Econom\'ia y Competitividad under grant
  \emph{AYA2013-45347P}. 
  The authors would like to thank Robert. L. Kurucz, Sergio Sousa, Yoichi Takeda,
  Chris Sneden, and L\'eo Girardi, for making their codes publicly available. Jean Schneider
  is also acknowledged for maintaining the Extrasolar Planets Encyclopedia.  
  This research has also made use of the Exoplanet Orbit Database
  and the Exoplanet Data Explorer at exoplanets.org,
  the VizieR catalogue access tool, CDS, Strasbourg, France,
  as well as the NASA's Astrophysics Data System Bibliographic Services.

\end{acknowledgements}


\bibliographystyle{aa}
\bibliography{chemical_debris.bib}

\begin{thebibliography}{104}
\expandafter\ifx\csname natexlab\endcsname\relax\def\natexlab#1{#1}\fi

\bibitem[{{Adibekyan} {et~al.}(2013){Adibekyan}, {Figueira}, {Santos},
  {Mortier}, {Mordasini}, {Delgado Mena}, {Sousa}, {Correia}, {Israelian}, \&
  {Oshagh}}]{2013A&A...560A..51A}
{Adibekyan}, V.~Z., {Figueira}, P., {Santos}, N.~C., {et~al.} 2013, \aap, 560,
  A51

\bibitem[{{Adibekyan} {et~al.}(2014){Adibekyan}, {Gonz{\'a}lez Hern{\'a}ndez},
  {Delgado Mena}, {Sousa}, {Santos}, {Israelian}, {Figueira}, \& {Bertran de
  Lis}}]{2014A&A...564L..15A}
{Adibekyan}, V.~Z., {Gonz{\'a}lez Hern{\'a}ndez}, J.~I., {Delgado Mena}, E.,
  {et~al.} 2014, \aap, 564, L15

\bibitem[{{Allende Prieto} {et~al.}(2004){Allende Prieto}, {Barklem},
  {Lambert}, \& {Cunha}}]{s4n}
{Allende Prieto}, C., {Barklem}, P.~S., {Lambert}, D.~L., \& {Cunha}, K. 2004,
  \aap, 420, 183

\bibitem[{{Asplund} {et~al.}(2009){Asplund}, {Grevesse}, {Sauval}, \&
  {Scott}}]{2009ARA&A..47..481A}
{Asplund}, M., {Grevesse}, N., {Sauval}, A.~J., \& {Scott}, P. 2009, \araa, 47,
  481

\bibitem[{{Aumann} {et~al.}(1984){Aumann}, {Beichman}, {Gillett}, {de Jong},
  {Houck}, {Low}, {Neugebauer}, {Walker}, \& {Wesselius}}]{1984ApJ...278L..23A}
{Aumann}, H.~H., {Beichman}, C.~A., {Gillett}, F.~C., {et~al.} 1984, \apjl,
  278, L23

\bibitem[{{Backman} \& {Paresce}(1993)}]{1993prpl.conf.1253B}
{Backman}, D.~E. \& {Paresce}, F. 1993, in Protostars and Planets III, ed.
  E.~H. {Levy} \& J.~I. {Lunine}, 1253--1304

\bibitem[{{Beichman} {et~al.}(2005){Beichman}, {Bryden}, {Rieke}, {Stansberry},
  {Trilling}, {Stapelfeldt}, {Werner}, {Engelbracht}, {Blaylock}, {Gordon},
  {Chen}, {Su}, \& {Hines}}]{2005ApJ...622.1160B}
{Beichman}, C.~A., {Bryden}, G., {Rieke}, G.~H., {et~al.} 2005, \apj, 622, 1160

\bibitem[{{Boisse} {et~al.}(2012){Boisse}, {Pepe}, {Perrier}, {Queloz},
  {Bonfils}, {Bouchy}, {Santos}, {Arnold}, {Beuzit}, {D{\'{\i}}az}, {Delfosse},
  {Eggenberger}, {Ehrenreich}, {Forveille}, {H{\'e}brard}, {Lagrange}, {Lovis},
  {Mayor}, {Moutou}, {Naef}, {Santerne}, {S{\'e}gransan}, {Sivan}, \&
  {Udry}}]{2012A&A...545A..55B}
{Boisse}, I., {Pepe}, F., {Perrier}, C., {et~al.} 2012, \aap, 545, A55

\bibitem[{{Bonsor} {et~al.}(2014){Bonsor}, {Kennedy}, {Wyatt}, {Johnson}, \&
  {Sibthorpe}}]{2014MNRAS.437.3288B}
{Bonsor}, A., {Kennedy}, G.~M., {Wyatt}, M.~C., {Johnson}, J.~A., \&
  {Sibthorpe}, B. 2014, \mnras, 437, 3288

\bibitem[{{Bressan} {et~al.}(2012){Bressan}, {Marigo}, {Girardi}, {Salasnich},
  {Dal Cero}, {Rubele}, \& {Nanni}}]{2012MNRAS.427..127B}
{Bressan}, A., {Marigo}, P., {Girardi}, L., {et~al.} 2012, \mnras, 427, 127

\bibitem[{{Bryden} {et~al.}(2009){Bryden}, {Beichman}, {Carpenter}, {Rieke},
  {Stapelfeldt}, {Werner}, {Tanner}, {Lawler}, {Wyatt}, {Trilling}, {Su},
  {Blaylock}, \& {Stansberry}}]{2009ApJ...705.1226B}
{Bryden}, G., {Beichman}, C.~A., {Carpenter}, J.~M., {et~al.} 2009, \apj, 705,
  1226

\bibitem[{{Buchhave} {et~al.}(2012){Buchhave}, {Latham}, {Johansen},
  {Bizzarro}, {Torres}, {Rowe}, {Batalha}, {Borucki}, {Brugamyer}, {Caldwell},
  {Bryson}, {Ciardi}, {Cochran}, {Endl}, {Esquerdo}, {Ford}, {Geary},
  {Gilliland}, {Hansen}, {Isaacson}, {Laird}, {Lucas}, {Marcy}, {Morse},
  {Robertson}, {Shporer}, {Stefanik}, {Still}, \&
  {Quinn}}]{2012Natur.486..375B}
{Buchhave}, L.~A., {Latham}, D.~W., {Johansen}, A., {et~al.} 2012, \nat, 486,
  375

\bibitem[{{Casagrande} {et~al.}(2010){Casagrande}, {Ram{\'{\i}}rez},
  {Mel{\'e}ndez}, {Bessell}, \& {Asplund}}]{2010A&A...512A..54C}
{Casagrande}, L., {Ram{\'{\i}}rez}, I., {Mel{\'e}ndez}, J., {Bessell}, M., \&
  {Asplund}, M. 2010, \aap, 512, A54

\bibitem[{{Casagrande} {et~al.}(2011){Casagrande}, {Sch{\"o}nrich}, {Asplund},
  {Cassisi}, {Ram{\'{\i}}rez}, {Mel{\'e}ndez}, {Bensby}, \&
  {Feltzing}}]{2011A&A...530A.138C}
{Casagrande}, L., {Sch{\"o}nrich}, R., {Asplund}, M., {et~al.} 2011, \aap, 530,
  A138

\bibitem[{{Cassan} {et~al.}(2012){Cassan}, {Kubas}, {Beaulieu}, {Dominik},
  {Horne}, {Greenhill}, {Wambsganss}, {Menzies}, {Williams}, {J{\o}rgensen},
  {Udalski}, {Bennett}, {Albrow}, {Batista}, {Brillant}, {Caldwell}, {Cole},
  {Coutures}, {Cook}, {Dieters}, {Prester}, {Donatowicz}, {Fouqu{\'e}}, {Hill},
  {Kains}, {Kane}, {Marquette}, {Martin}, {Pollard}, {Sahu}, {Vinter},
  {Warren}, {Watson}, {Zub}, {Sumi}, {Szyma{\'n}ski}, {Kubiak}, {Poleski},
  {Soszynski}, {Ulaczyk}, {Pietrzy{\'n}ski}, \&
  {Wyrzykowski}}]{2012Natur.481..167C}
{Cassan}, A., {Kubas}, D., {Beaulieu}, J.-P., {et~al.} 2012, \nat, 481, 167

\bibitem[{{Chavero} {et~al.}(2006){Chavero}, {G{\'o}mez}, {Whitney}, \&
  {Saffe}}]{2006A&A...452..921C}
{Chavero}, C., {G{\'o}mez}, M., {Whitney}, B.~A., \& {Saffe}, C. 2006, \aap,
  452, 921

\bibitem[{{Currie}(2009)}]{2009ApJ...694L.171C}
{Currie}, T. 2009, \apjl, 694, L171

\bibitem[{{da Silva} {et~al.}(2006){da Silva}, {Girardi}, {Pasquini},
  {Setiawan}, {von der L{\"u}he}, {de Medeiros}, {Hatzes}, {D{\"o}llinger}, \&
  {Weiss}}]{2006A&A...458..609D}
{da Silva}, L., {Girardi}, L., {Pasquini}, L., {et~al.} 2006, \aap, 458, 609

\bibitem[{{Ecuvillon} {et~al.}(2006){Ecuvillon}, {Israelian}, {Santos},
  {Mayor}, \& {Gilli}}]{2006A&A...449..809E}
{Ecuvillon}, A., {Israelian}, G., {Santos}, N.~C., {Mayor}, M., \& {Gilli}, G.
  2006, \aap, 449, 809

\bibitem[{{Eiroa} {et~al.}(2010){Eiroa}, {Fedele}, {Maldonado},
  {Gonz{\'a}lez-Garc{\'{\i}}a}, {Rodmann}, {Heras}, {Pilbratt}, {Augereau},
  {Mora}, {Montesinos}, {Ardila}, {Bryden}, {Liseau}, {Stapelfeldt},
  {Launhardt}, {Solano}, {Bayo}, {Absil}, {Ar{\'e}valo}, {Barrado},
  {Beichmann}, {Danchi}, {Del Burgo}, {Ertel}, {Fridlund}, {Fukagawa},
  {Guti{\'e}rrez}, {Gr{\"u}n}, {Kamp}, {Krivov}, {Lebreton}, {L{\"o}hne},
  {Lorente}, {Marshall}, {Mart{\'{\i}}nez-Arn{\'a}iz}, {Meeus}, {Montes},
  {Morbidelli}, {M{\"u}ller}, {Mutschke}, {Nakagawa}, {Olofsson}, {Ribas},
  {Roberge}, {Sanz-Forcada}, {Th{\'e}bault}, {Walker}, {White}, \&
  {Wolf}}]{2010A&A...518L.131E}
{Eiroa}, C., {Fedele}, D., {Maldonado}, J., {et~al.} 2010, \aap, 518, L131

\bibitem[{{Eiroa} {et~al.}(2013){Eiroa}, {Marshall}, {Mora}, {Montesinos},
  {Absil}, {Augereau}, {Bayo}, {Bryden}, {Danchi}, {del Burgo}, {Ertel},
  {Fridlund}, {Heras}, {Krivov}, {Launhardt}, {Liseau}, {L{\"o}hne},
  {Maldonado}, {Pilbratt}, {Roberge}, {Rodmann}, {Sanz-Forcada}, {Solano},
  {Stapelfeldt}, {Th{\'e}bault}, {Wolf}, {Ardila}, {Ar{\'e}valo}, {Beichmann},
  {Faramaz}, {Gonz{\'a}lez-Garc{\'{\i}}a}, {Guti{\'e}rrez}, {Lebreton},
  {Mart{\'{\i}}nez-Arn{\'a}iz}, {Meeus}, {Montes}, {Olofsson}, {Su}, {White},
  {Barrado}, {Fukagawa}, {Gr{\"u}n}, {Kamp}, {Lorente}, {Morbidelli},
  {M{\"u}ller}, {Mutschke}, {Nakagawa}, {Ribas}, \&
  {Walker}}]{2013A&A...555A..11E}
{Eiroa}, C., {Marshall}, J.~P., {Mora}, A., {et~al.} 2013, \aap, 555, A11

\bibitem[{{Fischer} \& {Valenti}(2005)}]{2005ApJ...622.1102F}
{Fischer}, D.~A. \& {Valenti}, J. 2005, \apj, 622, 1102

\bibitem[{{Frandsen} \& {Lindberg}(1999)}]{1999anot.conf...71F}
{Frandsen}, S. \& {Lindberg}, B. 1999, in Astrophysics with the NOT, ed.
  {H.~Karttunen \& V.~Piirola}, 71

\bibitem[{{G{\"a}nsicke} {et~al.}(2012){G{\"a}nsicke}, {Koester}, {Farihi},
  {Girven}, {Parsons}, \& {Breedt}}]{2012MNRAS.424..333G}
{G{\"a}nsicke}, B.~T., {Koester}, D., {Farihi}, J., {et~al.} 2012, \mnras, 424,
  333

\bibitem[{{Ghezzi} {et~al.}(2010){Ghezzi}, {Cunha}, {Smith}, {de Ara{\'u}jo},
  {Schuler}, \& {de la Reza}}]{2010ApJ...720.1290G}
{Ghezzi}, L., {Cunha}, K., {Smith}, V.~V., {et~al.} 2010, \apj, 720, 1290

\bibitem[{{Gonzalez}(1997)}]{1997MNRAS.285..403G}
{Gonzalez}, G. 1997, \mnras, 285, 403

\bibitem[{{Gonzalez}(1998)}]{1998A&A...334..221G}
{Gonzalez}, G. 1998, \aap, 334, 221

\bibitem[{{Gonzalez}(2011)}]{2011MNRAS.416L..80G}
{Gonzalez}, G. 2011, \mnras, 416, L80

\bibitem[{{Gonzalez} {et~al.}(2010){Gonzalez}, {Carlson}, \&
  {Tobin}}]{2010MNRAS.407..314G}
{Gonzalez}, G., {Carlson}, M.~K., \& {Tobin}, R.~W. 2010, \mnras, 407, 314

\bibitem[{{Gonz{\'a}lez Hern{\'a}ndez} {et~al.}(2013){Gonz{\'a}lez
  Hern{\'a}ndez}, {Delgado-Mena}, {Sousa}, {Israelian}, {Santos}, {Adibekyan},
  \& {Udry}}]{2013A&A...552A...6G}
{Gonz{\'a}lez Hern{\'a}ndez}, J.~I., {Delgado-Mena}, E., {Sousa}, S.~G.,
  {et~al.} 2013, \aap, 552, A6

\bibitem[{{Gonz{\'a}lez Hern{\'a}ndez} {et~al.}(2010){Gonz{\'a}lez
  Hern{\'a}ndez}, {Israelian}, {Santos}, {Sousa}, {Delgado-Mena}, {Neves}, \&
  {Udry}}]{2010ApJ...720.1592G}
{Gonz{\'a}lez Hern{\'a}ndez}, J.~I., {Israelian}, G., {Santos}, N.~C., {et~al.}
  2010, \apj, 720, 1592

\bibitem[{{Gratton} {et~al.}(2001){Gratton}, {Bonanno}, {Bruno}, {Cali},
  {Claudi}, {Cosentino}, {Desidera}, {Diego}, {Farisato}, {Martorana},
  {Rebeschini}, \& {Scuderi}}]{sarg}
{Gratton}, R.~G., {Bonanno}, G., {Bruno}, P., {et~al.} 2001, Experimental
  Astronomy, 12, 107

\bibitem[{{Gray}(2008)}]{2008oasp.book.....G}
{Gray}, D.~F. 2008, {The Observation and Analysis of Stellar Photospheres}

\bibitem[{{Greaves} {et~al.}(2006){Greaves}, {Fischer}, \&
  {Wyatt}}]{2006MNRAS.366..283G}
{Greaves}, J.~S., {Fischer}, D.~A., \& {Wyatt}, M.~C. 2006, \mnras, 366, 283

\bibitem[{{Grevesse} {et~al.}(2014){Grevesse}, {Scott}, {Asplund}, \&
  {Sauval}}]{2014arXiv1405.0288G}
{Grevesse}, N., {Scott}, P., {Asplund}, M., \& {Sauval}, A.~J. 2014, ArXiv
  e-prints

\bibitem[{{Haywood}(2009)}]{2009ApJ...698L...1H}
{Haywood}, M. 2009, \apjl, 698, L1

\bibitem[{{Hubickyj} {et~al.}(2005){Hubickyj}, {Bodenheimer}, \&
  {Lissauer}}]{2005Icar..179..415H}
{Hubickyj}, O., {Bodenheimer}, P., \& {Lissauer}, J.~J. 2005, \icarus, 179, 415

\bibitem[{{Ida} \& {Lin}(2004)}]{2004ApJ...616..567I}
{Ida}, S. \& {Lin}, D.~N.~C. 2004, \apj, 616, 567

\bibitem[{{Jewitt} {et~al.}(2009){Jewitt}, {Moro-Mart{\`i}n}, \&
  {Lacerda}}]{2009and..book...53J}
{Jewitt}, D., {Moro-Mart{\`i}n}, A., \& {Lacerda}, P. 2009, {The Kuiper Belt
  and Other Debris Disks}, ed. H.~A. {Thronson}, M.~{Stiavelli}, \&
  A.~{Tielens}, 53

\bibitem[{{Kiselman}(1993)}]{1993A&A...275..269K}
{Kiselman}, D. 1993, \aap, 275, 269

\bibitem[{{Kiselman}(2001)}]{2001NewAR..45..559K}
{Kiselman}, D. 2001, \nar, 45, 559

\bibitem[{{K{\'o}sp{\'a}l} {et~al.}(2009){K{\'o}sp{\'a}l}, {Ardila},
  {Mo{\'o}r}, \& {{\'A}brah{\'a}m}}]{2009ApJ...700L..73K}
{K{\'o}sp{\'a}l}, {\'A}., {Ardila}, D.~R., {Mo{\'o}r}, A., \&
  {{\'A}brah{\'a}m}, P. 2009, \apjl, 700, L73

\bibitem[{{Kupka} {et~al.}(1999){Kupka}, {Piskunov}, {Ryabchikova}, {Stempels},
  \& {Weiss}}]{1999A&AS..138..119K}
{Kupka}, F., {Piskunov}, N., {Ryabchikova}, T.~A., {Stempels}, H.~C., \&
  {Weiss}, W.~W. 1999, \aaps, 138, 119

\bibitem[{{Kurucz}(1993)}]{1993KurCD..13.....K}
{Kurucz}, R. 1993, ATLAS9 Stellar Atmosphere Programs and 2 km/s grid.~Kurucz
  CD-ROM No.~13.~ Cambridge, Mass.: Smithsonian Astrophysical Observatory,
  1993., 13

\bibitem[{{Lestrade} {et~al.}(2012){Lestrade}, {Matthews}, {Sibthorpe},
  {Kennedy}, {Wyatt}, {Bryden}, {Greaves}, {Thilliez}, {Moro-Mart{\'{\i}}n},
  {Booth}, {Dent}, {Duch{\^e}ne}, {Harvey}, {Horner}, {Kalas}, {Kavelaars},
  {Phillips}, {Rodriguez}, {Su}, \& {Wilner}}]{2012A&A...548A..86L}
{Lestrade}, J.-F., {Matthews}, B.~C., {Sibthorpe}, B., {et~al.} 2012, \aap,
  548, A86

\bibitem[{{Lodders}(2003)}]{2003ApJ...591.1220L}
{Lodders}, K. 2003, \apj, 591, 1220

\bibitem[{{Luhman} {et~al.}(2007){Luhman}, {Patten}, {Marengo}, {Schuster},
  {Hora}, {Ellis}, {Stauffer}, {Sonnett}, {Winston}, {Gutermuth}, {Megeath},
  {Backman}, {Henry}, {Werner}, \& {Fazio}}]{2007ApJ...654..570L}
{Luhman}, K.~L., {Patten}, B.~M., {Marengo}, M., {et~al.} 2007, \apj, 654, 570

\bibitem[{{Maldonado} {et~al.}(2012){Maldonado}, {Eiroa}, {Villaver},
  {Montesinos}, \& {Mora}}]{2012A&A...541A..40M}
{Maldonado}, J., {Eiroa}, C., {Villaver}, E., {Montesinos}, B., \& {Mora}, A.
  2012, \aap, 541, A40

\bibitem[{{Maldonado} {et~al.}(2010){Maldonado}, {Mart{\'{\i}}nez-Arn{\'a}iz},
  {Eiroa}, {Montes}, \& {Montesinos}}]{2010A&A...521A..12M}
{Maldonado}, J., {Mart{\'{\i}}nez-Arn{\'a}iz}, R.~M., {Eiroa}, C., {Montes},
  D., \& {Montesinos}, B. 2010, \aap, 521, A12

\bibitem[{{Maldonado} {et~al.}(2013){Maldonado}, {Villaver}, \&
  {Eiroa}}]{2013A&A...554A..84M}
{Maldonado}, J., {Villaver}, E., \& {Eiroa}, C. 2013, \aap, 554, A84

\bibitem[{{Maldonado} {et~al.}(2014){Maldonado}, {Villaver}, \&
  {Eiroa}}]{IWSSL}
{Maldonado}, J., {Villaver}, E., \& {Eiroa}, C. 2014, in Astronomical Society
  of India Conference Series, Vol.~11, Astronomical Society of India Conference
  Series, 167--174

\bibitem[{{Mamajek} \& {Hillenbrand}(2008)}]{2008ApJ...687.1264M}
{Mamajek}, E.~E. \& {Hillenbrand}, L.~A. 2008, \apj, 687, 1264

\bibitem[{{Marshall} {et~al.}(2011){Marshall}, {L{\"o}hne}, {Montesinos},
  {Krivov}, {Eiroa}, {Absil}, {Bryden}, {Maldonado}, {Mora}, {Sanz-Forcada},
  {Ardila}, {Augereau}, {Bayo}, {Del Burgo}, {Danchi}, {Ertel}, {Fedele},
  {Fridlund}, {Lebreton}, {Gonz{\'a}lez-Garc{\'{\i}}a}, {Liseau}, {Meeus},
  {M{\"u}ller}, {Pilbratt}, {Roberge}, {Stapelfeldt}, {Th{\'e}bault}, {White},
  \& {Wolf}}]{2011A&A...529A.117M}
{Marshall}, J.~P., {L{\"o}hne}, T., {Montesinos}, B., {et~al.} 2011, \aap, 529,
  A117

\bibitem[{{Marshall} {et~al.}(2014){Marshall}, {Moro-Mart{\'{\i}}n}, {Eiroa},
  {Kennedy}, {Mora}, {Sibthorpe}, {Lestrade}, {Maldonado}, {Sanz-Forcada},
  {Wyatt}, {Matthews}, {Horner}, {Montesinos}, {Bryden}, {del Burgo},
  {Greaves}, {Ivison}, {Meeus}, {Olofsson}, {Pilbratt}, \&
  {White}}]{2014A&A...565A..15M}
{Marshall}, J.~P., {Moro-Mart{\'{\i}}n}, A., {Eiroa}, C., {et~al.} 2014, \aap,
  565, A15

\bibitem[{{Mart{\'{\i}}nez-Arn{\'a}iz}
  {et~al.}(2010){Mart{\'{\i}}nez-Arn{\'a}iz}, {Maldonado}, {Montes}, {Eiroa},
  \& {Montesinos}}]{2010A&A...520A..79M}
{Mart{\'{\i}}nez-Arn{\'a}iz}, R., {Maldonado}, J., {Montes}, D., {Eiroa}, C.,
  \& {Montesinos}, B. 2010, \aap, 520, A79

\bibitem[{{Matthews} {et~al.}(2014){Matthews}, {Krivov}, {Wyatt}, {Bryden}, \&
  {Eiroa}}]{2014arXiv1401.0743M}
{Matthews}, B.~C., {Krivov}, A.~V., {Wyatt}, M.~C., {Bryden}, G., \& {Eiroa},
  C. 2014, ArXiv e-prints

\bibitem[{{Mayor} {et~al.}(2011){Mayor}, {Marmier}, {Lovis}, {Udry},
  {S{\'e}gransan}, {Pepe}, {Benz}, {Bertaux}, {Bouchy}, {Dumusque}, {Lo Curto},
  {Mordasini}, {Queloz}, \& {Santos}}]{2011arXiv1109.2497M}
{Mayor}, M., {Marmier}, M., {Lovis}, C., {et~al.} 2011, ArXiv e-prints

\bibitem[{{Mel{\'e}ndez} {et~al.}(2009){Mel{\'e}ndez}, {Asplund}, {Gustafsson},
  \& {Yong}}]{2009ApJ...704L..66M}
{Mel{\'e}ndez}, J., {Asplund}, M., {Gustafsson}, B., \& {Yong}, D. 2009, \apjl,
  704, L66

\bibitem[{{Mordasini} {et~al.}(2009){Mordasini}, {Alibert}, \&
  {Benz}}]{2009A&A...501.1139M}
{Mordasini}, C., {Alibert}, Y., \& {Benz}, W. 2009, \aap, 501, 1139

\bibitem[{{Mordasini} {et~al.}(2012){Mordasini}, {Alibert}, {Benz}, {Klahr}, \&
  {Henning}}]{2012A&A...541A..97M}
{Mordasini}, C., {Alibert}, Y., {Benz}, W., {Klahr}, H., \& {Henning}, T. 2012,
  \aap, 541, A97

\bibitem[{{Moro-Martin}(2013)}]{2013pss3.book..431M}
{Moro-Martin}, A. 2013, {Dusty Planetary Systems}, ed. T.~D. {Oswalt}, L.~M.
  {French}, \& P.~{Kalas}, 431

\bibitem[{{Moro-Mart{\'{\i}}n} {et~al.}(2007){Moro-Mart{\'{\i}}n}, {Carpenter},
  {Meyer}, {Hillenbrand}, {Malhotra}, {Hollenbach}, {Najita}, {Henning}, {Kim},
  {Bouwman}, {Silverstone}, {Hines}, {Wolf}, {Pascucci}, {Mamajek}, \&
  {Lunine}}]{2007ApJ...658.1312M}
{Moro-Mart{\'{\i}}n}, A., {Carpenter}, J.~M., {Meyer}, M.~R., {et~al.} 2007,
  \apj, 658, 1312

\bibitem[{{Moro-Mart{\'{\i}}n} {et~al.}(2015){Moro-Mart{\'{\i}}n}, {Marshall},
  {Kennedy}, {Sibthorpe}, {Matthews}, {Eiroa}, {Wyatt}, {Lestrade},
  {Maldonado}, {Rodriguez}, {Greaves}, {Montesinos}, {Mora}, {Booth},
  {Duchene}, {Wilner}, \& {Horner}}]{2015arXiv150103813M}
{Moro-Mart{\'{\i}}n}, A., {Marshall}, J.~P., {Kennedy}, G., {et~al.} 2015,
  ArXiv e-prints

\bibitem[{{Mortier} {et~al.}(2013){Mortier}, {Santos}, {Sousa}, {Fernandes},
  {Adibekyan}, {Delgado Mena}, {Montalto}, \&
  {Israelian}}]{2013A&A...558A.106M}
{Mortier}, A., {Santos}, N.~C., {Sousa}, S.~G., {et~al.} 2013, \aap, 558, A106

\bibitem[{{Neves} {et~al.}(2009){Neves}, {Santos}, {Sousa}, {Correia}, \&
  {Israelian}}]{2009A&A...497..563N}
{Neves}, V., {Santos}, N.~C., {Sousa}, S.~G., {Correia}, A.~C.~M., \&
  {Israelian}, G. 2009, \aap, 497, 563

\bibitem[{{Perryman} \& {ESA}(1997)}]{1997ESASP1200.....P}
{Perryman}, M.~A.~C. \& {ESA}, eds. 1997, ESA Special Publication, Vol. 1200,
  {The HIPPARCOS and TYCHO catalogues. Astrometric and photometric star
  catalogues derived from the ESA HIPPARCOS Space Astrometry Mission}

\bibitem[{{Pfeiffer} {et~al.}(1998){Pfeiffer}, {Frank}, {Baumueller},
  {Fuhrmann}, \& {Gehren}}]{foces}
{Pfeiffer}, M.~J., {Frank}, C., {Baumueller}, D., {Fuhrmann}, K., \& {Gehren},
  T. 1998, \aaps, 130, 381

\bibitem[{{Piskunov} {et~al.}(1995){Piskunov}, {Kupka}, {Ryabchikova}, {Weiss},
  \& {Jeffery}}]{1995A&AS..112..525P}
{Piskunov}, N.~E., {Kupka}, F., {Ryabchikova}, T.~A., {Weiss}, W.~W., \&
  {Jeffery}, C.~S. 1995, \aaps, 112, 525

\bibitem[{{Pollack} {et~al.}(1996){Pollack}, {Hubickyj}, {Bodenheimer},
  {Lissauer}, {Podolak}, \& {Greenzweig}}]{1996Icar..124...62P}
{Pollack}, J.~B., {Hubickyj}, O., {Bodenheimer}, P., {et~al.} 1996, \icarus,
  124, 62

\bibitem[{{Queloz} {et~al.}(2000){Queloz}, {Mayor}, {Weber}, {Bl{\'e}cha},
  {Burnet}, {Confino}, {Naef}, {Pepe}, {Santos}, \&
  {Udry}}]{2000A&A...354...99Q}
{Queloz}, D., {Mayor}, M., {Weber}, L., {et~al.} 2000, \aap, 354, 99

\bibitem[{{Ram{\'{\i}}rez} {et~al.}(2010){Ram{\'{\i}}rez}, {Asplund},
  {Baumann}, {Mel{\'e}ndez}, \& {Bensby}}]{2010A&A...521A..33R}
{Ram{\'{\i}}rez}, I., {Asplund}, M., {Baumann}, P., {Mel{\'e}ndez}, J., \&
  {Bensby}, T. 2010, \aap, 521, A33

\bibitem[{{Ram{\'{\i}}rez} {et~al.}(2009){Ram{\'{\i}}rez}, {Mel{\'e}ndez}, \&
  {Asplund}}]{2009A&A...508L..17R}
{Ram{\'{\i}}rez}, I., {Mel{\'e}ndez}, J., \& {Asplund}, M. 2009, \aap, 508, L17

\bibitem[{{Ram{\'{\i}}rez} {et~al.}(2014){Ram{\'{\i}}rez}, {Mel{\'e}ndez}, \&
  {Asplund}}]{2014A&A...561A...7R}
{Ram{\'{\i}}rez}, I., {Mel{\'e}ndez}, J., \& {Asplund}, M. 2014, \aap, 561, A7

\bibitem[{{Raskin} {et~al.}(2011){Raskin}, {van Winckel}, {Hensberge},
  {Jorissen}, {Lehmann}, {Waelkens}, {Avila}, {de Cuyper}, {Degroote},
  {Dubosson}, {Dumortier}, {Fr{\'e}mat}, {Laux}, {Michaud}, {Morren}, {Perez
  Padilla}, {Pessemier}, {Prins}, {Smolders}, {van Eck}, \&
  {Winkler}}]{2011A&A...526A..69R}
{Raskin}, G., {van Winckel}, H., {Hensberge}, H., {et~al.} 2011, \aap, 526, A69

\bibitem[{{Raymond} {et~al.}(2011){Raymond}, {Armitage}, {Moro-Mart{\'{\i}}n},
  {Booth}, {Wyatt}, {Armstrong}, {Mandell}, {Selsis}, \&
  {West}}]{2011A&A...530A..62R}
{Raymond}, S.~N., {Armitage}, P.~J., {Moro-Mart{\'{\i}}n}, A., {et~al.} 2011,
  \aap, 530, A62

\bibitem[{{Raymond} {et~al.}(2012){Raymond}, {Armitage}, {Moro-Mart{\'{\i}}n},
  {Booth}, {Wyatt}, {Armstrong}, {Mandell}, {Selsis}, \&
  {West}}]{2012A&A...541A..11R}
{Raymond}, S.~N., {Armitage}, P.~J., {Moro-Mart{\'{\i}}n}, A., {et~al.} 2012,
  \aap, 541, A11

\bibitem[{{Santos} {et~al.}(2004){Santos}, {Israelian}, \&
  {Mayor}}]{2004A&A...415.1153S}
{Santos}, N.~C., {Israelian}, G., \& {Mayor}, M. 2004, \aap, 415, 1153

\bibitem[{{Schneider} {et~al.}(2011){Schneider}, {Dedieu}, {Le Sidaner},
  {Savalle}, \& {Zolotukhin}}]{2011A&A...532A..79S}
{Schneider}, J., {Dedieu}, C., {Le Sidaner}, P., {Savalle}, R., \&
  {Zolotukhin}, I. 2011, \aap, 532, A79

\bibitem[{{Schuler} {et~al.}(2011){Schuler}, {Flateau}, {Cunha}, {King},
  {Ghezzi}, \& {Smith}}]{2011ApJ...732...55S}
{Schuler}, S.~C., {Flateau}, D., {Cunha}, K., {et~al.} 2011, \apj, 732, 55

\bibitem[{{Scott} {et~al.}(2014{\natexlab{a}}){Scott}, {Asplund}, {Grevesse},
  {Bergemann}, \& {Sauval}}]{2014arXiv1405.0287S}
{Scott}, P., {Asplund}, M., {Grevesse}, N., {Bergemann}, M., \& {Sauval}, A.~J.
  2014{\natexlab{a}}, ArXiv e-prints

\bibitem[{{Scott} {et~al.}(2014{\natexlab{b}}){Scott}, {Grevesse}, {Asplund},
  {Sauval}, {Lind}, {Takeda}, {Collet}, {Trampedach}, \&
  {Hayek}}]{2014arXiv1405.0279S}
{Scott}, P., {Grevesse}, N., {Asplund}, M., {et~al.} 2014{\natexlab{b}}, ArXiv
  e-prints

\bibitem[{{Sneden}(1973)}]{1973PhDT.......180S}
{Sneden}, C.~A. 1973, PhD thesis, THE UNIVERSITY OF TEXAS AT AUSTIN.

\bibitem[{{Sousa} {et~al.}(2011{\natexlab{a}}){Sousa}, {Santos}, {Israelian},
  {Lovis}, {Mayor}, {Silva}, \& {Udry}}]{2011A&A...526A..99S}
{Sousa}, S.~G., {Santos}, N.~C., {Israelian}, G., {et~al.} 2011{\natexlab{a}},
  \aap, 526, A99

\bibitem[{{Sousa} {et~al.}(2007){Sousa}, {Santos}, {Israelian}, {Mayor}, \&
  {Monteiro}}]{2007A&A...469..783S}
{Sousa}, S.~G., {Santos}, N.~C., {Israelian}, G., {Mayor}, M., \& {Monteiro},
  M.~J.~P.~F.~G. 2007, \aap, 469, 783

\bibitem[{{Sousa} {et~al.}(2011{\natexlab{b}}){Sousa}, {Santos}, {Israelian},
  {Mayor}, \& {Udry}}]{2011A&A...533A.141S}
{Sousa}, S.~G., {Santos}, N.~C., {Israelian}, G., {Mayor}, M., \& {Udry}, S.
  2011{\natexlab{b}}, \aap, 533, A141

\bibitem[{{Sousa} {et~al.}(2008){Sousa}, {Santos}, {Mayor}, {Udry},
  {Casagrande}, {Israelian}, {Pepe}, {Queloz}, \&
  {Monteiro}}]{2008A&A...487..373S}
{Sousa}, S.~G., {Santos}, N.~C., {Mayor}, M., {et~al.} 2008, \aap, 487, 373

\bibitem[{{Sozzetti}(2004)}]{2004MNRAS.354.1194S}
{Sozzetti}, A. 2004, \mnras, 354, 1194

\bibitem[{{Sozzetti} {et~al.}(2007){Sozzetti}, {Torres}, {Charbonneau},
  {Latham}, {Holman}, {Winn}, {Laird}, \& {O'Donovan}}]{2007ApJ...664.1190S}
{Sozzetti}, A., {Torres}, G., {Charbonneau}, D., {et~al.} 2007, \apj, 664, 1190

\bibitem[{{Su} {et~al.}(2006){Su}, {Rieke}, {Stansberry}, {Bryden},
  {Stapelfeldt}, {Trilling}, {Muzerolle}, {Beichman}, {Moro-Martin}, {Hines},
  \& {Werner}}]{2006ApJ...653..675S}
{Su}, K.~Y.~L., {Rieke}, G.~H., {Stansberry}, J.~A., {et~al.} 2006, \apj, 653,
  675

\bibitem[{{Takeda}(2003)}]{2003A&A...402..343T}
{Takeda}, Y. 2003, \aap, 402, 343

\bibitem[{{Takeda} \& {Honda}(2005)}]{2005PASJ...57...65T}
{Takeda}, Y. \& {Honda}, S. 2005, \pasj, 57, 65

\bibitem[{{Takeda} {et~al.}(2002{\natexlab{a}}){Takeda}, {Ohkubo}, \&
  {Sadakane}}]{2002PASJ...54..451T}
{Takeda}, Y., {Ohkubo}, M., \& {Sadakane}, K. 2002{\natexlab{a}}, \pasj, 54,
  451

\bibitem[{{Takeda} {et~al.}(2005){Takeda}, {Ohkubo}, {Sato}, {Kambe}, \&
  {Sadakane}}]{2005PASJ...57...27T}
{Takeda}, Y., {Ohkubo}, M., {Sato}, B., {Kambe}, E., \& {Sadakane}, K. 2005,
  \pasj, 57, 27

\bibitem[{{Takeda} {et~al.}(2002{\natexlab{b}}){Takeda}, {Sato}, {Kambe},
  {Sadakane}, \& {Ohkubo}}]{2002PASJ...54.1041T}
{Takeda}, Y., {Sato}, B., {Kambe}, E., {Sadakane}, K., \& {Ohkubo}, M.
  2002{\natexlab{b}}, \pasj, 54, 1041

\bibitem[{{Takeda} {et~al.}(2008){Takeda}, {Sato}, \&
  {Murata}}]{2008PASJ...60..781T}
{Takeda}, Y., {Sato}, B., \& {Murata}, D. 2008, \pasj, 60, 781

\bibitem[{{Torres} {et~al.}(2012){Torres}, {Fischer}, {Sozzetti}, {Buchhave},
  {Winn}, {Holman}, \& {Carter}}]{2012ApJ...757..161T}
{Torres}, G., {Fischer}, D.~A., {Sozzetti}, A., {et~al.} 2012, \apj, 757, 161

\bibitem[{{Trilling} {et~al.}(2008){Trilling}, {Bryden}, {Beichman}, {Rieke},
  {Su}, {Stansberry}, {Blaylock}, {Stapelfeldt}, {Beeman}, \&
  {Haller}}]{Trilling08}
{Trilling}, D.~E., {Bryden}, G., {Beichman}, C.~A., {et~al.} 2008, \apj, 674,
  1086

\bibitem[{{Tsantaki} {et~al.}(2013){Tsantaki}, {Sousa}, {Adibekyan}, {Santos},
  {Mortier}, \& {Israelian}}]{2013A&A...555A.150T}
{Tsantaki}, M., {Sousa}, S.~G., {Adibekyan}, V.~Z., {et~al.} 2013, \aap, 555,
  A150

\bibitem[{van Leeuwen(2007)}]{Leeuwen}
van Leeuwen, F.~v. 2007, Hipparcos, the New Reduction of the Raw Data (XXXII,
  449 p., Hardcover, ISBN: 978-1-4020-6341-1: Astrophysics and Space Science
  Library , Vol. 350)

\bibitem[{{Villaver} \& {Livio}(2009)}]{2009ApJ...705L..81V}
{Villaver}, E. \& {Livio}, M. 2009, \apjl, 705, L81

\bibitem[{{Villaver} {et~al.}(2014){Villaver}, {Livio}, {Mustill}, \&
  {Siess}}]{2014ApJ...794....3V}
{Villaver}, E., {Livio}, M., {Mustill}, A.~J., \& {Siess}, L. 2014, \apj, 794,
  3

\bibitem[{{Wright} {et~al.}(2011){Wright}, {Fakhouri}, {Marcy}, {Han}, {Feng},
  {Johnson}, {Howard}, {Fischer}, {Valenti}, {Anderson}, \&
  {Piskunov}}]{2011PASP..123..412W}
{Wright}, J.~T., {Fakhouri}, O., {Marcy}, G.~W., {et~al.} 2011, \pasp, 123, 412

\bibitem[{{Wright} {et~al.}(2009){Wright}, {Upadhyay}, {Marcy}, {Fischer},
  {Ford}, \& {Johnson}}]{2009ApJ...693.1084W}
{Wright}, J.~T., {Upadhyay}, S., {Marcy}, G.~W., {et~al.} 2009, \apj, 693, 1084

\bibitem[{{Wyatt} {et~al.}(2012){Wyatt}, {Kennedy}, {Sibthorpe},
  {Moro-Mart{\'{\i}}n}, {Lestrade}, {Ivison}, {Matthews}, {Udry}, {Greaves},
  {Kalas}, {Lawler}, {Su}, {Rieke}, {Booth}, {Bryden}, {Horner}, {Kavelaars},
  \& {Wilner}}]{2012MNRAS.424.1206W}
{Wyatt}, M.~C., {Kennedy}, G., {Sibthorpe}, B., {et~al.} 2012, \mnras, 424,
  1206

\end{thebibliography}

\begin{appendix}

\section{Data from the ESO Science Archive Facility}
\label{apendiceA}

 To get full credits to data used in this paper coming from the ESO/ST-ECF
 Science Archive Facility \footnote{http://archive.eso.org/cms/},
 and the pipeline processed FEROS and HARPS data archive
 \footnote{http://archive.eso.org/wdb/wdb/eso/repro/form},
 the corresponding ESO programme IDs are listed in Table~\ref{eso_data_ids}.


\begin{table}
\centering
\caption{ESO/ST-ECF
 Science Archive Facility data used in this work}
\label{eso_data_ids}
\begin{scriptsize}
\begin{tabular}{lr|lr|lr}
\hline\noalign{\smallskip}
HIP &   OBS PROG ID   & HIP &  OBS PROG ID &  HIP &  OBS PROG ID \\
\hline 
490	&	079.A-9017(A)	&	33690	&       078.A-9059(A)	&	95149	&	083.A-9013(A)	\\
522	&	083.A-9011(B)	&	34065	&	079.C-0681(A)	&	97546	&	083.A-9013(A)	\\
910	&	184.C-0815(F)	&	37853	&	076.D-0103(A)	&	97675	&	079.A-9014(A)	\\
3170	&	072.C-0488(E)	&	39903	&	184.C-0815(C)	&	98959	&	072.C-0488(E)	\\
3185	&	077.C-0192(A)	&	42282	&	084.A-9004(B)	&	101983	&	 60.A-9036(A)	\\
3497	&	072.C-0488(E)	&	52462	&	079.A-9007(A)	&	103389	&	083.A-9013(A)	\\
3559	&	083.A-9011(B)	&	53252	&	184.C-0815(F)	&	105312	&	073.A-9008(A)	\\
5862	&	074.C-0135(A)	&	57507	&	072.C-0488(E)	&	105388	&	079.A-9007(A)	\\
7978	&	072.A-9006(A)	&	63033	&	072.D-0707(A)	&	106696	&	083.A-9011(B)	\\
8486	&	083.A-9011(B)	&	64408	&	080.D-2002(A)	&	107022	&	072.C-0488(E)	\\
10306	&	184.C-0815(F)	&	67275	&	083.A-9003(A)	&	107350	&	082.C-0446(A)	\\
11072	&	074.C-0037(A)	&	69090	&	072.C-0488(E)	&	107649	&	 60.A-9122(B)	\\
19855	&	083.A-9011(A)	&	72567	&	079.A-9009(A)	&	109422	&	074.D-0008(B)	\\
23816	&	087.C-0831(A)	&	77372	&	072.C-0488(E)	&	109821	&	072.C-0488(E)	\\
27887	&	079.A-9007(A)	&	86796	&	083.A-9013(A)	&	113044	&	083.A-9011(B)	\\
29568	&	078.A-9059(A)	&	89042	&	072.C-0033(A)	&	114236	&	083.A-9011(B)	\\
30503	&	 60.A-9036(A)	&	90485	&	083.A-9013(A)	&	114948	&	083.A-9013(A)	\\
31711	&	078.C-0378(A)	&	93858	&	072.C-0488(E) 	&	116250	&	072.C-0488(E)	\\
32970	&	084.A-9004(B)	&	94050	&	083.A-9013(A)	&	116745	&	076.B-0416(A)	\\
32984	&	079.A-9007(A)	&	94858	&	083.A-9013(A)	&	116906	&	072.C-0488(E)	\\
\hline
\end{tabular}
\end{scriptsize}
\end{table}

\section{The metallicity distribution of stars with cool/hot Jupiters}
\label{apendiceB}

\begin{figure*}
\centering
\begin{minipage}{0.48\linewidth}
\includegraphics[angle=270,scale=0.45]{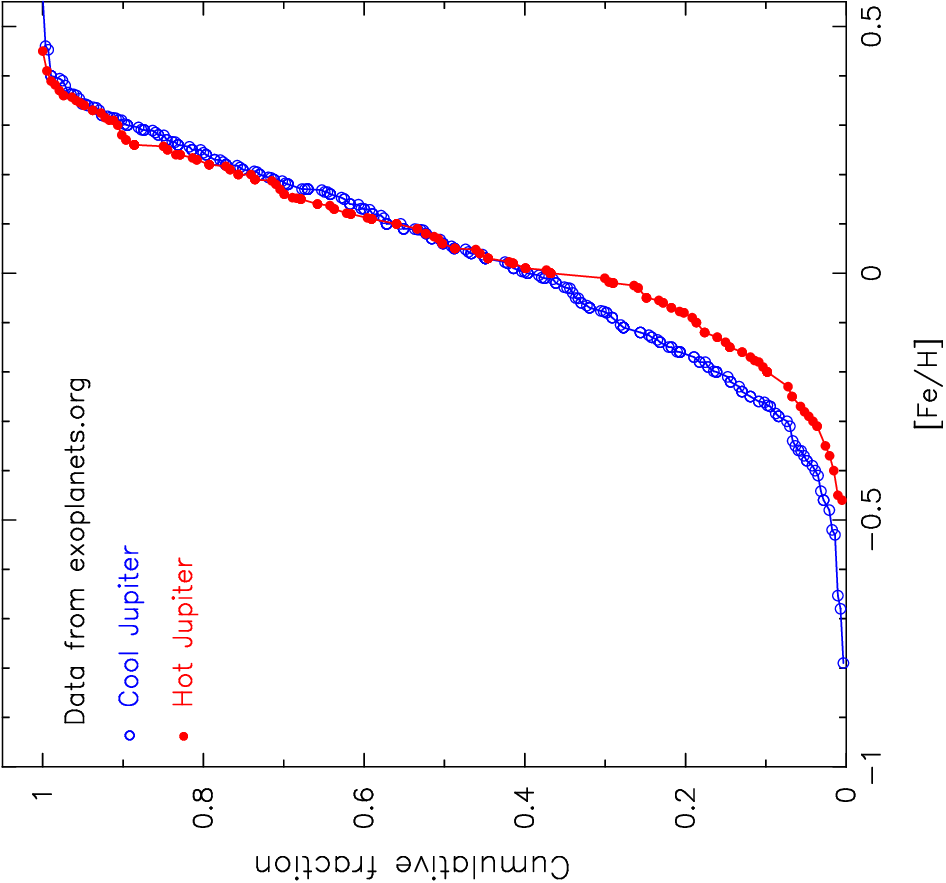}
\end{minipage}
\begin{minipage}{0.48\linewidth}
\includegraphics[angle=270,scale=0.45]{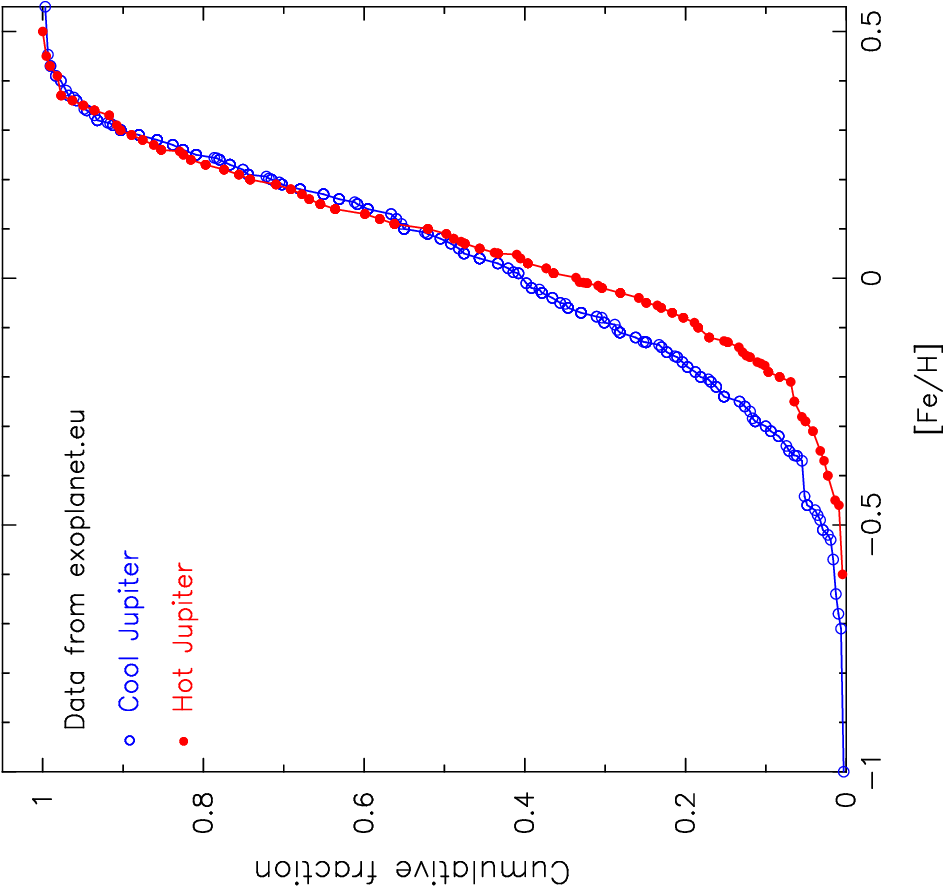}
\end{minipage}
\caption{Histogram of [Fe/H] cumulative frequencies for all stars with cool (blue) and
hot (red) Jupiter planet listed in exoplanets.org (left) and exoplanet.eu (right).}
\label{metallicity_hot_cool}
\end{figure*}

  As seen in Section~\ref{analysis_metallicity_distributions} we find that
  stars hosting close-in hot Jupiters tend to show higher metallicities than stars
  hosting more distant planets. Given the low number of stars considered in this work
  we performed an additional check by considering the metallicity distribution of
  all stars known to harbour planets as listed\footnote{Up to June 14, 2014} in 
  the Extrasolar Planets Encyclopaedia \citep{2011A&A...532A..79S} and
  the Exoplanet Orbit Database \citep{2011PASP..123..412W} databases. 
  All planet hosts with available values of planetary mass, semmimajor axis, and
  metallicity were considered. Stars with low-mass planets
  (M$_{\rm P}$$\sin i$ $<$ 30 M$_{\oplus}$) were discarded. No further selection
  criteria were applied. Stars with multiple planets are classified as ``hot'' if
  at least one of the planets has a semmimajor axis lower than 0.1 au. 
  The resulting metallicity distributions are  
  shown in Figure~\ref{metallicity_hot_cool}, while some statistic diagnostics
  are given in Table~\ref{metal_cool_hot}.

   
  We can see from the figure that at high metallicities (greater than +0.0/+0.1 dex)
  the metallicity distribution of stars hosting cool and star hosting hot planets
  are nearly the same. However, they differ at low-metallicities, being the distribution
  of hot Jupiters slightly shifted towards higher metallicities. We also note that
  there are no hot Jupiters harbouring stars with metallicities below -0.50/-0.60 dex,
  while cool Jupiters can be found around stars as metal-poor as -1.00 dex.
  A K-S test shows that we can not rule out the possibility of both distributions
  (hot/cool stellar hosts) being drawn from the same parent population. However,
  the derived $p$-values are considerably low, only 0.15 when data from exoplanets.org
  is considered (n$_{\rm eff}$ $\sim$ 115, $D$ $\sim$ 0.11) and 0.06 if
  the data from exoplanet.eu is employed (n$_{\rm eff}$ $\sim$ 127, $D$ $\sim$ 0.12).


\begin{table}
\centering
\caption{[Fe/H] statistics of cool/hot Jupiter's host stellar samples.}
\label{metal_cool_hot}
\begin{scriptsize}
\begin{tabular}{lcccccc}
\hline\noalign{\smallskip}
\multicolumn{7}{c}{\bf Data from exoplanets.org}\\
$Sample$  &  $Mean$ & $ Median$ & $Deviation$  &  $Min$&  $Max$ & $N$ \\
\hline\noalign{\smallskip}
 Cool Jupiters    &  +0.04  & +0.06  & 0.23 & -0.79 & +0.56 &  285 \\
 Hot Jupiters     &  +0.06  & +0.06  & 0.18 & -0.46 & +0.45 &  193 \\
\hline\noalign{\smallskip}
\multicolumn{7}{c}{\bf Data from exoplanet.eu}\\
$Sample$  &  $Mean$ & $ Median$ & $Deviation$  &  $Min$&  $Max$ & $N$ \\
\hline\noalign{\smallskip}
 Cool Jupiters    &  +0.03  & +0.08  & 0.25 & -1.00 & +0.56 & 309 \\
 Hot Jupiters     &  +0.07  & +0.10  & 0.19 & -0.60 & +0.50 & 217 \\
\noalign{\smallskip}\hline\noalign{\smallskip}
\end{tabular}
\end{scriptsize}
\end{table}

  These results are in agreement with previous works 
  (see references in Section~\ref{analysis_metallicity_distributions})
  which point towards a paucity of short period planets around  
  metal poor stars. While this trend could in principle suggest a 
  metallicity dependency of migration rates, further monitoring
  of metal-poor stars are required to confirm it 
  \citep{2004MNRAS.354.1194S}.

\section{Abundance ratios as a function of the stellar metallicity}
\label{apendiceC}

\begin{figure*}
\centering
\includegraphics[angle=270,scale=0.45]{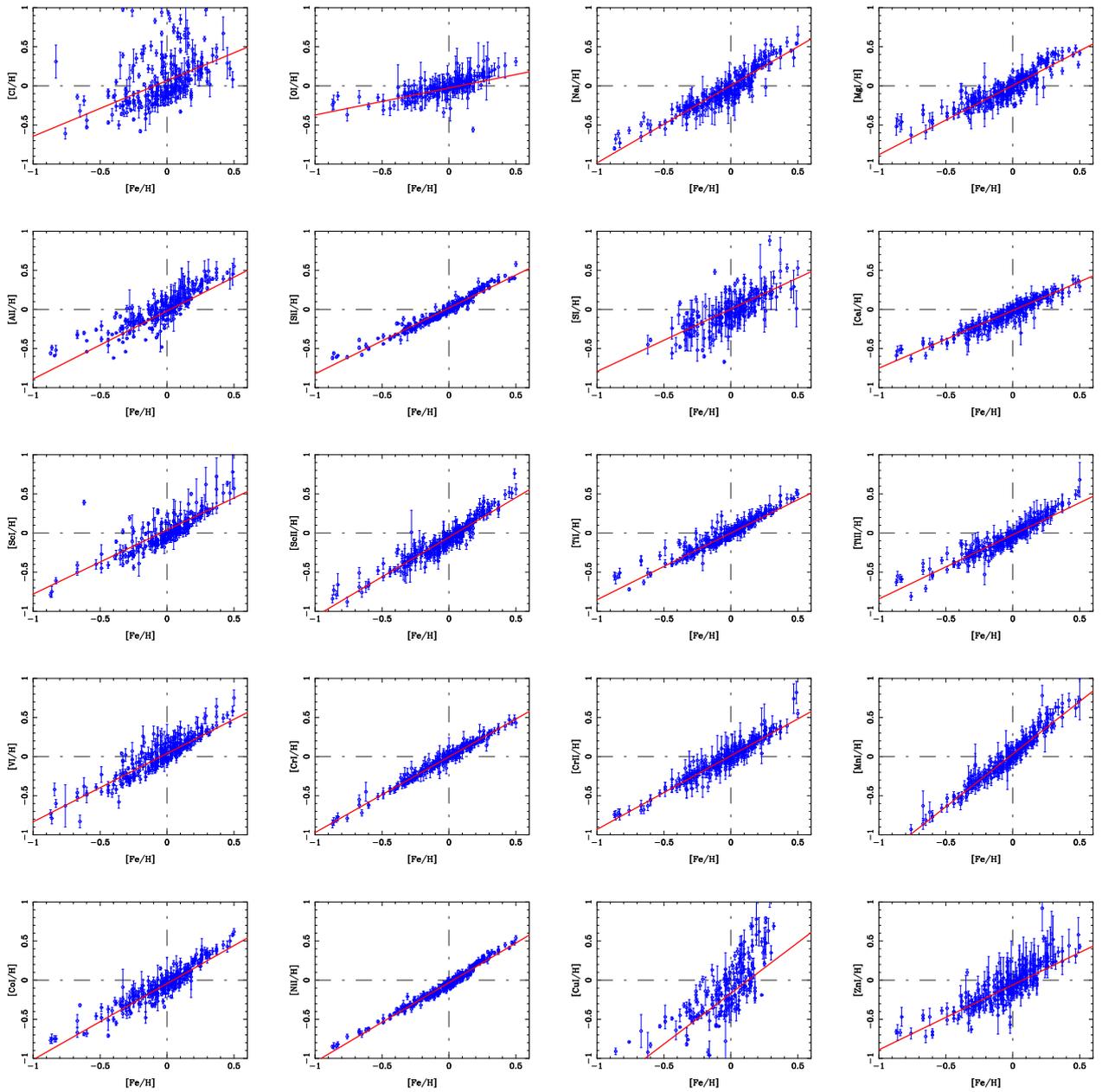}
\caption{Chemical abundance ratios of [X/H] as a function of the stellar metallicity.
 The red line shows the best linear fit.} 
\label{swds_swods_xh_vs_feh}
\end{figure*}

\begin{figure*}
\centering
\includegraphics[angle=270,scale=0.45]{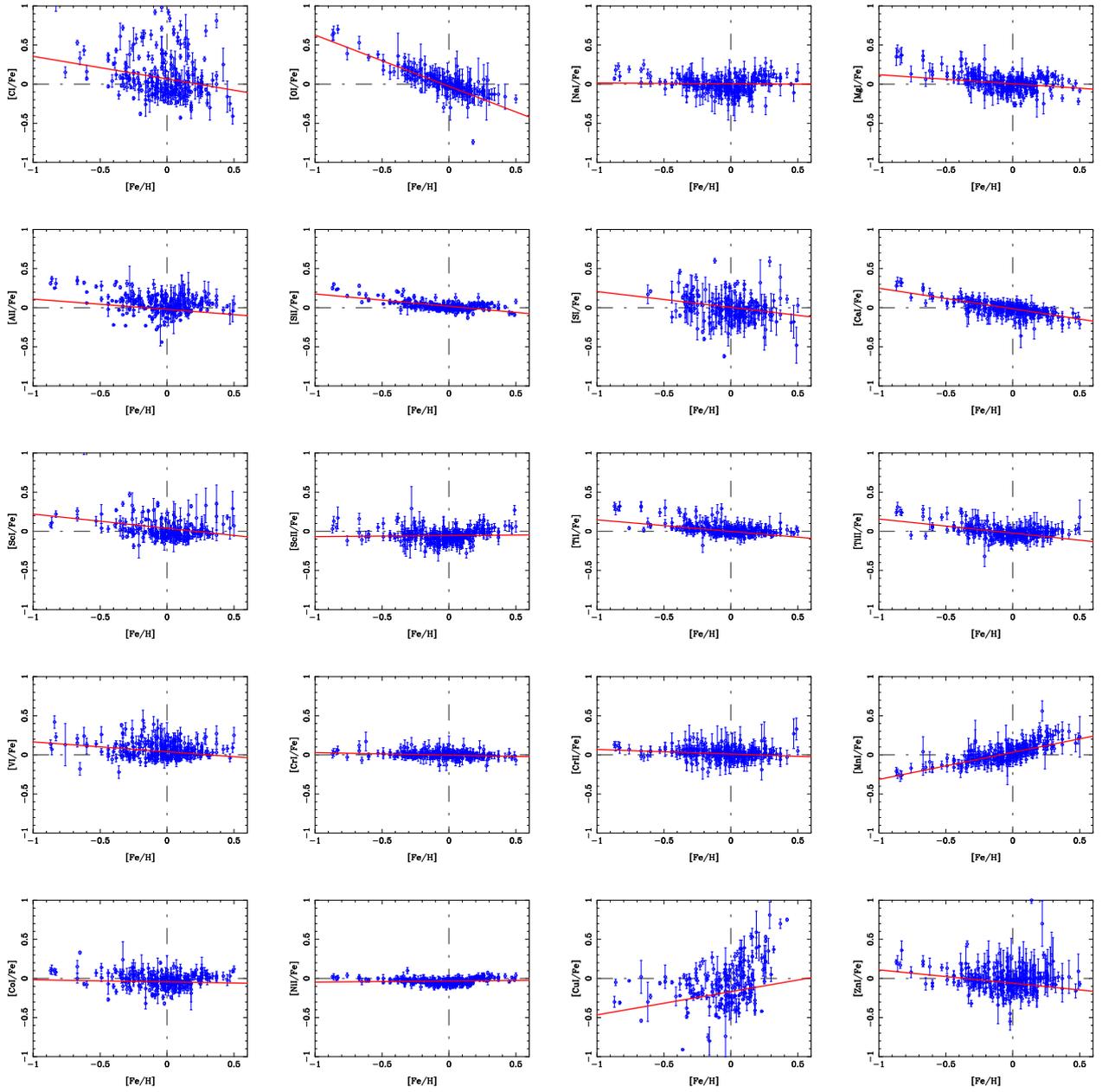}
\caption{Chemical abundance ratios of [X/Fe] as a function of the stellar metallicity.
 The red line shows the best linear fit.}
\label{swds_swods_xfe_vs_feh}
\end{figure*}

\end{appendix}
\Online
\section*{Online material}
\label{tables}

  Results produced in the framework of this work 
  are only available 
  in the electronic version of the corresponding paper or
  at the CDS via anonymous ftp to cdsarc.u-strasbg.fr (130.79.128.5)
  or via {\tt http://cdsweb.u-strasbg.fr/cgi-bin/qcat?J/A+A/}

  Table~\ref{parameters_table_full} lists all the stars analyzed in this work,
  classified according to the presence/absence of discs and/or planets.
  The table provides: 
  HIP number (column 1); HD number (column 2);
  effective temperature in kelvin (column 3); logarithm of the surface gravity in cms$^{\rm -2}$ (column 4);
  microturbulent velocity in kms$^{\rm -1}$ (column 5); final metallicity in dex (column 6);
  mean iron abundance derived from Fe~{\sc I} lines (column 7) in the usual scale
  ($ A(Fe) = \log[(N_{Fe}/N_{H}) + 12$]); number of Fe~{\sc I} lines used (column 8);
  mean iron abundance derived from Fe~{\sc II} lines (column 9);
  number of Fe~{\sc II} lines used (column 10); and spectrograph (column 11).
  Each measured quantity is accompanied by its corresponding uncertainty.

  Table~\ref{lista_unificada_lines_full}
  provides the wavelength, excitation potential (EP), and oscillator strength $\log (gf)$
  for the lines selected in the present work.
  References are also given. Data for HFS computations is from 
  \cite{2014A&A...561A...7R} and is not included in this list.

  Table~\ref{abundance_table_full}   gives the
  abundances of C~{\sc I}, O~{\sc I} (nLTE corrected), Na~{\sc I}, Mg~{\sc I}, Al~{\sc I}, Si~{\sc I},
  S~{\sc I}, Ca~{\sc I}, Sc~{\sc I}, Sc~{\sc II}, Ti~{\sc I}, Ti~{\sc II},
  V~{\sc I} (HFS taken into account), Cr~{\sc I},  Cr~{\sc II}
  Mn~{\sc I}, Co~{\sc I} (HFS taken into account) , Ni~{\sc I}, Cu~{\sc I} (HFS considered,
  and Zn~{\sc I} 
  They are expressed relative to the
  solar value, i.e
  $[X/H]=\log(N_{X}/N_{H}) - \log(N_{X}/N_{H})_{\odot}$.
  For each star abundances are given in the first row, whilst uncertainties
  are given in the second row.

\setcounter{table}{1}
\onllongtab{
\begin{longtable}{llcccccccccl}
\caption{
 Spectroscopic parameters with uncertainties
 for the stars measured in this work. Columns 7 and 9
 give the mean iron abundance derived from Fe~{\sc I} and Fe~{\sc  II} lines,
 respectively, while columns 8 and 10 give the corresponding number of lines.
 The rest of the columns are self-explanatory. 
  }\label{parameters_table_full}\\
\hline
\hline
HIP &  HD &  T$_{\rm eff}$ &  $\log g$      & $\xi_{t}$     & [Fe/H] &  $\left\langle A({\rm Fe}~{\rm I}) \right\rangle$ & n$_{\rm I}$ &
 $\left\langle A({\rm Fe}~{\rm II}) \right\rangle$  & n$_{\rm II}$ &  Spec.$^{\dag}$ \\
    &     &   (K)   & (${\rm cm s^{-2}}$)  & (${\rm km s^{-1}}$) & dex  &      &      &      &      &       \\
 (1)& (2) &  (3)    & (4)                  & (5)                 & (6)  & (7)  & (8)  & (9)  & (10) & (11)  \\
\hline
\endfirsthead
\caption{Continued.} \\
\hline
HIP &  HD &  T$_{\rm eff}$ &  $\log g$      & $\xi_{t}$     & [Fe/H] &  $\left\langle A({\rm Fe}~{\rm I}) \right\rangle$ & n$_{\rm I}$ &
 $\left\langle A({\rm Fe}~{\rm II}) \right\rangle$  & n$_{\rm II}$ &  Spec.$^{\dag}$ \\
    &     &   (K)   & (${\rm cm s^{-2}}$)  & (${\rm km s^{-1}}$) & dex  &      &      &      &      &       \\
 (1)& (2) &  (3)    & (4)                  & (5)                 & (6)  & (7)  & (8)  & (9)  & (10) & (11)  \\
\hline
\endhead
\hline
\endfoot
\hline
\endlastfoot
\multicolumn{2}{c}{Sun} & 5784 $\pm$ 15 & 4.51 $\pm$ 0.03 & 1.01 $\pm$ 0.09 &  0.02 $\pm$ 0.01  & 7.52 $\pm$ 0.02 & 253 & 7.52 $\pm$ 0.02 & 26 & 5 \\
\hline\noalign{\smallskip}
\multicolumn{11}{c}{Stars with known debris discs}\\
\hline\noalign{\smallskip}
171	&	224930	&	5354	$\pm$	15	&	4.32	$\pm$	0.04	&	0.74	$\pm$	0.12	&	-0.83	$\pm$	0.01	&	6.67	$\pm$	0.02	&	208	&	6.67	$\pm$	0.02	&	18	&	5	\\
490	&	105	&	5967	$\pm$	35	&	4.52	$\pm$	0.08	&	1.76	$\pm$	0.26	&	-0.16	$\pm$	0.03	&	7.35	$\pm$	0.04	&	151	&	7.34	$\pm$	0.05	&	19	&	7	\\
544	&	166	&	5584	$\pm$	20	&	4.73	$\pm$	0.05	&	1.22	$\pm$	0.16	&	0.15	$\pm$	0.02	&	7.65	$\pm$	0.03	&	256	&	7.65	$\pm$	0.03	&	21	&	5	\\
1598	&	1562	&	5768	$\pm$	20	&	4.56	$\pm$	0.04	&	1.12	$\pm$	0.13	&	-0.27	$\pm$	0.02	&	7.23	$\pm$	0.02	&	205	&	7.23	$\pm$	0.03	&	17	&	1	\\
1599	&	1581	&	5877	$\pm$	20	&	4.25	$\pm$	0.04	&	1.13	$\pm$	0.14	&	-0.24	$\pm$	0.02	&	7.26	$\pm$	0.02	&	221	&	7.26	$\pm$	0.03	&	23	&	6	\\
4148	&	5133	&	4962	$\pm$	23	&	4.76	$\pm$	0.06	&	0.80	$\pm$	0.25	&	-0.12	$\pm$	0.03	&	7.38	$\pm$	0.03	&	263	&	7.38	$\pm$	0.04	&	18	&	6	\\
\end{longtable}
\tablefoot{$^{\dag}$Spectrograph: {\bf(1)} CAHA/FOCES; {\bf(2)} TNG/SARG; {\bf(3)} NOT/FIES; {\bf(4)} MERCATOR/HERMES; {\bf(5)} S$^{4}$N-McD;
{\bf(6)} S$^{4}$N-FEROS; {\bf(7)} ESO/FEROS; {\bf(8)} ESO/HARPS\\
$^{\ddag}$ It also hosts cool and hot Jupiters planets.\\
$^{\star}$ It also hosts cool Jupiters planets.}
}

\onllongtab{
\begin{longtable}{lcccc}
\caption{
  Wavelength, excitation potential (EP), and oscillator strength $\log (gf)$
  for the lines selected in the present work.
  }\label{lista_unificada_lines_full}\\
\hline
\hline
 Ion      &   Wavelength (\AA) & EP (eV)  &  $\log (gf)$ & Ref. \\
\hline
\endfirsthead
\caption{Continued.} \\
\hline
 Ion      &   Wavelength (\AA) & EP (eV)  &  $\log (gf)$ & Ref. \\
\hline
\endhead
\hline
\endfoot
\hline
\endlastfoot
C~{\sc i}	& 6587.61	 & 8.54	 & -1.021	 & RA14	\\
C~{\sc i}	& 7111.47	 & 8.64	 & -1.074	 & RA14	\\
C~{\sc i}	& 7113.18	 & 8.65	 & -0.762	 & RA14	\\
C~{\sc i}	& 7115.17	 & 8.65	 & -0.930	 & RA14	\\
C~{\sc i}	& 7116.96	 & 8.65	 & -0.910	 & RA14	\\
O~{\sc i}	& 7771.94	 & 9.15	 &  0.352	 & RA14	\\
O~{\sc i}	& 7774.16	 & 9.15	 &  0.223	 & RA14	\\
O~{\sc i}	& 7775.39	 & 9.15	 &  0.002	 & RA14	\\
\end{longtable}
\tablefoot{RA14: \cite{2014A&A...561A...7R}, NE09: \cite{2009A&A...497..563N}
 VALD: \cite{1995A&AS..112..525P,1999A&AS..138..119K},    
 TA05: \cite{2005PASJ...57...65T} }
}

\setcounter{table}{10}

\onllongtab{
\begin{longtab}
\begin{landscape}
\begin{longtable}{llrrrrrrrrrrrrrrrrrrrr}
\caption{\label{abundance_table_full} Derived abundances [X/H]}\\
\hline\hline
 HIP	& Sample & C~{\sc i}  & O~{\sc i}  & Na~{\sc i} & Mg~{\sc i} & Al~{\sc i} &
                   Si~{\sc i} & S~{\sc i}  & Ca~{\sc i} & Sc~{\sc i} & Sc~{\sc ii} &
                   Ti~{\sc i} & Ti~{\sc ii} & V~{\sc i} & Cr~{\sc i} & Cr~{\sc ii} & 
                   Mn~{\sc i} & Co~{\sc i}  & Ni~{\sc i}  &Cu~{\sc i} & Zn~{\sc i}\\
\hline
\endfirsthead
\caption{continued.}\\
\hline\hline
 HIP    & Sample & C~{\sc i}  & O~{\sc i}  & Na~{\sc i} & Mg~{\sc i} & Al~{\sc i} &
                   Si~{\sc i} & S~{\sc i}  & Ca~{\sc i} & Sc~{\sc i} & Sc~{\sc ii} &
                   Ti~{\sc i} & Ti~{\sc ii} & V~{\sc i} & Cr~{\sc i} & Cr~{\sc ii} &
                   Mn~{\sc i} & Co~{\sc i}  & Ni~{\sc i}  &Cu~{\sc i} & Zn~{\sc i}\\
\hline
\endhead
\hline
\endfoot
171 &     SWDs &       0.31 &    -0.13 &   -0.73 &   -0.46 &   -0.52 &   -0.59 &          &  -0.51 &   -0.61 &   -0.66 &   -0.51 &   -0.59 &   -0.60 &   -0.77 &   -0.74 &   -1.09 &   -0.75 &   -0.82 &   -1.14 &   -0.47 \\ 
              &                &  0.21 &           0.05 &           0.06 &           0.10 &           0.06 &           0.01 &                       &    0.05 &           0.04 &           0.14 &           0.06 &           0.04 &           0.05 &           0.05 &           0.07 &           0.05 &           0.04 &           0.03 &           0.02 &           0.12 \\         
490 &     SWDs &       -0.13 &   -0.04 &   -0.17 &   -0.28 &          &  -0.13 &   -0.08 &   -0.17 &   -0.01 &   -0.26 &   -0.16 &   -0.26 &   -0.20 &   -0.15 &   -0.27 &   -0.07 &          &  -0.22 &          &  -0.36 \\ 
              &                &  0.12 &           0.03 &           0.07 &           0.03 &                       &    0.03 &           0.22 &           0.06 &           0.02 &           0.06 &           0.06 &           0.08 &           0.11 &           0.07 &           0.05 &           0.05 &                       &    0.04 &                       &    0.05 \\         
522 &     SWDPs &      -0.14 &   -0.06 &   0.12 &    -0.05 &   0.17 &    0.06 &    0.00 &    0.01 &    0.14 &    -0.10 &   -0.07 &   -0.07 &   0.09 &    -0.02 &   -0.06 &   0.00 &           &  -0.02 &          &  -0.16 \\ 
              &                &  0.07 &           0.03 &           0.07 &           0.10 &           0.01 &           0.02 &           0.20 &           0.05 &           0.03 &           0.05 &           0.04 &           0.05 &           0.09 &           0.04 &           0.07 &           0.05 &                       &    0.03 &                       &    0.04 \\
\end{longtable}
\tablefoot{SWDs: stars with discs; SWODs: stars without known discs;
SWDPs: stars harbouring simultaneously debris discs and planets; Cool: stars hosting cool (a $>$ 0.1 au) 
gas-giant planets; Hot: stars hosting at least one close-in (a $<$ 0.1 au) gas-giant;
Low-mass: stars hosting at least one low-mas  (M$_{\rm p}$$\sin i$ $\lesssim$ 30M$_{\oplus}$)
planet. $^{\dag}$: Giant star.\\
$^{\ddag}$ It also hosts cool and hot Jupiters planets.\\
$^{\star}$ It also hosts cool Jupiters planets.
} 
\end{landscape}
\end{longtab}
}


\end{document}